\documentclass[a4paper,11pt]{article}
\pdfoutput=1 

\usepackage{jheppub} 

\usepackage[T1]{fontenc} 

\usepackage{amsmath}
\usepackage{dsfont}
\usepackage{slashed}
\usepackage{bbold}
\usepackage{psfrag}
\PassOptionsToPackage{caption=false}{subfig}
\usepackage{subfig}
\usepackage{multirow}
\usepackage{booktabs}
\usepackage{hyperref}
\usepackage{feynmp}
\usepackage{color}
\usepackage[normalem]{ulem} 

\usepackage[utf8]{inputenc}

\usepackage{enumitem}

\newcommand{\todo}[1]{{\color{red} \ifmmode\else[todo]\fi #1}}
\usepackage[usenames,dvipsnames]{xcolor}
     
\newcommand{\xe}{XENON1T }
\newcommand{\units}[1]{{\rm \,#1}}

\def\beq{\begin{equation}}
\def\eeq{\end{equation}}

\graphicspath{{figures/}}

\begin{document}
\title{Exploring New Physics with O(keV) Electron Recoils in Direct Detection Experiments}

\author[a]{Itay M. Bloch,}
\author[b]{Andrea Caputo,}
\author[c]{Rouven Essig,}
\author[d,e]{Diego Redigolo,}
\author[c]{Mukul Sholapurkar,}
\author[a]{Tomer Volansky}

\affiliation[a]{
School of Physics and Astronomy, Tel-Aviv University, Tel-Aviv 69978, Israel}
\affiliation[b]{Instituto de Fisica Corpuscular, Universidad de Valencia and CSIC,
Edificio Institutos Investigacion, Catedratico Jose Beltran 2, Paterna, 46980 Spain}
\affiliation[c]{C.N. Yang Institute for Theoretical Physics, Stony Brook University, Stony Brook, NY 11794}
\affiliation[d]{CERN, Theory Division, CH-1211 Geneva 23, Switzerland}
\affiliation[e]{INFN Sezione di Firenze, Via G. Sansone 1, I-50019 Sesto Fiorentino, Italy and Department of
Physics and Astronomy, University of Florence, Italy}

\emailAdd{itay.bloch.m@gmail.com}
\emailAdd{andrea.caputo@uv.es}
\emailAdd{rouven.essig@stonybrook.edu}
\emailAdd{d.redigolo@gmail.com}
\emailAdd{mukul.sholapurkar@stonybrook.edu}
\emailAdd{tomerv@post.tau.ac.il}

\date{\today}

\abstract{Motivated by the recent XENON1T results, we explore various new physics models that can be discovered through searches for electron recoils in  ${\cal O}({\rm keV})$-threshold direct-detection experiments. First, we consider the absorption of axion-like particles, dark photons, and scalars, either as dark matter relics or being produced directly in the Sun. In the latter case, we find that keV mass bosons produced in the Sun provide an adequate fit to the data but are excluded by stellar cooling constraints. We address this tension by introducing a novel Chameleon-like axion model, which can explain the excess while evading the stellar bounds. We find that absorption of bosonic dark matter provides a viable explanation for the excess only if the dark matter is a dark photon or an axion. In the latter case, photophobic axion couplings are necessary to avoid X-ray constraints. Second, we analyze models of dark matter-electron scattering to determine which models might explain the excess. Standard scattering of dark matter with electrons is generically in conflict with data from lower-threshold experiments. Momentum-dependent interactions with a heavy mediator can fit the data with dark matter mass heavier than a GeV but are generically in tension with collider constraints. Next, we consider dark matter consisting of two (or more) states that have a small mass splitting.  The exothermic (down)scattering of the heavier state to the lighter state can fit the data for keV mass splittings.  Finally, we consider a subcomponent of dark matter that is accelerated by scattering off cosmic rays, finding that dark matter interacting though an $\mathcal{O}$(100~keV)-mass mediator can fit the data. The cross sections required in this scenario are, however, typically challenged by complementary probes of the light mediator. Throughout our study, we implement an unbinned Monte Carlo analysis and use an improved energy reconstruction of the XENON1T events. 
\\[0.2cm]

\noindent
\textit{Preprint: YITP-SB-2020-17}}
\maketitle

\section{Introduction}\label{sec:intro}

The quest to identify the particle nature of dark matter (DM) by detecting DM in terrestrial experiments has been ongoing for more than three decades.  
Despite numerous searches at direct-detection, indirect-detection, and collider experiments, no convincing signal for DM 
has been found to date.  
Given the profound implications for our understanding of the DM particle's properties if we were to find it in the laboratory, any claim for a possible DM signal in one of these experiments deserves to be studied carefully. 

The \xe collaboration has recently observed an unexplained excess of electronic recoil events with an energy of $\mathcal{O}$(keV)~\cite{Aprile:2020tmw}.  
While the most likely explanation is a neglected background source or a statistical fluctuation, the possibility that the excess could be the 
first sign of new physics (not necessarily even a sign of DM) is intriguing.  
The excess of events does not appear in the traditional search for nuclear recoils from elastic DM-nucleus scattering.  
Rather, it appears as an excess in a search for electron recoils (ER).  
The \xe search has an exposure of  $2.36\times 10^5 \units{kg-day}$ in the $1-30\units{keV}$ energy range. The background rate is reported to be $76\pm2\units{events / tonne\, year\, keV}$ implying a total of $\sim 1476$  background events.  An excess of 53 events has been observed at the $1-7\units{keV}$ low energy region (corresponding to roughly a $3\sigma$ excess), with the excess mainly located in the 2-keV and 3-keV energy bins. 

In this paper, we explore several possibilities for the origin of this signal.  
We will focus mostly on the possibility that the origin is attributable to DM, but will also consider bosonic particles (pseudo-scalar, scalar and vector) produced in the Sun, which do not necessarily have to be a DM component.  
We discuss in the context of the XENON1T excess several models previously considered in the literature: $\mathcal{O}$(keV) bosonic 
DM that is absorbed by an electron in the xenon atom~\cite{Dimopoulos:1985tm,PhysRevD.35.2752,Pospelov:2008jk,Derevianko:2010kz,Arisaka:2012pb,An:2013yua,Bloch:2016sjj,Hochberg:2016sqx}, bosonic DM that is emitted from the Sun~\cite{Raffelt:1996wa,Redondo:2008aa,An:2014twa,Redondo:2013lna,Budnik:2019olh}, 
and DM scattering off electrons in xenon~\cite{Essig:2011nj,Essig:2012yx,Essig:2015cda,Essig:2017kqs,Kopp:2009et}. 
For the absorption of bosonic DM, we show that only the dark photon or a ``photophobic'' axion-like particle can fit the \xe hint.  For light bosons produced in the Sun, bosons with a mass near 1 keV provide a better fit to the XENON1T data than massless bosons. The tension with star cooling constraints can be ameliorated in models where the shape of the scalar potential is substantially modified in dense environments (for a similar effect see e.g.~\cite{Khoury:2003aq,Masso:2005ym,Masso:2006gc,Jaeckel:2006xm,Ganguly:2006ki,Kim:2007wj,Brax:2007ak,Redondo:2008tq}). Here we present a model in which a pseudo-scalar is produced in the Sun and explains the \xe excess, but a density-dependent coupling between the pseudo-scalar and electrons avoids stellar cooling bounds. 

We also discuss, DM-electron scattering with different form factors, ``exothermic'' DM scattering off electrons (for previous work focused on nuclear scattering see~\cite{Essig:2010ye,Graham:2010ca} and focused on electron scattering see~\cite{Bernal:2017mqb}), and cosmic-ray accelerated DM that here interacts with electrons through an intermediate-mass mediator (for previous work focused on heavy mediators or light mediators interacting with nuclei see~\cite{Bringmann:2018cvk,Ema:2018bih,Cappiello:2019qsw,Bondarenko:2019vrb}; see also~\cite{Bringmann:2020}).  
These models deserve further study in future dedicated papers, but we provide their 
salient features focusing on the XENON1T excess. 

This paper is organized as follows.  
In \S\ref{sec:models}, we describe the requirements that new physics needs to satisfy in order to explain the \xe excess, and also detail the models that we will discuss.  
In \S\ref{sec:xenon}, we describe important features of the \xe data, our method for reconstructing the energy, and our statistical analysis.  
In \S\ref{sec:absorption}, we focus on the absorption of bosonic particles that are either (non-relativistic) DM particles in our halo or 
emitted from the Sun.  
\S\ref{sec:Chameleon} investigates how a density-dependent potential can be used to circumvent the stellar cooling bound.  
In \S\ref{sec:scattering}, we discuss DM-electron scattering, reviewing the ``standard'' case and then focusing on multi-component DM with small mass splittings.  We will see that ``exothermic'' DM scattering off electrons has a rich phenomenology.  
\S\ref{sec:acceleratedDM} considers a subdominant DM component that is accelerated by scattering off cosmic rays. 

\section{Models and Summary}
\label{sec:models}

The XENON1T excess motivates us to consider various known as well as novel new physics scenarios, focused mostly, but not solely, 
on DM models that can be discovered via a high-threshold ($\gtrsim{\rm keV}$) ER searches.   
We first summarize the relevant features of the excess and then identify possible mechanisms that may explain it. 

The following considerations are important when studying   a prospective new physics signal:
\begin{itemize}
\item The excess events have an energy of 2--3~keV.  
The measured spectrum suggests that a potential signal should contribute to more than a single bin.   However, the finite energy resolution of the experiment makes this (statistically weak) observation less sharp, allowing for rather narrow spectra to provide a reasonable fit.
The 1~keV bin and the bins with energies $\ge$4~keV are approximately consistent with background expectations. See Fig.~\ref{fig:spec} left. 
\item Low-threshold direct-detection experiments searching for electron recoils provide additional constraints on any signal that also produces sub-keV electron recoils~\cite{Essig:2012yx,Essig:2017kqs,Agnes:2018oej,Tiffenberg:2017aac,Abramoff:2019dfb,Crisler:2018gci,Romani:2017iwi,Agnese:2018col,An:2017ojc,Emken:2017hnp,Aprile:2019xxb,Barak:2020fql,Amaral:2020ryn}.  For energies of order 100's of eV, the \xe S2-only analysis is especially constraining~\cite{Aprile:2019xxb}.  
\item Numerous direct-detection experiments place stringent constraints on any accompanying nuclear recoil signal (for a recent compilation of low-mass DM limits on nuclear interactions see~\cite{Essig:2019xkx}).  Models that predict such a signal must evade these bounds.  
\item New physics that couples to electrons  is constrained by various collider and beam-dump experiments (see~\cite{Battaglieri:2017aum} for a compilation), as well as from astrophysical observations, such as the cooling of the Sun~\cite{Gondolo:2008dd,Redondo:2013wwa}, White Dwarfs (WD)~\cite{Raffelt:1985nj,Bertolami:2014wua,Giannotti:2017hny,Corsico:2019nmr}, Red Giants (RG)~\cite{Raffelt:1994ry,Viaux:2013lha,Straniero:2018fbv}, Horizontal Branch (HB) stars~\cite{Raffelt:1985nk}, and Supernovae (SN)~\cite{Calibbi:2020jvd}. 
\end{itemize}

Prospective models that could produce the observed excess and satisfy its features can be separated into models that predict an absorption signal (\S\ref{sec:absorption} and~\S\ref{sec:Chameleon}) and those that predict a scattering signal (\S\ref{sec:scattering} and~\S\ref{sec:acceleratedDM}).  We consider several scenarios: 
\begin{enumerate}
\item {\bf Absorption.}  We will consider the case that an electron absorbs a bosonic particle: pseudo-scalar (axion), a scalar, or a vector.  The boson may be either non-relativistic or relativistic.  The former may occur if the particle constitutes a component of the DM; in this case the ER spectrum is peaked at the mass of the DM, and can fit the data only due to the experiment's finite energy resolution. We find that a vector and a pseudo-scalar can explain the \xe excess, while a scalar is in conflict with stellar cooling constraints.  Next, light bosons may be produced in the Sun, which has a temperature of around 2~keV. A non-zero mass around 1.5-2.5~keV depending on the production mechanism could also cut the solar emission kinematically, providing the best fit to the data. However, for bosons produced in the Sun strong constraints arise from stellar cooling, strongly disfavoring the couplings needed to explain the \xe excess for the vanilla axion and dark photon models.  

\item {\bf Chameleons.}  The stellar cooling constraints on light bosons may be evaded if the couplings of SM particles to the corresponding bosons are screened inside high-density or high-temperature stellar objects.    Such chameleon-like particles have a rich phenomenology and can revive the Solar explanation of the \xe hint.  

\item {\bf DM scattering.}  The DM-electron scattering rate depends on the momentum-transfer-dependent atomic form-factor.  This steeply-falling function is highly suppressed for momenta $q \gg 1/a_0 = \alpha_{\rm EM} m_e$, where $a_0$ is the Bohr radius, $\alpha_{\rm EM}$ is the fine structure constant and $m_e$ is the electron's mass.   As a consequence, DM scattering through a light mediator or a velocity-independent heavy mediator predict a steeply rising spectrum at sub-keV energies and are thus disfavored. 

\item {\bf Velocity-suppressed DM scattering.}  Models that exhibit velocity- or momentum-dependent heavy-particle-mediated DM-electron scattering are allowed by experimental data at lower energies and provide an adequate fit to the \xe excess. 
However, such models are likely in tension with collider bounds on new particles that generate this operator~\cite{Fox:2011fx,Essig:2013vha}.  

\item {\bf Exothermic DM.}  An unsuppressed high-energy spectrum from DM-electron scattering may stem from an exothermic scattering of DM off electrons, the result of DM consisting of two or more states whose masses are slightly split by an amount denoted as $\delta$.  The atomic form factor, together with the scattering kinematics, imply a rather narrow electron recoil spectrum that is peaked near $|\delta|$ for a wide DM mass range and can explain the \xe excess for $|\delta|\sim \mathcal{O}$(keV).  The spectrum can be broadened if the DM-electron interaction increases with increasing momentum transfer $q$, if there are three or more DM states whose mass is split by different amount of $\mathcal{O}$(keV), or if the DM mass is well below the GeV-scale.  

\item {\bf Accelerated DM.}   A small subcomponent of DM may be accelerated through its interactions in the Sun~\cite{An:2017ojc,Emken:2017hnp} or with Cosmic Rays (CRs)~\cite{Bringmann:2018cvk, Ema:2018bih, Cappiello:2019qsw}.   While we find that the component accelerated from the Sun cannot explain the \xe excess without being in conflict with lower-threshold direct-detection searches, we find that CR scattering of DM with non-trivial momentum-dependent form factor can address the \xe excess while evading other direct-detection 
constraints.  However, in the scenario we consider here, direct constraints on the mediator exclude robustly this explanation.

\end{enumerate}

\begin{figure}[t]
\centering
\includegraphics[width=0.8\textwidth]{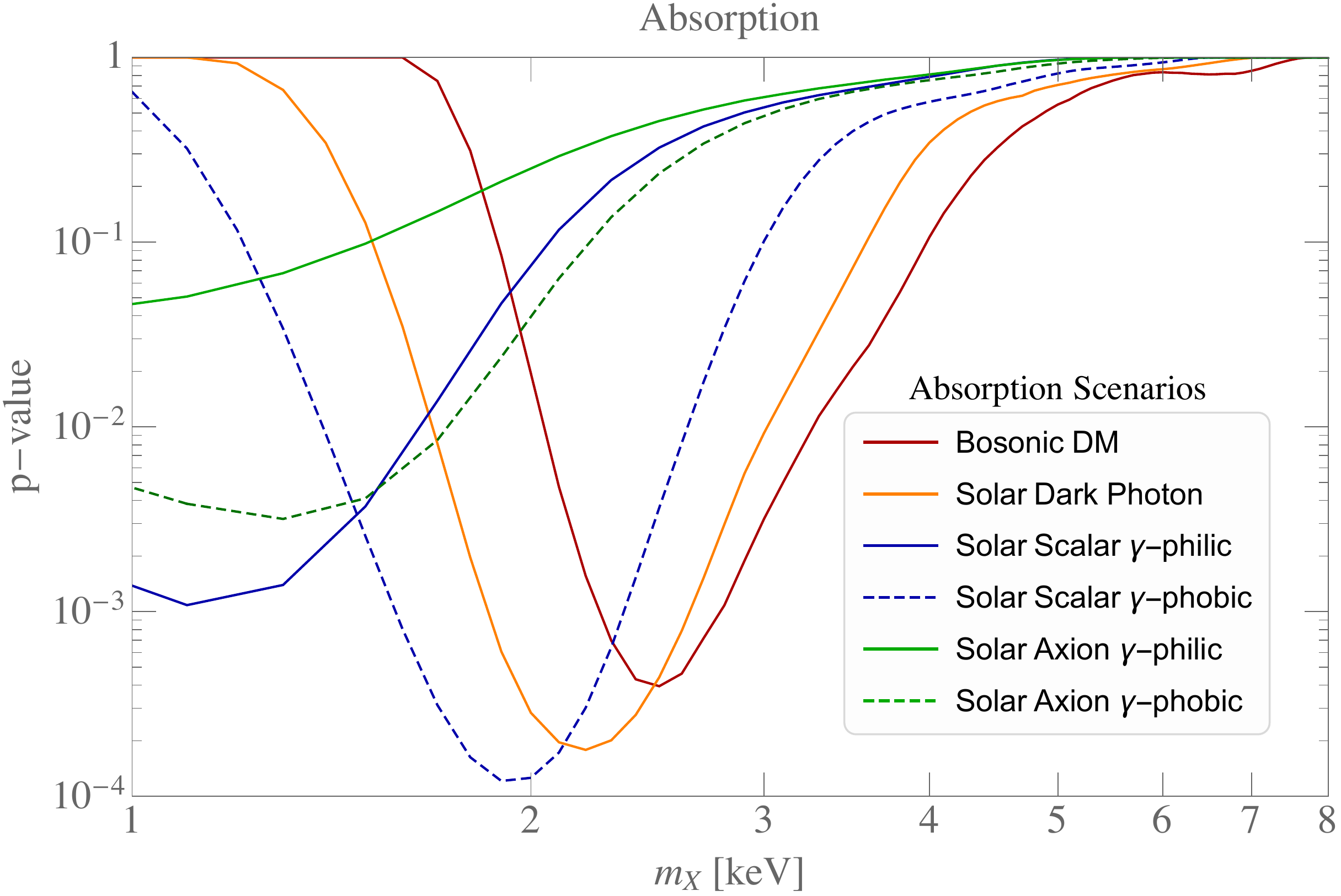}
\caption{Summary of the absorption scenarios considered here, with their p-value as a function of the mass. We show in {\bf dark red} all the cases of bosonic DM: axion-like particles in Sec.~\ref{sec:axionDM}, light scalars in Sec.~\ref{sec:scalarDM} and dark photon in Sec.~\ref{sec:darkphotonDM}. We also discuss all the solar scenarios: in {\bf green} axion-like particles discussed in Sec.~\ref{sec:solaraxion}, in {\bf blue} light scalars in Sec.~\ref{sec:solarscalar} and in {\bf orange} dark photon in Sec.~\ref{sec:solardarkphoton}.}\label{fig:modelabs}
\end{figure}

In Fig.~\ref{fig:modelabs}, we summarize the goodness-of-fit of the various absorption scenarios discussed above to the \xe measurement.  We see that the bosonic DM scenarios (red curve) can fit the data well with a predicted mass of $m_X=2.5\text{ keV}$ and coupling to electrons. Among these, the scalar DM case is excluded by stellar constraints while the dark photon and the axion are good explanation of the \xe excess. In the latter case the anomalous axion coupling to photon should be set to zero to avoid X-rays constraints. 

In all the solar cases, the adddition of a non zero mass ameliorates the fit by cutting off the spectrum kinematically, in better agreement with the 1 keV bin being consistent with the background prediction.  For pure electron coupling the axion explanation is disfavored compared to the scalar or the dark photon. The reason is that the peaks in the spectrum of the axion ABC production~\cite{Redondo:2013wwa} are not observed in the data. If the solar production happens through the  Primakoff process~\cite{Raffelt:1985nk} the scalar provides a very good fit of the data while the axion explanation is disfavored. As we will discuss, the reason can be traced back to the different energy dependences of the axion and scalar absorption rates in xenon. The scalar rate grows fast at low energies for very light masses but a good fit can be obtained for a scalar mass of $m_\phi=1.9\text{ keV}$, which cuts the sharp rise towards low energies and hence generates a bump between 2 and 3~keV. On the other hand, the axion absorption rate is suppressed at low energies and the resulting spectrum is too flat at energies above 3 keV to provide a good fit to the excess, independently of the mass of the axion. Therefore, fitting to a massive axion consistently tends to prefer a very light, even massless, axion.  

\begin{figure}[t]
\centering
\includegraphics[width=0.8\textwidth]{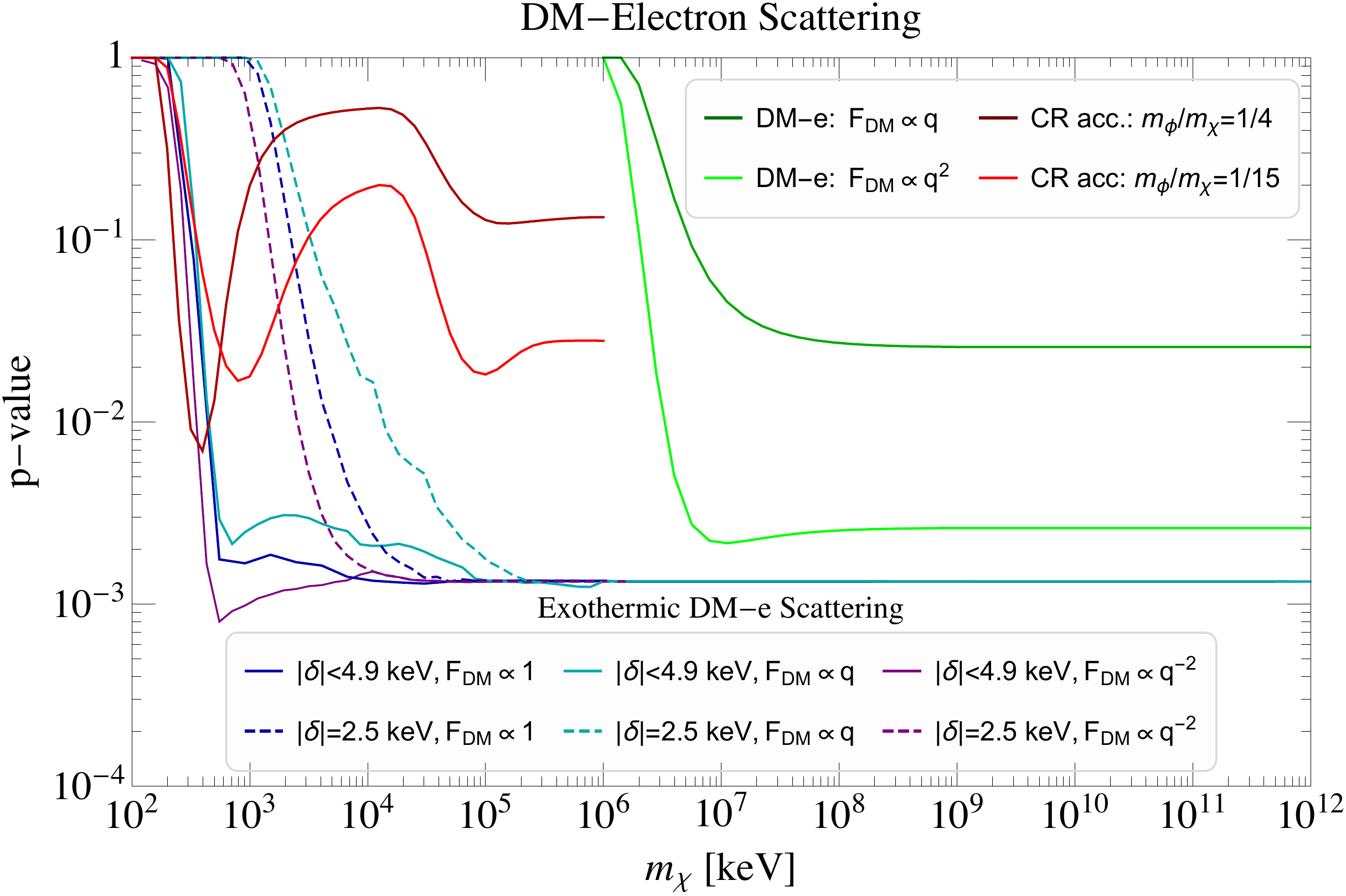}
\caption{We show in {\bf green} momentum suppressed DM-electron scattering with form factors $F(q)\propto q$ {\bf (dark green)} and $F(q)\propto q^2$ {\bf (light green)}  discussed in Sec.~\ref{sec:elscattering}. We show accelerated DM by cosmic rays scattering for fixed mass ratios between the DM and the mediator $m_\phi/m_{\chi}=1/4$ ({\bf dark red}) and $m_\phi/m_{\chi}=1/15$ ({\bf light red}), where $\phi$ can be a scalar or a vector with scalar interactions on the SM side and axial interactions on the DM side. This is discussed in Sec.~\ref{sec:acceleratedDM}. We also include different scenarios for exothermic scattering: light mediator $F(q)\propto 1/q^2$ {\bf (purple)}, heavy mediator $F(q)\propto 1$ {\bf (dark blue)} , and momentum suppressed form factors $F(q)\propto q$ {\bf (cyan)} discussed Sec.~\ref{sec:EXO}.  The splitting $\vert\delta\vert$ between the heavy state and the light state in the dark sector is marginalized to minimize the p-value for $\vert\delta\vert<4.9\text{ keV}$ (as discussed in the text, this range of splittings may not capture the entire possible parameter space, and should thus be treated with caution). The {\bf dashed lines} have fixed $\vert\delta\vert=2.5\text{ keV}$, which implies a lower bound on the DM mass.}\label{fig:modelscatter}
\end{figure}  

In Fig.~\ref{fig:modelscatter}, we summarize the goodness-of-fit of the different scattering scenarios presented above. First, we notice that elastic scattering cannot explain the \xe hint for $F(q)\propto q^n$ form factors with $n\leq0$, since the electron recoil spectrum rises at low energy, in tension with complementary direct-detection experiments at lower energy thresholds. However, for $F(q)\propto q^n$ with $n>0$ the spectrum falls fast enough towards lower energies and provides an adequate fit to the \xe data. 

Second, we show that exothermic scattering can fit well the data when the heavy and light DM states are split in mass by a few keV. Once the splitting is marginalized to the best fit value, the p-value is essentially independent of the DM mass as long as it is heavier than the splitting itself. The spectrum is peaked near 2~keV and fits well the data, without being trivially excluded by complementary direct detection experiments. In concrete models, the rich phenomenology of these DM scenario could provide other handles of testing them at beam dump experiments or in nuclear recoil. We also show that for a fixed splitting, a lower bound on the DM mass can be derived, which varies depending on the nature of the form factor.

Third, we discuss accelerated DM by scattering with cosmic rays. In such a case, the challenge is again to find a scenario where the accelerated spectrum falls sufficiently rapidly at energies lower than 2~keV. We achieve this by considering axial-scalar interactions between the accelerated DM and the SM, mediated by a light new mediator with mass around 100~keV.  Just as other models of accelerated DM, this scenario is likely to be challenged by other observation probes. We leave a more in depth study of this scenario for future work.     

We now present our data analysis framework, before discussing each of these model scenarios in detail. 

\section{XENON1T}
\label{sec:xenon}

In this section, we review the relevant aspects of the \xe experimental apparatus and the electron recoil analysis, with a focus on describing our treatment of the 
energy reconstruction and  statistical analysis that is used throughout this work. 

\subsection{Energy Reconstruction Method}\label{subsec:energy-reconstruction} 
The experiment utilizes a dual-phase xenon Time Projection Chamber~\cite{Dolgoshein1970,Alner:2007ja,Aprile:2017iyp,Aprile:2017aty,Aprile:2018dbl,Aprile:2018cxk,Aprile:2019dbj,Aprile:2019dme,Aprile:2019xxb,Aprile:2019jmx,Aprile:2020yad}, to search for weakly interacting particles. When one of the xenon atoms in the Liquid Xenon (LXe) phase recoils or is ionized due to a collision, photons are emitted and detected by photomultiplier tubes (PMTs). This signal is called the prompt scintillation signal (S1). In addition to the photons emitted close to the interaction point, ionized electrons drift inside the detector due to an external electric field. When the electrons reach the Gaseous Xenon layer (GXe) at the top of the detector, they are extracted across the liquid-gas interface, collide with xenon atoms, and produce a proportional scintillation light, known as the S2 signal, which is also measured by the PMTs. 

The ratio of S2/S1 provides a handle that enables one to differentiate between Nuclear Recoil (NR) and ER events. Further information about a given event can be inferred by its location inside the PMTs, the time difference between the arrival of the S1 and S2 signals, and the S1 and S2 signal shapes. This complementary information is taken into account in the analysis by the \xe collaboration, however, it is not publicly available.  When the \xe collaboration reports their data, they use the corrected S1 (cS1) and corrected S2 (cS2), which takes into account this additional information. 

In their analysis of the Science Run~1 (SR1) data, the \xe collaboration provides a scatter plot of ${\rm (cS1,cS2_b)}$ (the `b' subscript signifies that only the PMTs at the bottom of the detector were used for the S2 reconstruction).  
Rather than using their reconstructed keV-binned energy spectrum, we will use the data from this scatter plot to reconstruct the energies for each event.  We do this, since the keV-binned data results in a loss of information, as the \xe detector resolution is as low as $\sim 0.3~\rm{keV}$~\cite{Aprile:2020yad} at their analysis threshold $\sim {\rm keV}$.  
In order to reconstruct the energies, we use the procedures laid out by the \xe collaboration in~\cite{Aprile:2019dme} (which uses detector modeling techniques created by the NEST collaboration~\cite{Szydagis_2011}, and with additional data taken from~\cite{XENON1T-talk2020,Aprile:2019bbb}), which allows us to simulate the detector response and the effects of reconstructing the signal. We use a Monte Carlo (MC) simulation to determine how an ER with a given energy is distributed on the (cS1,cS2$_b$) plane, and use a maximum likelihood estimator to find the energy of the event. 
Below, we refer to this way of reconstructing the energy as ``our method'', even though it is based on information provided in previous 
\xe papers; we do so to differentiate it from the way the energy was reconstructed by 
the \xe collaboration in their ER analysis paper~\cite{Aprile:2020tmw}, where they simply use 
\begin{equation}
\label{eq:simplifiedE}
E^{\rm\xe}_{\rm reconstructed}=\left(\frac{cS_1}{g_1}+\frac{cS2_b}{g_2}\right) W \,,
\end{equation}
where $g_1=0.142$ and $g_2=11.4$ are the probabilities for one photon to be detected as a photo-electron in the PMT and the charge amplification factor, respectively, and the mean energy to produce a detectable quanta is $W=13.8~{\rm eV}$. 
For additional discussion of the possible problems with this simplified energy reconstruction formula, see~\cite{Szydagis:2020isq}.   

\begin{figure}[t]
\centering
\includegraphics[width=0.8\textwidth]{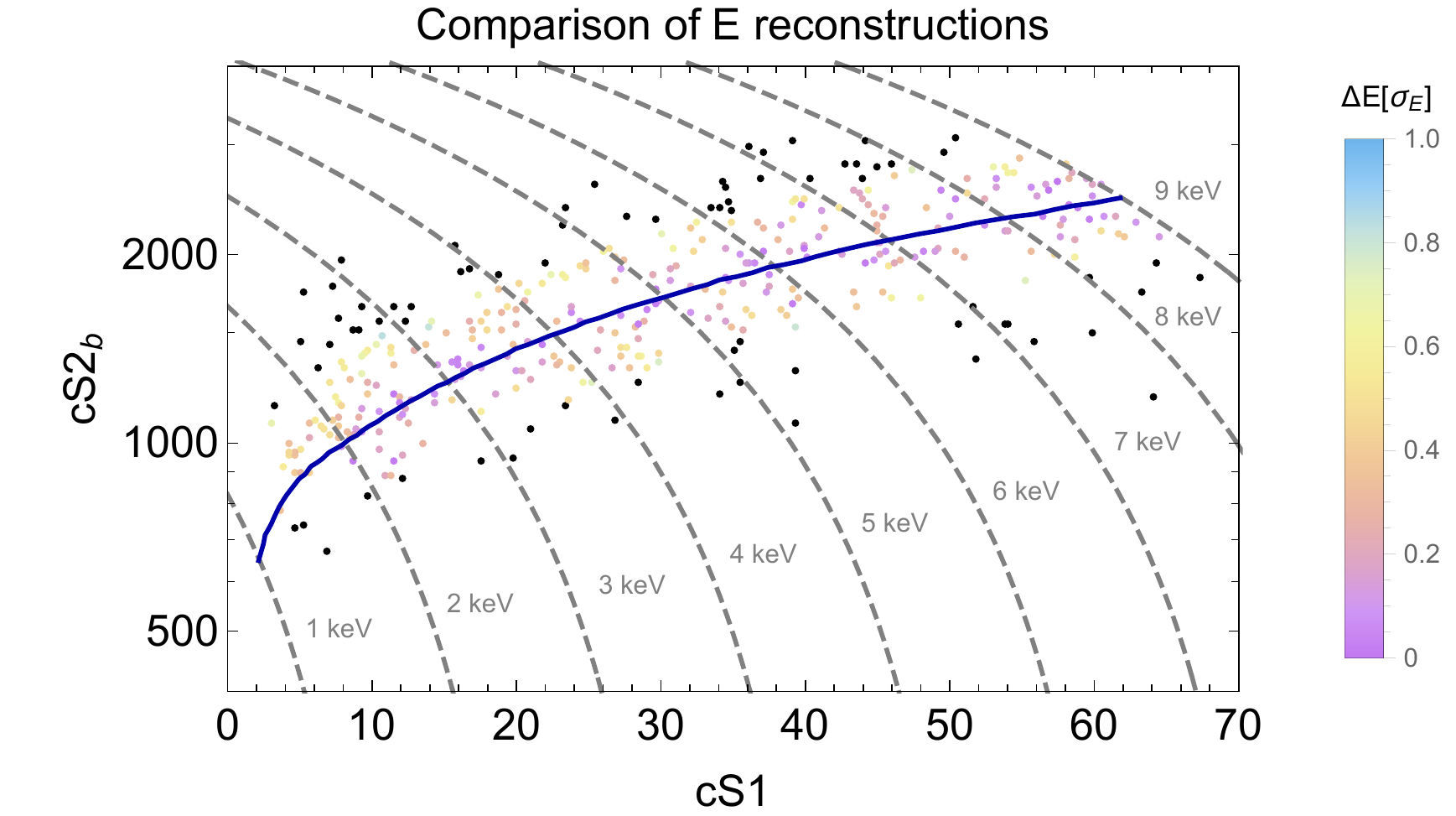}
\caption{Observed events by the \xe collaboration~\cite{Aprile:2020tmw} in the (cS1,cS2$_b$) plane, 
for events they tagged as having an energy $\leq 9~{\rm keV}$. 
The colors of the points correspond to the difference in the reconstructed energy (in units of the energy resolution) between our energy reconstruction calculation and the simplified equation used by \xe (Eq.~\eqref{eq:simplifiedE}).  The energy resolution is estimated using our energy reconstruction calculation.  
The {\bf gray dashed} lines show constant energy lines using the simplified energy reconstruction given in Eq.~\eqref{eq:simplifiedE}. 
In {\bf blue} is the expectation value for the energy interval $\left[{\rm keV},9{\rm keV}\right]$.
The {\bf black points} are more than $2\sigma$ away from this expectation value, and we did not sample the parameter space finely enough with our MC to reliably reconstruct their energy; for these points, we assume simply that their energy is given by Eq.~\eqref{eq:simplifiedE}.}\label{fig:scatter}
\end{figure}

In Fig.~\ref{fig:scatter}, we reproduce the (cS1,cS2$_b$) scatter plot for events tagged in~\cite{Aprile:2020tmw} to have an energy below $9~{\rm keV}$ (above this energy, the resolution is $\sim$keV so the binning leads to only marginal information loss; moreover, the excess is concentrated below this energy, so we will not be concerned with events at higher energies). 
The color of the points shows the difference between the energy reconstructed by our method and the simplified formula used in~\cite{Aprile:2020tmw}, Eq.~\eqref{eq:simplifiedE}. The colored points on the plot are in units of the energy resolution, calculated with our method (see below).  
Due to the finite size of our MC sample, large numerical errors may occur in rare cases where the calculated likelihood for the reconstructed energy of a given event is small ($\leq 5\%$ C.L.) for all energies.   To avoid such errors, in those cases we use  the simplified reconstruction method,  Eq.~\eqref{eq:simplifiedE}.

\begin{figure}[t]
	\centering
	\includegraphics[width=0.48\textwidth]{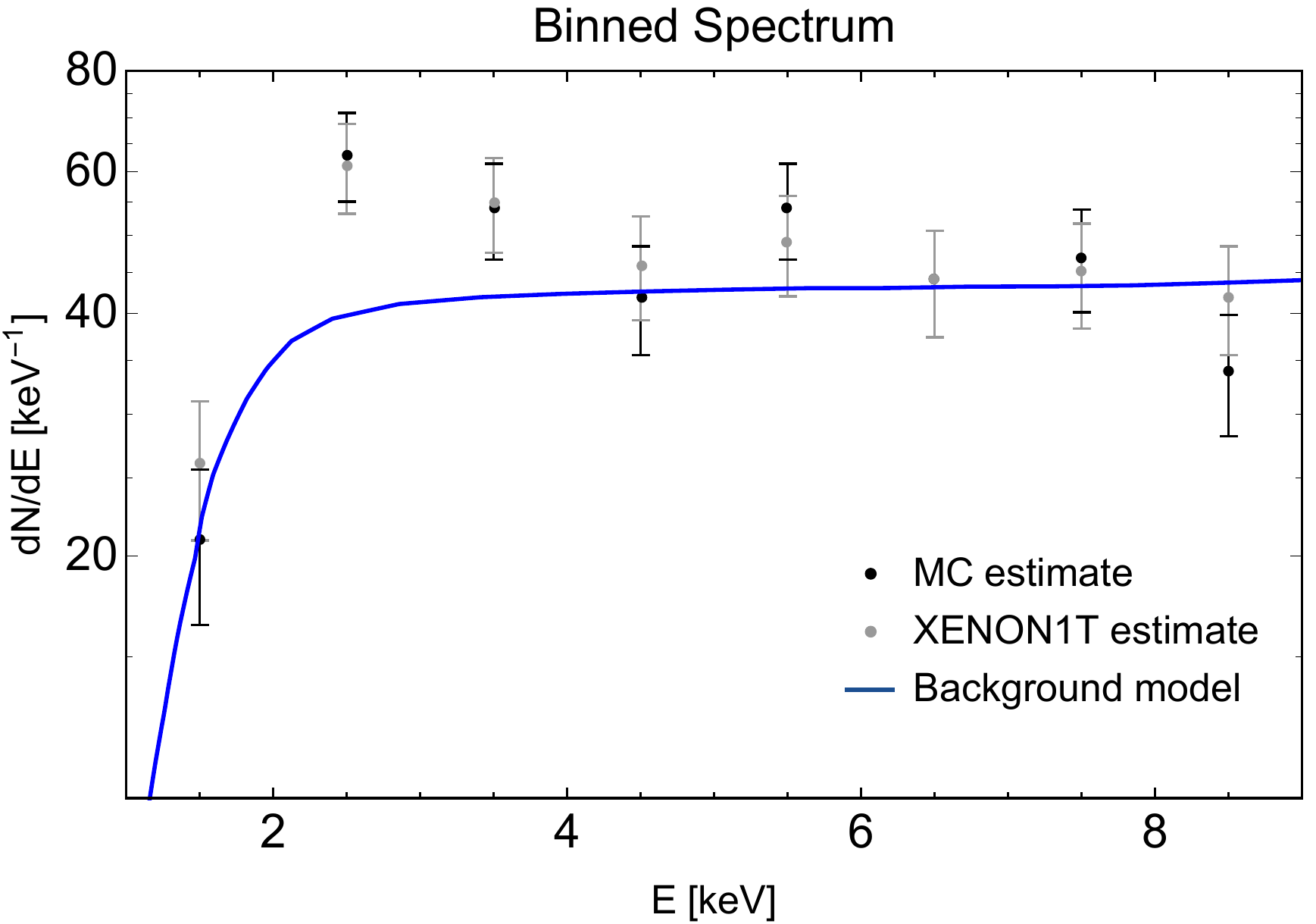}
	\includegraphics[width=0.48\textwidth]{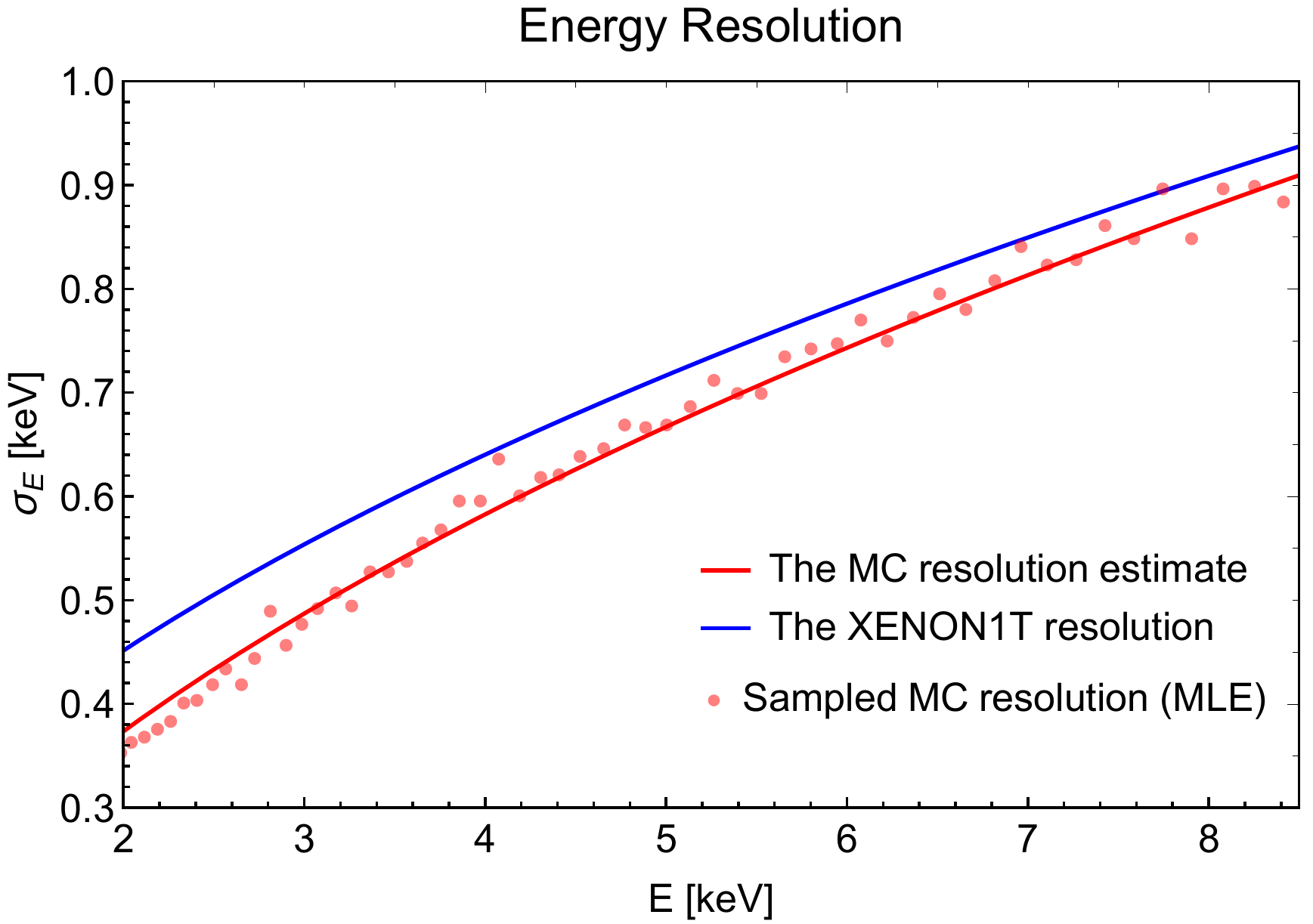}
	\caption{{\bf Left:} Comparison of the naive spectrum reconstructed, and the one by the MC, for energies below $9\text{ keV}$.   In blue is the background model from~\cite{Aprile:2020tmw}.
	While the biggest disagreement, at the lowest-energy bin,  seems to be a $\geq 1\sigma$ disagreement, we note that this is misleading,  as many of the points were reconstructed on the edge of the bin causing small differences to be magnified.  An overall good agreement between our MC method and the one used by the \xe collaboration is observed, enabling us to use the full unbinned energy information throughout this paper.  As an aside, we also note that our binned energy spectrum does not have the same monotonically decreasing spectrum in the $\sim$$5-10$~keV energy bins.  {\bf Right:} Energy resolution estimated from our MC (black points), a fit to these MC data ($\sigma_E~[{\rm keV}]=-0.21+ 0.39\sqrt{E}$, red line), and the energy resolution estimated by the \xe collaboration in~\cite{Aprile:2020tmw} ($\sigma_E~[{\rm keV}]=0.3171 \sqrt{E} + 0.0015 E~[{\rm keV}]$, blue line). 
} \label{fig:spec}
\end{figure}

We show in Fig.~\ref{fig:spec} (left) our calculation of the keV-wide binned energy spectrum and compare it with the \xe spectrum.  The 
two spectra are nearly identical.  This provides confidence in our energy reconstruction method, and allows us to use the full unbinned energy information for our new physics analyses below. 
We also include in this plot the background model from~\cite{Aprile:2020tmw}. 

For the formula Eq.~\eqref{eq:simplifiedE} to be the best estimator for the energy, the variables cS1 and cS2$_b$ should be anti-correlated. While this has been validated for high energies, preliminary measurements appear to suggest that there is only a weak anti-correlation for low energies.  In particular, this can be seen from measurements of the $^{37}$Ar line at 2.83~keV presented 
in~\cite{XENON1T-talk2020}.  Our MC simulation of 2.8~keV events (assuming a uniform distribution in $z$) agrees well with the contours in the (cS1, cS2$_b$) plane for the $^{37}$Ar data found in~\cite{XENON1T-talk2020}: our agreement is better than $\lesssim 3\%$ for the central value, and we can find even better agreement if we change the 
simulation parameters slightly from those given by the \xe collaboration in~\cite{Aprile:2019dme}  within their error margins.  Our simulation also agrees well with the observed weak correlation between cS1 and cS2$_b$.  
This provides further confidence in our energy reconstruction method, especially at the $\mathcal{O}$(keV) energies relevant for the excess events.  
Fig.~\ref{fig:spec} (right) shows the energy resolution estimated from our MC (black points), a fit to these MC data (red line), and the 
energy resolution estimated in~\cite{Aprile:2020tmw} (blue line).  As can be seen, the energy resolution estimated from the MC is slightly better than that used in~\cite{Aprile:2020tmw}. As the actual smearing of appears to be not entirely symmetric, an asymmetric resolution might provide an even more accurate description than the symmetric one used here (see also Ref.~\cite{Szydagis:2020isq}), however, this does not appear to change of the results significantly.\footnote{We thank Matthew Szydagis, for helping us verify our results with the more detailed calculation done by the NEST code~\cite{Szydagis_2011}.}.

\subsection{Statistical Method}

For our analyses, we use a likelihood ratio test, with unbinned likelihoods. For each signal model, $s$, that depends on parameters ${\bf \theta}_s$, we find the likelihood of the signal+background hypothesis for the data as a function of the model parameters,
  \begin{equation}
 \label{eq:likelyhood}
 \mathcal{L}(s+b)=\frac{e^{-\mu_s-\mu_b}}{n!}\prod_{i=1}^n \frac{d(N_s+N_b)}{dE}\large(E_i|{\bf \theta }_s\large),
 \end{equation}
where $E_i$ are the reconstructed energies, $n$ is the number of observed events,  $dN_{b}/dE$ ($dN_{s}/dE$) is the background spectrum (signal spectrum), and $\mu_b=\int dN_b/dE$ ($\mu_s=dN_s/dE$) are the total expected background (signal) events.  We maximize the likelihood to find the best fit points. In order to estimate the significance and quality of our fits, we assume the asymptotic formulas found in~\cite{Cowan:2010js}; we therefore assume that twice the log-likelihood-ratio of the signal+background hypothesis compared to the background-only hypothesis is distributed according to a $\chi^2$ distribution, with the number of degrees of freedom set equal to the number of model parameters,

\begin{equation}
{\rm p-value}=Q\left(\frac{{\rm \# D.O.F.}}{2},\log\left(\frac{ \mathcal{L}(\hat{s}+b)}{ \mathcal{L}(b)}\right)\right),
\end{equation}
where $\mathcal{L}(\hat{s}+b)$ ($\mathcal{L}(b)$) is the likelihood of the best fit for the signal+background (background-only) hypothesis. ${\rm \# D.O.F }$ is the number of degrees of freedom for the signal hypothesis, and $Q$ is the regularized incomplete gamma function (one minus the p-value gives the cumulative distribution function for the $\chi^2_{\rm \# D.O.F. }$ distribution).

To ease interpreting the ${\rm p-value}$ of an excess, we also present the more commonly used significance\footnote{Our p-value definition differs from the one of  Ref.~\cite{Cowan:2010js} by a factor of 2. A p-value of  $5\%$ would correspond to $2\sigma$ significance in our notation.}
\begin{equation}
{\rm Significance}= \sqrt{2}~{\rm erfc}^{-1}({\rm p-value}).
\end{equation}
Where ${\rm erfc}^{-1}$ is the inverse function to the complementary error function.

When presenting later 2D plots with $1-\sigma$ and $2-\sigma$ bands (see e.g. Fig.~\ref{fig:axionDM} left, for an example parameter space for the ALP DM hypothesis), each point on the graph is treated as an independent hypothesis (i.e. with a given coupling, mass, etc.). At such graphs, the $1-\sigma$ ($2-\sigma$) band presents the points that are $1-\sigma$ ($2-\sigma$) away from the best fit point on that graph (i.e. not necessarily the best fit point in general).  

In each of the following sections, we describe how to derive the spectrum of events.  
The measured spectrum will be modified by detector response effects.  In particular, for a given theoretically predicted signal, we modify the spectrum by the effective exposure, $\mathcal{E}(\omega)$, of the xenon detector.  We then smear the resulting spectrum by a gaussian with the resolution presented by the red line in Fig.~\ref{fig:spec} (right), and then calculate the likelihood. The effective exposure models the non-flat efficiency, and should in fact be applied during the MC stage, as it directly relates to the S1 signal, and not the energy. However, for simplicity, we have applied it as described in the text. Small variations on our methods yield changes only for signal models with a large rate at the 1-2 keV bin where the efficiency is not flat, and even for such models, the effect is not significant.

In our analysis, we will ignore any contribution to the background from, e.g., tritium decays (see e.g.~\cite{Aprile:2020tmw,Robinson:2020gfu}) and $^{37}$Ar (see e.g.~\cite{Szydagis:2020isq}).  We will also ignore the look-elsewhere effect which is important for determining the global significance of a particular model~\cite{Vitells:2011da}. While a formal calculation of the global significances for each model is beyond the scope of this work~(see Ref.~\cite{Vitells:2011da} for a thorough discussion), we briefly discuss here what we expect the importance of the look elsewhere effect to be for each of the models considered in this paper. 

For both standard DM-e scattering models considered here, as well as for the CR-accelerated DM presented, the look-elsewhere-effect is expected to be non-important. For the case of particles produced in the Sun, the look-elsewhere will have a mild importance, since changing the mass of the particle can lead to peaks in the signal spectrum at different energies, and yet the range of possible masses is limited by roughly the temperature of the solar core. For the case of the DM absorption, the look-elsewhere effect is expected to possibly be important, since it corresponds to the classical case of a highly-localized signal. Indeed, the reported local significance by the \xe collaboration is $4\sigma$, while the reported global one is $3\sigma$~\cite{Aprile:2020tmw}. For the case of exothermic-DM, the multi-dimensional parameter space can greatly affect both the location and width of the signal spectrum, and it is thus expected for the look-elsewhere to have the most drastic effect for these models.



\section{Absorption}
\label{sec:absorption}

We consider first models of bosonic DM, confronting them with the \xe measurement.   Three cases are considered: pseudo-scalar (axion), scalar, and vector bosons.   For each we explore the non-relativistic case, in which the boson constitutes the DM, and the relativistic case, for which the boson is produced in the Sun. 

In the case of bosonic DM, the rate of events in the \xe detector per unit energy~is 
 \begin{equation}
\frac{dR^{\text{DM}}_{\text{abs}}}{d\omega}=\Phi^{\text{DM}}\sigma^{I}_{\text{abs}}(\omega)\delta(\omega-m_I)\,,
\end{equation}
where the energy, $\omega$, is kinematically constrained   to equal the DM mass (ignoring small non-relativistic corrections of order the DM energy), and is then smeared to account for the detector resolution as described in \S\ref{sec:xenon}.  The total number of events in the relevant \xe energy window is then obtained by convolving the above rate with the effective exposure $\mathcal{E}(\omega)$ reproduced in \S\ref{sec:xenon} and integrating over energy. 

The DM flux, $\Phi^{\text{DM}}$, is the same for all bosons, and depends only on the DM relic density, $\rho_\chi$, and the mass of the light boson 
\begin{equation}
\Phi^{\text{DM}}=1.2\times 10^{13}\units{cm^{-2} sec^{-1}}\left(\frac{\rho_\chi}{0.4 \units{GeV cm^{-3}}}\right)\left(\frac{\text{keV}}{m_I}\right)\left(\frac{v}{10^{-3}}\right)\,.
\end{equation}
The absorption cross section, $\sigma^{I}_{\text{abs}}$, depends, however, on the interaction of a given light boson $I$ with the bounded electrons in the liquid xenon. Here we consider three cases: (i)~ALP DM absorption via the  axioelectric effect ($I=AE$; Sec.~\ref{sec:axion}), (ii) scalar absorption via the scalar-electric effect ($I=SE$; Sec.~\ref{sec:scalar}), and (iii) dark photon DM absorption via the photoelectric effect ($I=PE$; Sec.~\ref{sec:darkphoton}).

\begin{figure}[t]
\centering
\includegraphics[width=0.49\textwidth]{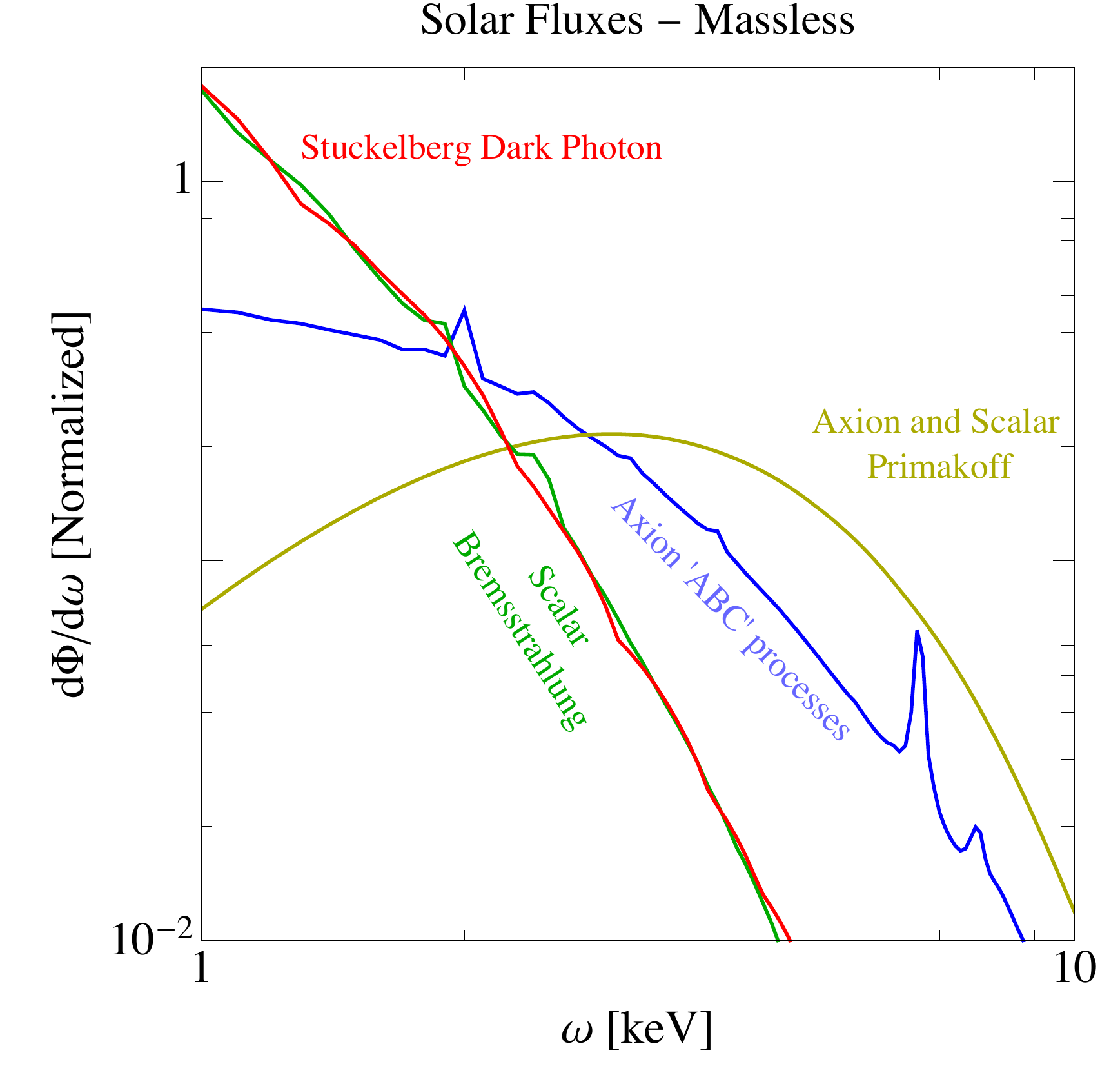}\hfill
\includegraphics[width=0.49\textwidth]{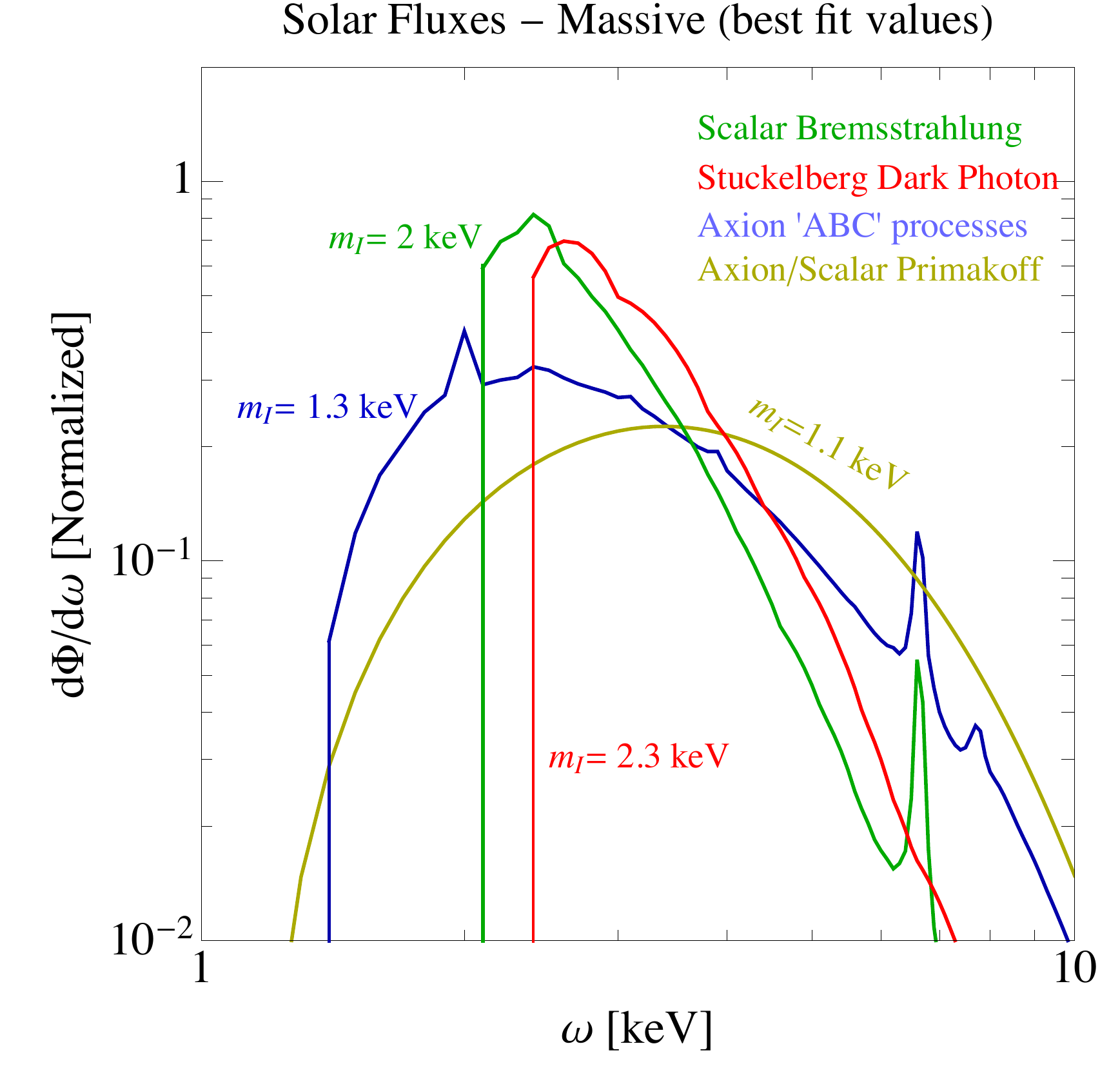}
\caption{Solar flux spectra for the axion production from ABC processes~\cite{Redondo:2013wwa} ({\bf blue}) and Primakoff production~\cite{Raffelt:1985nk} ({\bf yellow}), for scalars from  bremsstrahlung~\cite{Redondo:2008aa,Budnik:2019olh} ({\bf green}), and for dark photons ({\bf red}). On the {\bf left}, we assume a massless boson, while on the {\bf right} we included the kinematical threshold due to a finite boson mass, fixed to the best fit point for each case.
  To highlight the spectral features, the plots are normalized such that the total  integrated flux in the energy window of interest for \xe, $\omega\in [1,10]\units{keV}$ equals 1~event/cm$^2$/s.} 
\label{fig:sunfluxes}
\end{figure}

For light bosons produced in the Sun, the differential event rate per unit energy can be written as 
\begin{equation}
\frac{dR^{\text{Sun}}_{\text{abs}}}{d\omega}= \frac{d\Phi^{\text{Sun}}_{I}}{d \omega}\sigma^{I}_{\text{abs}}(\omega)\ ,
\end{equation}
where the differential solar flux $d\Phi^{\text{Sun}}_{I}/d \omega$ depends on the production mechanisms of the light bosons $I$ inside the Sun's environment and needs to be treated case by case. Below we discuss  solar axions in Sec.~\ref{sec:solaraxion},  solar scalars in Sec.~\ref{sec:solarscalar} and solar dark photons in Sec.~\ref{sec:solardarkphoton}.   In Fig.~\ref{fig:sunfluxes}, we show the relevant solar fluxes that are important for the derivation of the predicted signal's spectrum for the different cases. 

\subsection{Axion-Like Particles}\label{sec:axion}
We consider an axion-like particle (ALP) of arbitrary mass $m_a$ that couples to photons and electrons, 
\begin{equation}
\mathcal{L}_{\text{ALP}}=\frac{g_{a\gamma\gamma}}{4} aF_{\mu\nu}\tilde{F}^{\mu\nu}+\frac{g_{aee}}{2m_e} \partial_\mu a \bar{e}\gamma^\mu\gamma_5 e\ .\label{eq:axioncoupl}
\end{equation}
The ALP can be absorbed  inside the detector material leading to a ioniziation signal. The cross section for this so called axio-electric (AE) effect~\cite{Dimopoulos:1986kc,Avignone:1986vm,Pospelov:2008jk,Derevianko:2010kz} can be written as~\cite{Bloch:2016sjj} 
\begin{equation}
\sigma_{\text{AE}}(\omega_a)=\sigma_{\text{PE}}(\omega_a)\frac{3g_{aee}^2}{16\pi\alpha_{\text{EM}}v_a}\frac{\omega_a^2}{m_e^2}\left(1-\frac{1}{3}v_a^{2/3}\right)\,,
\label{eq:AExsec}
\end{equation}
where $\omega_a=\sqrt{m_a^2+k_a^2}$ is the energy of the ALP and $v_a$ is its velocity. We take the photoelectric cross section, $\sigma_{\text{PE}}(\omega_a)$, from~\cite{1993ADNDT..54..181H}, which agrees reasonably well with experimental data above 30~eV. The above formula is approximate, and chosen to correctly reproduce the results obtained in the non-relativistic limit, $v_a\ll1$, and in the relativistic limit, $v_a\to1$.

In what follows, we will derive the \xe best-fit regions for $g_{aee}$ as a function of the ALP mass.  Theoretically, however, $g_{aee}$ is often related to $g_{a\gamma\gamma}$ and for the ALP DM case,  X-rays measurements can then be used to exclude part of the parameter space.  It is therefore interesting to understand the theoretical relation between the two couplings, which will allow us to identify viable ALP models.   As we shall see below,   three conclusions can be drawn:
\begin{enumerate}
\item Fitting the data with QCD axion DM requires a high degree of fine tuning of its ultraviolet~(UV) couplings to electrons and the UV anomaly with respect to electromagnetism. 
\item More general ALP DM requires suppressed couplings to photons in the UV, which typically implies a non-anomalous global symmetry with respect to QED.
\item Standard solar ALPs could be the QCD axion but are excluded by stellar constraints, motivating chameleon-like ALPs to be discussed in Sec.~\ref{sec:Chameleon}.
\end{enumerate}
To understand these statements, let us briefly discuss the origin for $g_{aee}$ and $g_{a\gamma\gamma}$. The parametrization of Eq.~\eqref{eq:axioncoupl} can be mapped to concrete models where the pseudo-Nambu-Goldstone boson (pNGb) of a spontaneously broken global symmetry couples to the photons and  electrons. An arbitrarily small mass $m_a$ can be introduced as a soft breaking of the pNGb shift symmetry. More explicitly, we can write 
\begin{equation}
g_{a\gamma\gamma}=\frac{\alpha_{\text{EM}}}{2\pi f_a}E_{\text{eff}}\qquad ,\qquad g_{aee}= \frac{m_e}{f_a}C_{\text{eff}}\,,
\end{equation}
where $f_a$ is the ALP decay constant and $E_{\rm eff}$ parametrizes the effective coupling to photons, which is related to the UV parameters through 
\begin{equation}
E_{\text{eff}}=E_{\text{UV}}+C_{\rm UV}\mathcal{A}(x)\,.\label{eq:oneloop}
\end{equation}
Here $C_{\rm UV}$ is the UV coupling of the axion to electrons, while $E_{\text{UV}}$ is the UV anomaly with respect to electromagnetism, which is model dependent. ${\cal A}(x)$ parametrizes the electron loop function,  $\mathcal{A}(x)=x\text{ arctan}^2\frac{1}{\sqrt{x-1}}-1$
with $x=4m_e^2/m_a^2-i\epsilon$ which decouples as $m_a^2/m_e^2$ for $m_a\ll m_e$. This feature can be traced back to the fact that in the presence of a purely derivative coupling to electrons, only the effective operator $\partial^2 a F\tilde{F}$ is generated below the electron threshold~\cite{Nakayama:2014cza}. If $E_{\text{UV}}$ is non-zero, the electron coupling is modified by the running contribution induced by the photon coupling~\cite{Chang:1993gm}. At low energies, one finds  
\begin{equation}
C_{\text{eff}}=C_{\rm UV}+\frac{3\alpha_{\text{EM}}^2}{4\pi^2} E_{\text{UV}}\log\left(\frac{f_a}{m_e}\right)\ .
\end{equation}
For the QCD axion, the coupling to the gluon field strength gives further contributions to the effective photon and electron couplings generated by the mixing of the axion with the QCD mesons below the confinement scale~\cite{diCortona:2015ldu}, 
\begin{equation}
E_{\text{eff}}\stackrel{{\rm QCD}}{\longrightarrow} E_{\text{eff}}-1.92\qquad , \qquad C_{\text{eff}}\stackrel{{\rm QCD}}{\longrightarrow}  C_{\text{eff}}-\frac{2}{3}\frac{4 m_d+m_u}{m_u+m_d}\log \frac{\Lambda_{\text{QCD}}}{m_e}\,.
\end{equation}
The strong X-ray limits on $E_{\rm eff}$ together with the ${\cal O}(1)$ contribution from QCD explains why QCD axion DM must be tuned to address the anomaly. 

Various axion models have been studied, where the different hierarchies between the electron and photon couplings are realized:  
\begin{itemize}
\item {\bf DFSZ models}, where naturally $C_{\rm UV}\sim E_{\text{UV}}\sim \mathcal{O}(1)$~\cite{Zhitnitsky:1980tq,Dine:1981rt}.
\item {\bf KSVZ models}, where $C_{\rm UV}=0$, and the electron coupling is only generated from the photon coupling via the running~\cite{Kim:1979if,Shifman:1979if}. 
\item {\bf Photophobic models} where $E_{\text{UV}}=0$ and the electron coupling dominates the phenomenology. See~\cite{Craig:2018kne} for a general discussion of photophobic ALPs and the Majoron~\cite{Ibarra:2011xn,Heeck:2019guh} as a particularly motivated example of this coupling structure.  
\end{itemize}
Of the above, and in the absence of tuning, only the Photophobic ALPs can fit the \xe hint without being excluded, if they are DM.


\subsubsection{ALP Dark Matter}\label{sec:axionDM}

\begin{figure}[t]
\centering
\includegraphics[width=0.55\textwidth]{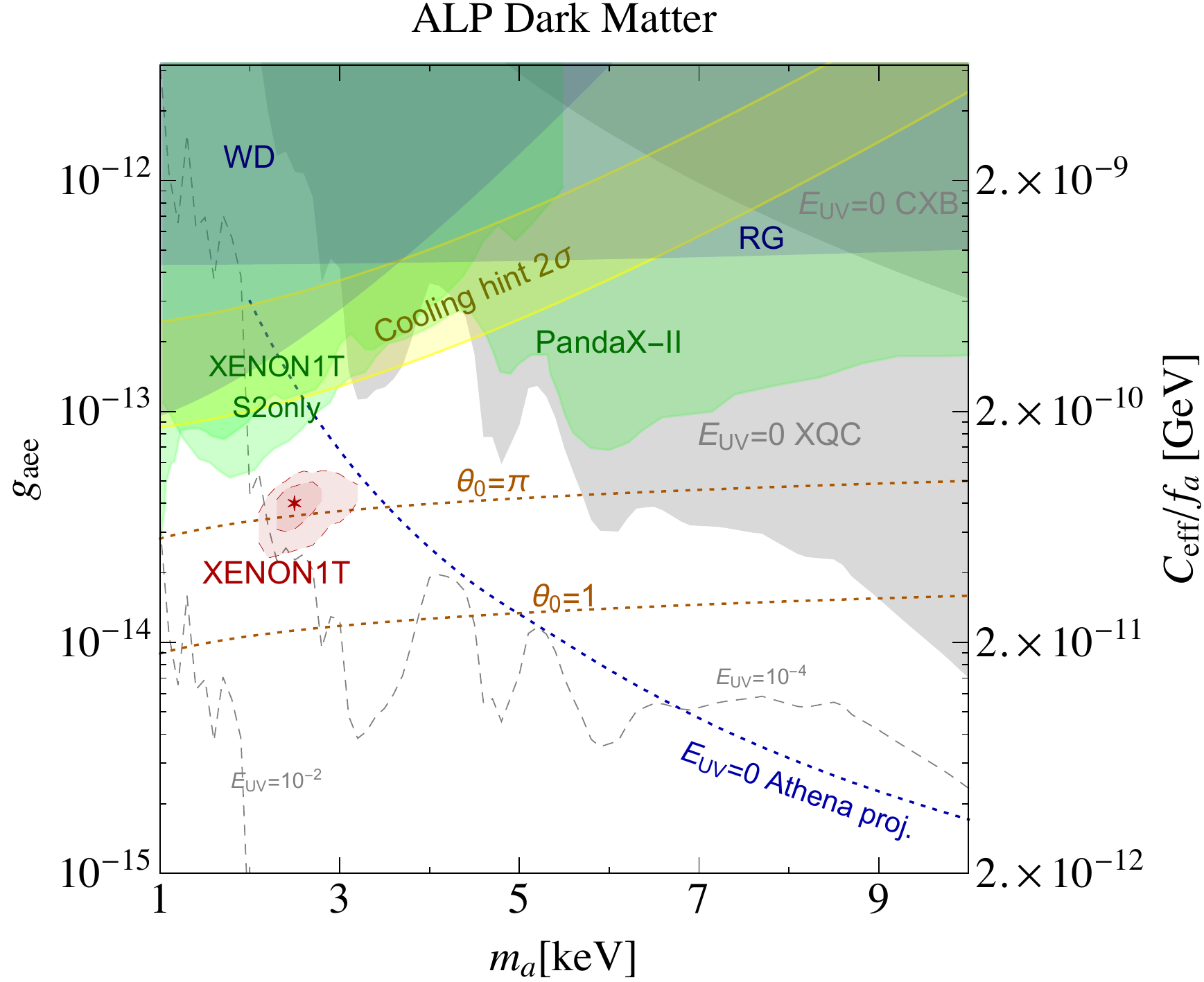}\hfill
\includegraphics[width=0.45\textwidth]{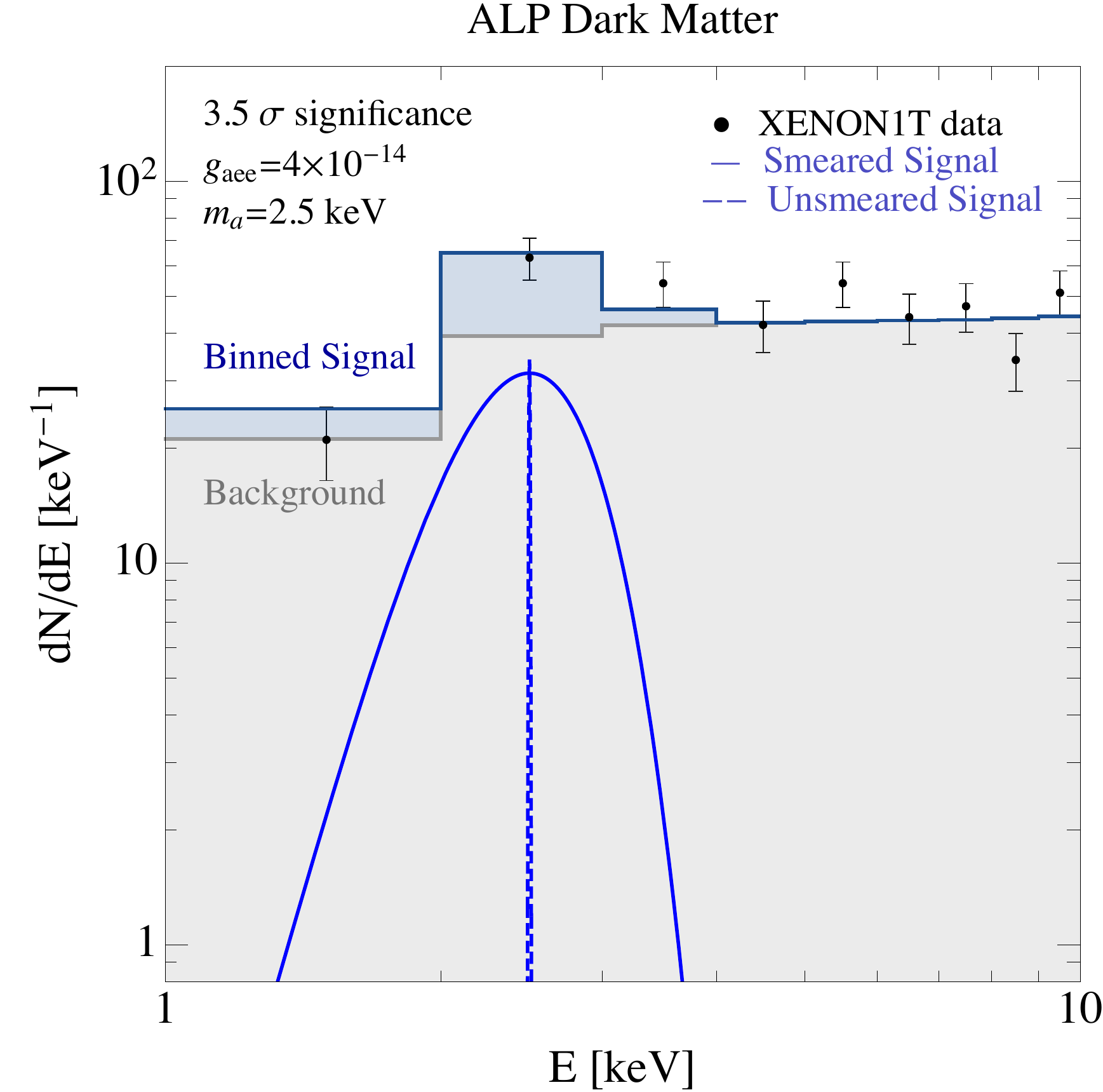}
\caption{
{\bf Left:} Allowed parameter space for ALP dark matter in the $(m_a,g_{aee})$ plane. The ALP decay constant is plotted on the right y-axis. The {\bf red star} is our best fit point in Eq.~\eqref{eq:bestfitALPDM} and and the {\bf dark red} regions are the $1\sigma$ and $2\sigma$ regions. In {\bf blue}, we show the bounds from star cooling of red giants~\cite{Raffelt:1994ry,Viaux:2013lha} and white dwarfs~\cite{Bertolami:2014wua}, and in {\bf green} the current direct detection constraints from Xenon1T and PandaX~\cite{Fu:2017lfc, Aprile:2019xxb}. The {\bf gray dotted} contours show the X-rays constraints from XQC~\cite{Boyarsky:2006hr,Figueroa-Feliciano:2015gwa} for different values of $E_{\text{UV}}$, the shaded region on the top right is excluded by XQC even for $E_{\text{UV}}=0$. We also show the weaker bound obtained from CXB~\cite{Hill:2018trh}.  {\bf Dashed brown} contours show the initial misalignment necessary to get the right DM relic abundance for $C_{\text{eff}}=1$. For completeness we show in {\bf yellow} the cooling hint $2\sigma$ band adapted from~\cite{Giannotti:2017hny}. {\bf Right:} Signal shape for the best fit point in Eq.~\eqref{eq:bestfitALPDM}. The {\bf black dots} are the \xe data, the {\bf gray shaded} region is the expected background, the {\bf blue solid/dashed} line is the signal shape after/before smearing and the {\bf blue shaded} region is the resulting signal plus background distribution.
\label{fig:axionDM}
}
\end{figure}

If the ALP is DM, the axio-electric effect should be treated in the non-relativistic limit with $E\simeq m_a$, and thus the energy absorbed by the bounded electron in the detector is equal to the axion mass. Consequently, in order to explain the \xe signal the ALP masses must be around $m_a\sim 1\text{ keV}$. 
The predicted spectrum is a narrow peak around the ALP mass, with the observed signal spreading into several bins from detector resolution 
effects that smear the predicted signal, as shown in Fig.~\ref{fig:axionDM} (right). 

The $1\sigma$ and $2\sigma$ bands of our likelihood fit is shown in red in Fig.~\ref{fig:axionDM} (left), where the best fit point is 
\begin{equation} 
m_a=2.5\text{ keV}\quad ,\quad g_{aee}=4\times 10^{-14}\quad ,\quad  2\text{log}(\mathcal{L}(S+B)/\mathcal{L}(B))=15.7\ ,\label{eq:bestfitALPDM}
\end{equation}
which corresponds to a $3.5 \sigma$ local significance. The number of signal events is given by,  
\begin{equation}
R_{\text{AE}}=33\left(\frac{\rho_{\text{DM}}}{0.4\units{GeV cm^{-3}}}\right)\left(\frac{m_a}{2.5\text{ keV}}\right)\left(\frac{g_{aee}^2}{4.\times 10^{-14}}\right)\left(\frac{\mathcal{E}}{200\text{ tonne-day}}\right)\,, 
\end{equation}
where we used that $\sigma_{\text{PE}}=1133\text{ cm}^2/\text{gram}$ and the effective \xe exposure, $\mathcal{E}(E)$ evaluated at the best fit mass . The predicted coupling to electrons fixes the decay constant to be $f_a/ C_{\rm eff}\simeq 10^{10}\text{ GeV}$ shown on the right y-axis.  
We further show constraints from white dwarfs~\cite{Bertolami:2014wua}  (dark blue) and red giants~\cite{Raffelt:1994ry,Viaux:2013lha}  (light blue) cooling as well as terrestrial limits from PandaX~\cite{Fu:2017lfc, Aprile:2019xxb} (light green) and the \xe S2-only analysis~\cite{Aprile:2019xxb} (darker green).  

If the coupling to photons is non-vanishing,  the ALP DM with the desired  range of masses and decay constants is severely challenged by its large decay rate into di-photons,
\begin{equation}
\Gamma_{\gamma\gamma}=\frac{g_{a\gamma\gamma}^2}{16\pi}m_a^3\ .
\end{equation} 
Imposing that the ALP is stable on timescales of our Universe we get
\begin{equation}
\label{eq:ALPDMstable}
\text{ALP stability:}\qquad \frac{E_{\text{eff}}}{C_{\rm eff}} \lesssim 23 \left(\frac{2.5\text{ keV}}{m_a}\right)^{3/2}\left(\frac{4.\times 10^{-14}}{g_{aee}}\right)\,,
\end{equation}
which gives already an upper bound on the coupling to photons in order for our best fit point to be stable. Even stronger constraints on the diphoton width come from observations of the cosmic X-ray background (CXB)~\cite{Hill:2018trh}. The best fit ALP is predicted to produce monochromatic photon-lines at frequency 
\begin{equation}
\nu_a=3\times 10^{17}\units{Hz}\left(\frac{m_a}{2.5\text{ keV}} \right)\ .
\end{equation}
A very conservative bound can be extracted by requiring the intensity of the photon line to be less than the measured CXB background at that frequency, which is $\nu_a I_{\nu_a}\simeq (2.3\pm0.2)\times10^{-11}\text{ W}\text{ m}^{-2}\text{ rad}^{-1}$. Using this procedure, we find 
\begin{equation}
\text{CXB bound:}\qquad \frac{E_{\text{eff}}}{C_{\rm eff}} \lesssim 1\times 10^{-3} \left(\frac{2.5\text{ keV}}{m_a}\right)^{3/2}\left(\frac{4\times 10^{-14}}{g_{aee}}\right)\,.
\end{equation}
This bound is very similar to the one obtained in~\cite{Arias:2012az} and could be substantially improved by looking at individual sources and performing background subtraction. For instance we consider the bounds obtained in~\cite{Boyarsky:2006hr,Figueroa-Feliciano:2015gwa} using the X-ray microcalorimeters in the XQC rocket. Using these bounds we find for the best fit value 
\begin{equation}
\text{XQC bound:}\qquad \frac{E_{\text{eff}}}{C_{\rm eff}} \lesssim 1.6\times 10^{-4} \left(\frac{2.5\text{ keV}}{m_a}\right)^{3/2}\left(\frac{4\times 10^{-14}}{g_{aee}}\right)\,.\label{eq:boundXQCD}
\end{equation}
On the left of Fig.~\ref{fig:axionDM}, we illustrate this limit with dashed gray lines for different values of $E_{\rm UV}$.  Interestingly, the X-ray bounds discussed so far can exclude a portion of the ALP parameter space even if $E_{\text{UV}}=0$, due to the irreducible one-loop contribution to the photon coupling in Eq.~\eqref{eq:oneloop}. The bound from CXB and the one from the XQC rocket are shown as shaded gray regions in Fig.~\ref{fig:axionDM} left. Future X-ray missions like Athena~\cite{Barret:2013bna}, as well as new techniques like line intensity mapping~\cite{Caputo:2019djj, Creque-Sarbinowski:2018ebl}, will further improve the X-ray bound in Eq.~\eqref{eq:boundXQCD} and could become important to test the ALP DM interpretation of the Xenon1T excess. In Fig.~\ref{fig:axionDM} we show that at the moment, even the more optimistic Athena prospects derived in \cite{Caputo:2019djj} are not enough to test the region of parameter space explaining the Xenon1T excess if $E_{\text{UV}}=0$.  

It is interesting to ask what are the conditions for an ALP DM addressing the anomaly to have the observed DM relic abundance. 
If one considers a generic axion-like particle with a non-dynamical mass $m_a$, the correct relic abundance can be generated in the region of interest via the misalignment mechanism~\cite{Preskill:1982cy,Abbott:1982af,Dine:1982ah,Arias:2012az} 
\begin{equation}
\Omega_a h^2=0.01\left(\frac{m_a}{\text{keV}}\right)^{1/2} \left(\frac{4\times 10^{-14}}{g_{aee}}\right)^2 \left(\frac{90}{g_*}\right)^{1/4}C_{\text{eff}}^2\ \theta_0^2\, ,
\label{eq:relicmaconst}
\end{equation}
On the left of Fig.~\ref{fig:axionDM}, we show in dotted brown lines two ${\cal O}(1)$ values for the misalignment angle, $\theta_0$, for which the observed DM relic abundance is obtained with $C_{\text{eff}}=1$.  We conclude that the standard misalignment mechanism, with no tuning of the ALP initial condition can  address the ALP DM relic density in the region of interest as long as $C_{\text{eff}}$ can be made sufficiently large.\footnote{In Eq.~\eqref{eq:relicmaconst} we assumed that the reheating temperature, $T_{\text{rh}}$, is larger than the temperature at which the ALP starts oscillating, $H(T_{\text{osc}})\simeq m_a$. A large reheating temperature enhances the thermal production of hot ALPs from the SM thermal bath. This hot DM component could become problematic if dominant compared to the cold one. However, it is easy to check that for an ALP coupled to electrons only there is a large parameter space where $T_{\text{rh}}\gtrsim T_{\text{osc}}$ and the ALP thermal production is suppressed~\cite{Arias:2012az,Nakayama:2014cza}.}    

All in all, we showed that a very small $E_{\rm UV}$ value is needed to explain the \xe anomaly, disfavoring most existing ALP models, and in particular the QCD axion, and hinting towards photophobic ALPs. Last, we comment on a particularly interesting example of photophobic ALP: the Majoron. In this case electron coupling are generated at loop level together with LFV couplings, after right-handed neutrinos are integrated out. A first consequence of this framework is that the \xe signal is correlated with future signals in $\mu^+\to e^+a$ and $\mu^+\to e^+\gamma a$ that could be seen at future high intensity muon facilities like MEGII and Mu3e (see~\cite{Calibbi:2020jvd} for further details). Depending on the actual seesaw scale one could  further explore the parameter space of this model by looking at $\mu\to e\gamma$ at MEGII~\cite{Heeck:2019guh}. Another interesting consequence is that since $C_{\text{eff}}\sim 1/16\pi^2$, non-minimal production mechanisms are required to enhance the Majoron relic abundance beyond the misalignment contribution~\cite{Hook:2019hdk,Arvanitaki:2019rax}.  

\subsubsection{Solar ALPs}\label{sec:solaraxion}

ALPs can also be produced in the Sun through processes involving the electron and photon couplings of Eq.~\eqref{eq:axioncoupl}. Here we study solar production, not making any assumptions on the ALPs relic density.  We consider both the relativistic case, $m_a\ll T_{\odot}$, for which the energy absorbed by the bounded electrons is independent of the ALP's mass, as well as the non-relativistic case, $m_a\gg T_{\odot}$, in which the spectrum is significantly modified, improving the fit to the \xe data.   See Fig.~\ref{fig:sunfluxes} for the different spectra with a massless (relativistic) and massive (non-relativistic) ALPs.

\begin{figure}[t]
\centering
\includegraphics[width=0.49\textwidth]{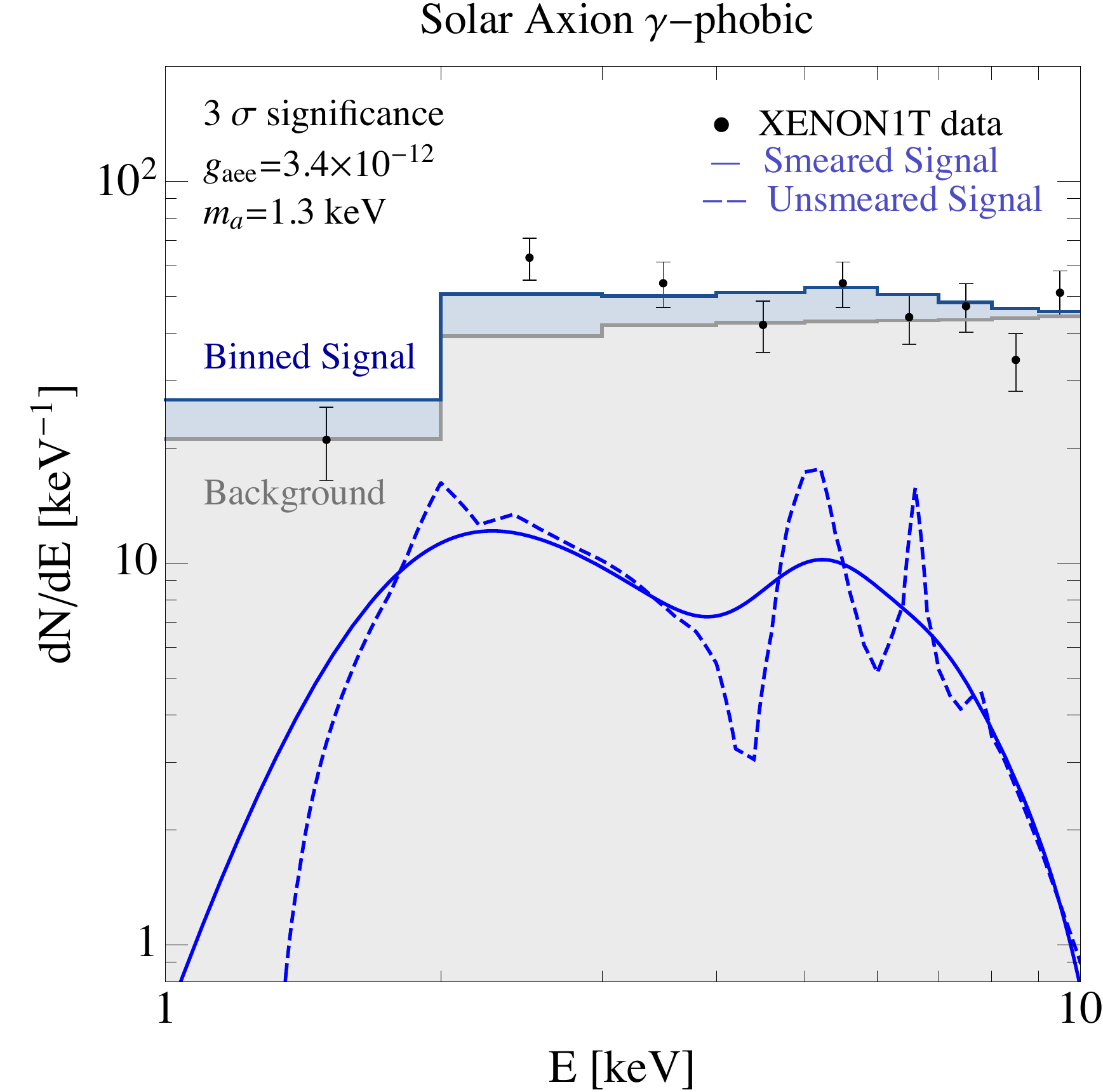}
\hfill\includegraphics[width=0.49\textwidth]{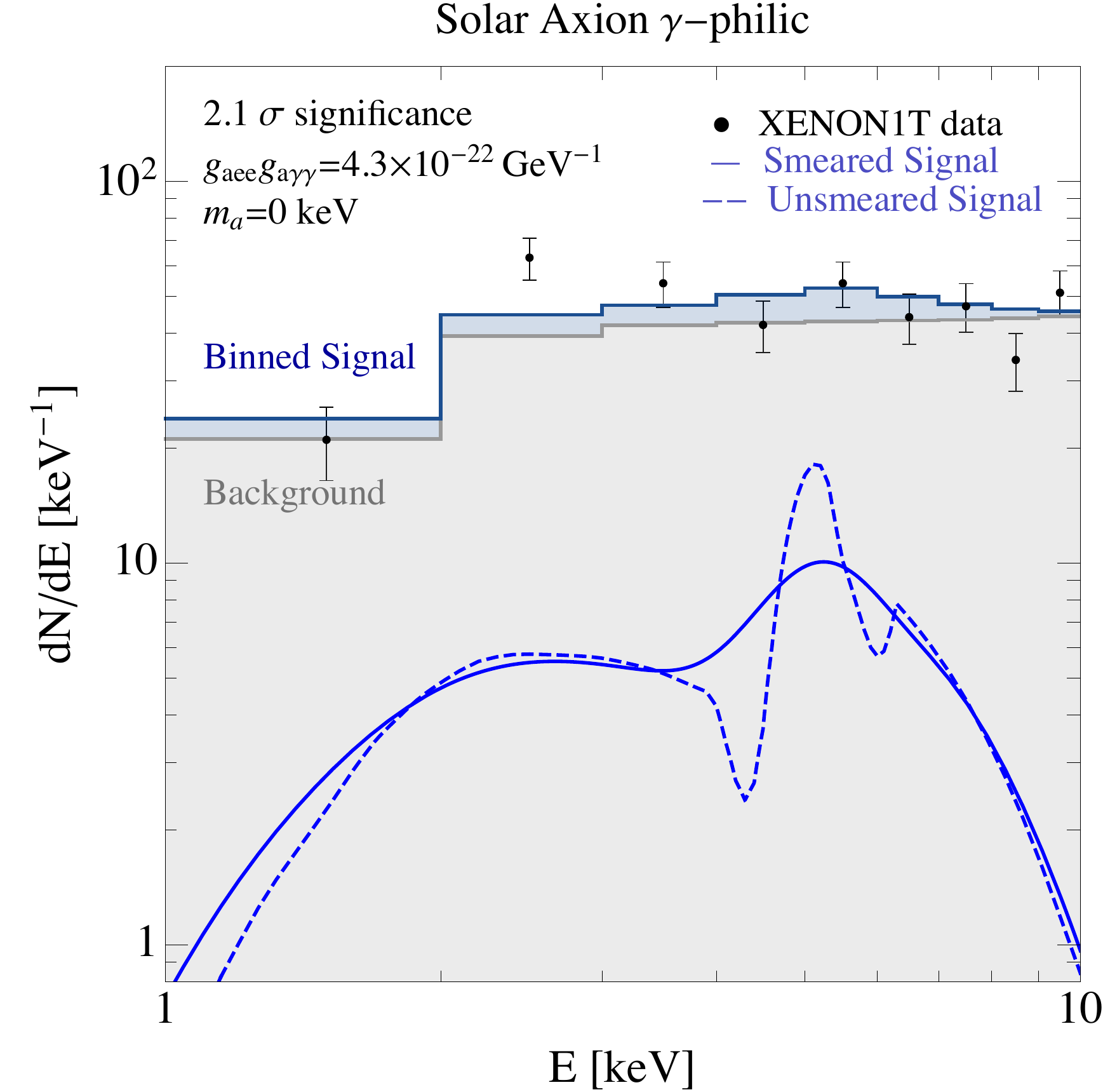}
\caption{
Predicted spectrum for the solar production of a  photophobic axion with a  best-fit value of $m_a=1.3 \units{keV}$ ({\bf left}) and a photophilic axion with a massless axion as the best fit model ({\bf right}).  The {\bf dashed} and {\bf solid} lines show the signal spectrum before and after detector smearing effects, respectively.   The measured \xe data is shown as {\bf black dots}, while the {\bf gray-shaded} region is the expected binned background, and the {\bf blue-shaded} region is the predicted binned signal.}\label{fig:axionALPsolarsignal}

\end{figure}

Two  production mechanisms are of interest: (i) the ``ABC'' processes: atomic recombination and de-excitation, bremsstrahlung, and Compton scattering, all depending on the value of $g_{aee}$~\cite{Redondo:2013wwa}.  (ii) The Primakoff process~\cite{Raffelt:1985nk}, which is the conversion of photons into axions  in the electromagnetic fields of the electrons and ions making up the solar plasma.  This is the dominant production mechanism in the energy range relevant for XENON1T, which depends on  $g_{a\gamma\gamma}$.

We discuss photophobic ALPs where both production and absorption are controlled by $g_{aee}$, so that the total signal rate scales as 
\begin{equation}
R_{\rm solar}^{\gamma-\text{phobic}}=61\, \left(\frac{\Phi_{\text{ABC, Xe}}}{2.5\times 10^{12} \text{ cm}^{-2}\text{sec}^{-1}}\right) \left(\frac{g_{aee}}{3.4\times 10^{-12}}\right)^4\left(\frac{\mathcal{E}}{200\text{ tonne-day}}\right)\,.
\end{equation}
Here $\Phi_{\text{ABC, Xe}}$ is the integrated ABC flux in the energy window that is relevant for the \xe experiment, calculated for $m_a=1.3\units{keV}$. We also consider photophilic ALP models, where the ALP coupling to photons contributes substantially to the production, while the ALP coupling to the electrons controls the absorption rate.  Here one finds 
\begin{equation}
R_{\text{solar}}^{\gamma-\text{philic}}=48\left(\frac{\Phi_{\text{P, Xe}}}{1.6\times 10^{14} \text{ cm}^{-2}\text{sec}^{-1}} \right)\left(\frac{g_{aee}}{10^{-13}}\right)^2\left(\frac{g_{a\gamma\gamma}}{4.3\times 10^{-9}\units{GeV^{-1}}}\right)^2
\left(\frac{\mathcal{E}}{200\text{ tonne-day}}\right)\ , 
\end{equation} 
where $\Phi_{\text{P, Xe}}$ is the integrated Primakoff flux, once again, in the energy window that is relevant for the \xe experiment, and with a massless ALP.
We find  that for the energy range of interest ($\omega \gtrsim \units{keV}$), the ABC productions are subdominant for $g_{aee}/ g_{a\gamma\gamma}\lesssim 16 \units{MeV}$, or equivalently, $E_{\text{eff}}\gtrsim 27\, C_{\rm eff}$.   This  is satisfied in many standard QCD axion models (see also~\cite{Farina:2016tgd} for an explicit model where $E_{\text{eff}}$ takes on very large values).   

\begin{figure}[t]
\centering
\includegraphics[width=0.33\textwidth]{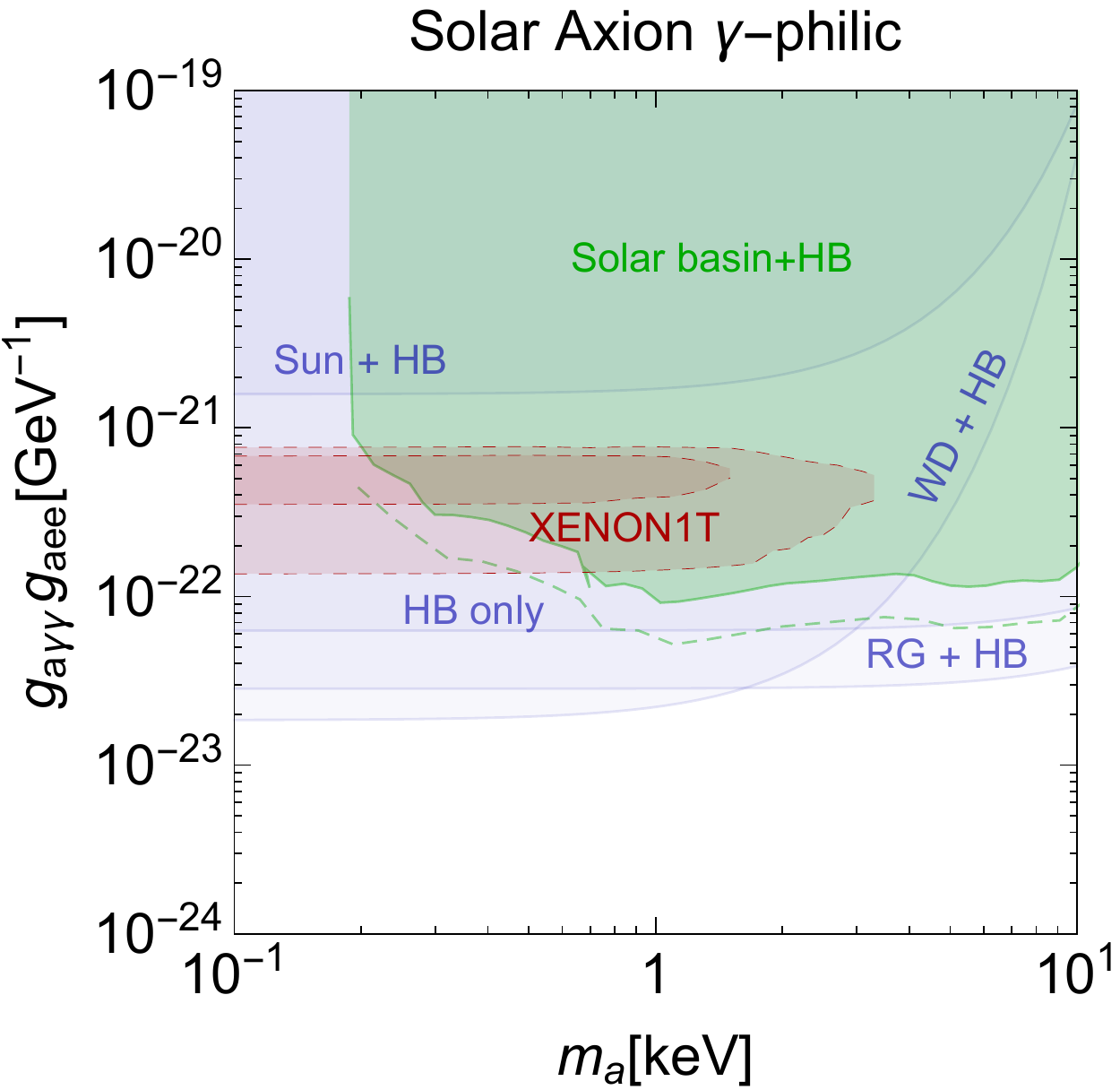}\hfill
\includegraphics[width=0.34\textwidth]{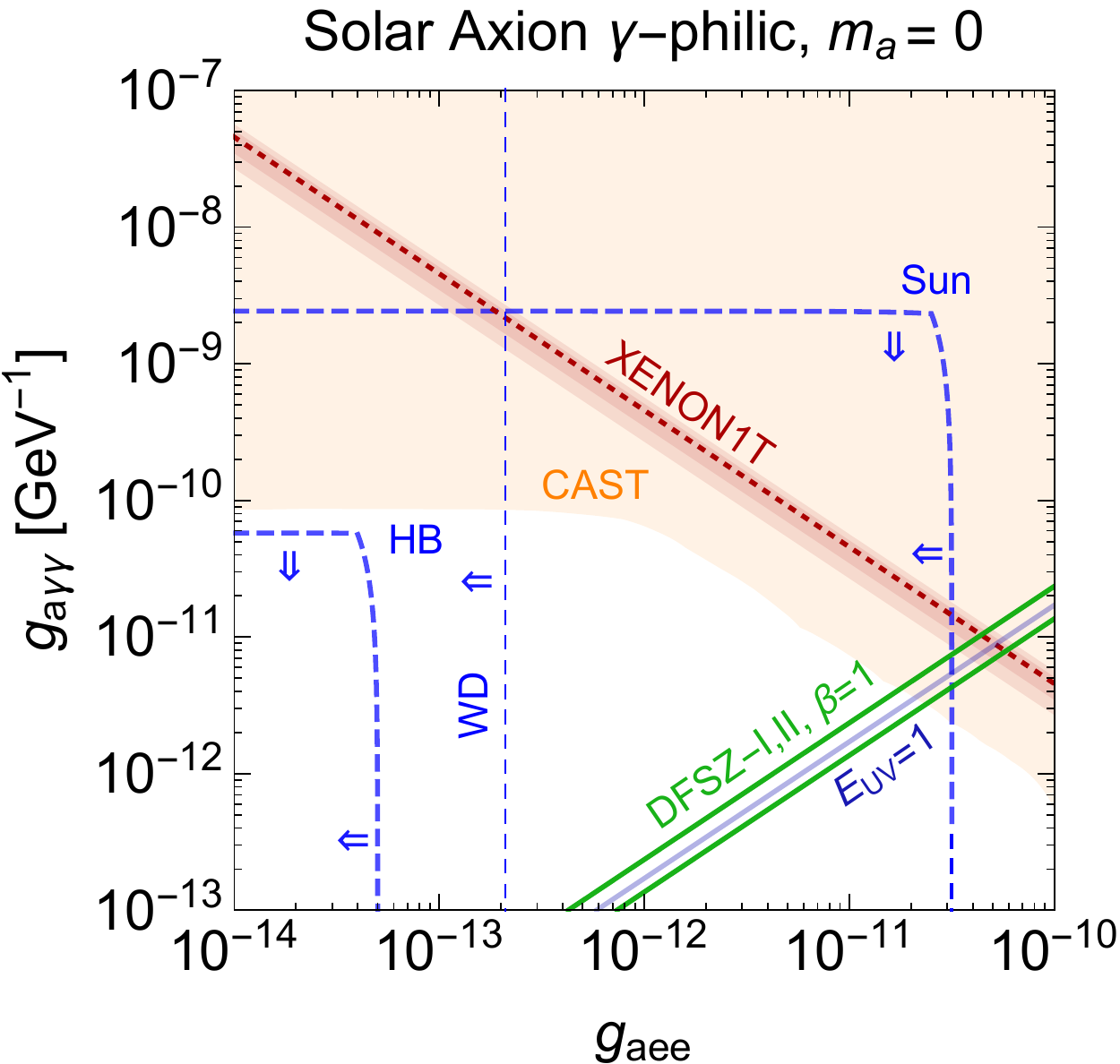}\hfill \includegraphics[width=0.33\textwidth]{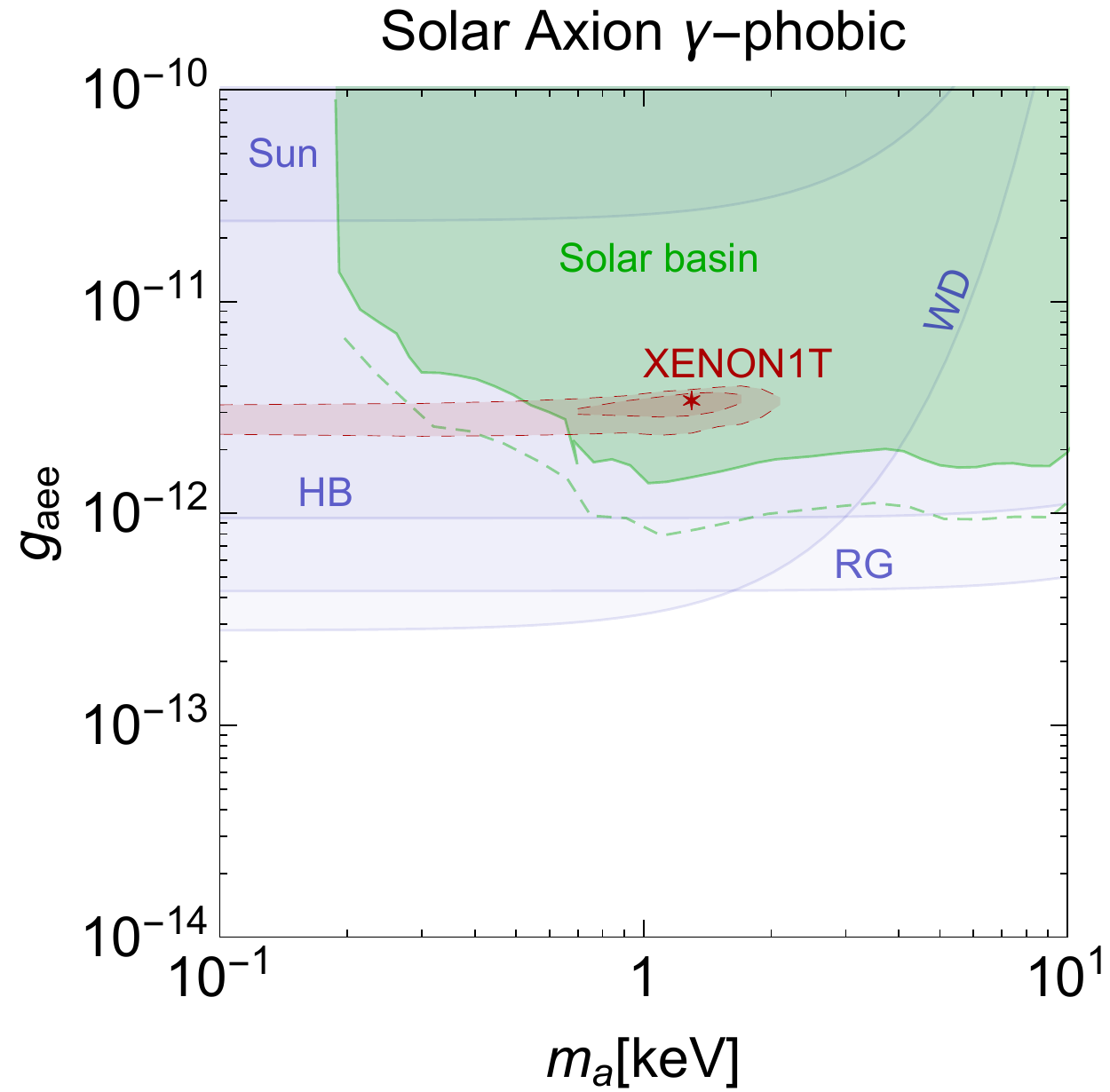}
\caption{{\bf Left:} $1\sigma$ and $2\sigma$ best-fit regions ({\bf red}) for the photophilic solar axion scenario.  The best fit point corresponds to the $m_a=0$ case and lies outside the plot.   White dwarfs (WD)~\cite{Bertolami:2014wua}, red giants (RG)~\cite{Raffelt:1994ry,Viaux:2013lha}, horizontal branch stars (HB)~\cite{Raffelt:1985nk} and Sun~\cite{Raffelt:1996wa} stellar cooling constraints on $g_{ee}$ are shown by the {\bf blue-shaded} regions in combination with the bound on $g_{a\gamma\gamma}$ from HB stars~\cite{Ayala:2014pea}, the Sun basin bound on $g_{aee}$~\cite{VanTilburg:2020jvl} is shown in {\bf green} in combination with HB stars. Different assumptions about the gravitational ejection timescale distinguish the shaded region and the dashed green line.  {\bf Middle}:  $1\sigma$ and $2\sigma$ best-fit regions as in the left plot ({\bf red}) for the best-fit massless photophilic solar axion model, but here shown in the $g_{a\gamma\gamma}-g_{aee}$ plane.  Stellar cooling constraints are indicated with {\bf dashed blue} lines, while limits from CAST~\cite{Anastassopoulos:2017ftl} are shaded in {\bf orange}; arrows point to regions that are allowed.  The theoretical axion model lines are shown in the bottom-right part of the plot.  {\bf Right}: Same as the left plot, but for the photophobic solar axion model.  The   {\bf red star} indicates the best fit point in this case. 
}\label{fig:SolarAxion}
\end{figure}

The best fit points in these scenarios are 
\begin{align}
& \gamma\text{-phobic}:\quad m_a=1.3\, \text{keV}\,, \quad g_{aee}=3.4\times10^{-12}\,,\quad  2\text{log}(\mathcal{L}(S+B)/\mathcal{L}(B))=11.5\ ,\label{eq:solarALP1}\\
& \gamma\text{-philic}:\quad m_a=0\,,\quad g_{a\gamma\gamma}g_{aee}=4.3\times10^{-22}\quad ,\quad  2\text{log}(\mathcal{L}(S+B)/\mathcal{L}(B))=6.9\ ,\label{eq:solarALP2}
\end{align}
corresponding to $3\,\sigma$ and $2.1\,\sigma$ local significance respectively. The spectrum for the two cases is shown on the top of Fig.~\ref{fig:axionALPsolarsignal}. The peaked structure of these signals is due to the convolution of the solar fluxes with the detector smearing and efficiency, suggesting that in principle, one may be able to differentiate between the two solar production mechanisms with more data. 

First, we discuss the case of a strong prior on a massless ALP. This prior can be justified as a theory bias, given that QCD axion models will typically predict an 
axion mass of $m_a= 5.70\mu\text{eV}\times(10^{12}\text{ GeV}/f_a)$~\cite{diCortona:2015ldu}, unless non-trivial dynamics modifies the behavior of QCD at high energies. The solar production of a massless photophobic axion does not reproduce well the spectral shape of the data. The reason can be traced back to Fig.~\ref{fig:sunfluxes} where one can clearly see that the ABC production does not shut off fast enough below $2.5~\text{keV}$, leaving an excess signal in the lowest energy bin. On the other hand, the massless photophilic model provides the best-fit one parameter model. The significance of the one parameter fit is $2.6\sigma$, obviously larger than the one in Eq.~\eqref{eq:solarALP2} where the ALP mass was left as a free parameter.  

Second, we comment on the case of a massive ALP. The ABC production fit can be sensibly improved by introducing an ALP mass of $1.3\text{ keV}$ shutting of kinematically the solar flux to ameliorate the agreement with the 1 keV bin. This is clearly shown in Fig.~\ref{fig:axionALPsolarsignal} left. We checked that introducing a mass does not ameliorate the Primakoff fit. Comparing Eq.~\eqref{eq:solarALP1} and \eqref{eq:solarALP2} we conclude that ABC production provides a slightly better fit to the data than Primakoff after a mass for the ALP is introduced.  

The parameter space for the solar production of the photophilic and photophobic axions is shown in Fig.~\ref{fig:SolarAxion}. On the left plot, we show in red the photophilic $1\sigma$ and $2\sigma$ best-fit regions in the $g_{a\gamma\gamma}g_{aee}$ versus $m_a$ plane.  As mentioned above, the best-fit value lies outside the plot at $m_a=0$.  Stellar cooling constraints~\cite{Raffelt:1994ry,Viaux:2013lha,Bertolami:2014wua,Raffelt:1996wa} are shown in blue.   For the same model with the best-fit value $m_a=0$, the middle plot shows constraints in the $g_{a\gamma\gamma}-g_{aee}$ plane.   The $1\sigma$ and $2\sigma$ best fit regions are shown in red, white dwarf (WD), horizontal branch (HB), and sun cooling limits are marked with dashed-blue lines, and limits form CAST are shown in orange.   We also show the predicted model lines for the DFSZ and KSVZ axion models.     Finally, on the right plot we show the $1\sigma$ and $2\sigma$ best-fit regions for the photophobic case in red and the stellar constraints in blue.  
We conclude that for all cases, the solar axion explanation to the \xe anomaly is in severe tension with stellar cooling constraints.  In Sec.~\ref{sec:Chameleon}, we discuss briefly a possible mechanism to circumvent these bounds.

\subsection{The Scalar}\label{sec:scalar}

Consider now a scalar, $\phi$, that couples to  photons and electrons 
\begin{equation}
\mathcal{L}_{\text{scalar}}=\frac{g_{\phi\gamma\gamma}}{4} \phi F_{\mu\nu}F^{\mu\nu}+g_{\phi ee}\phi  \bar{e} e\ .\label{eq:scalarcoupl}
\end{equation}
The cross section for scalar-electric (SE) effect can be written in terms of the photoelectric one as~\cite{Hardy:2016kme,Budnik:2019olh} 
\begin{equation}
\label{eq:scalarabsorption}
\sigma_{\text{SE}}(\omega_\phi)=\sigma_\text{PE}(\omega_\phi)\frac{g_{\phi ee}^2}{4\pi\alpha_{\text{EM}}v_\phi}\left(\frac{k_\phi}{\omega_\phi}\right)^2\ ,
\end{equation}
where $\omega_\phi=\sqrt{m_\phi^2+k_\phi^2}$ is the energy of the scalar $\phi$, $v_\phi$ its velocity, and $\sigma_{\text{PE}}(\omega_
\phi)$ is again the photoelectric cross section already used in Eq.~\eqref{eq:AExsec}. Notice that in the case of scalar DM, the expression above leads to a suppression of the absorption rate of $v^2_{\text{DM}}\simeq 10^{-6}$. 

The parametrization of Eq.~\eqref{eq:scalarcoupl} can be mapped to concrete models. Two particularly motivated scenarios are (i) a light SM singlet mixing with the SM Higgs doublet, and (ii) the dilaton from a spontaneously broken conformal-invariance.   Below we briefly review these models, pointing to the distinct nature of their photon and electron couplings.
\begin{figure}[t]
\centering
\includegraphics[width=0.55\textwidth]{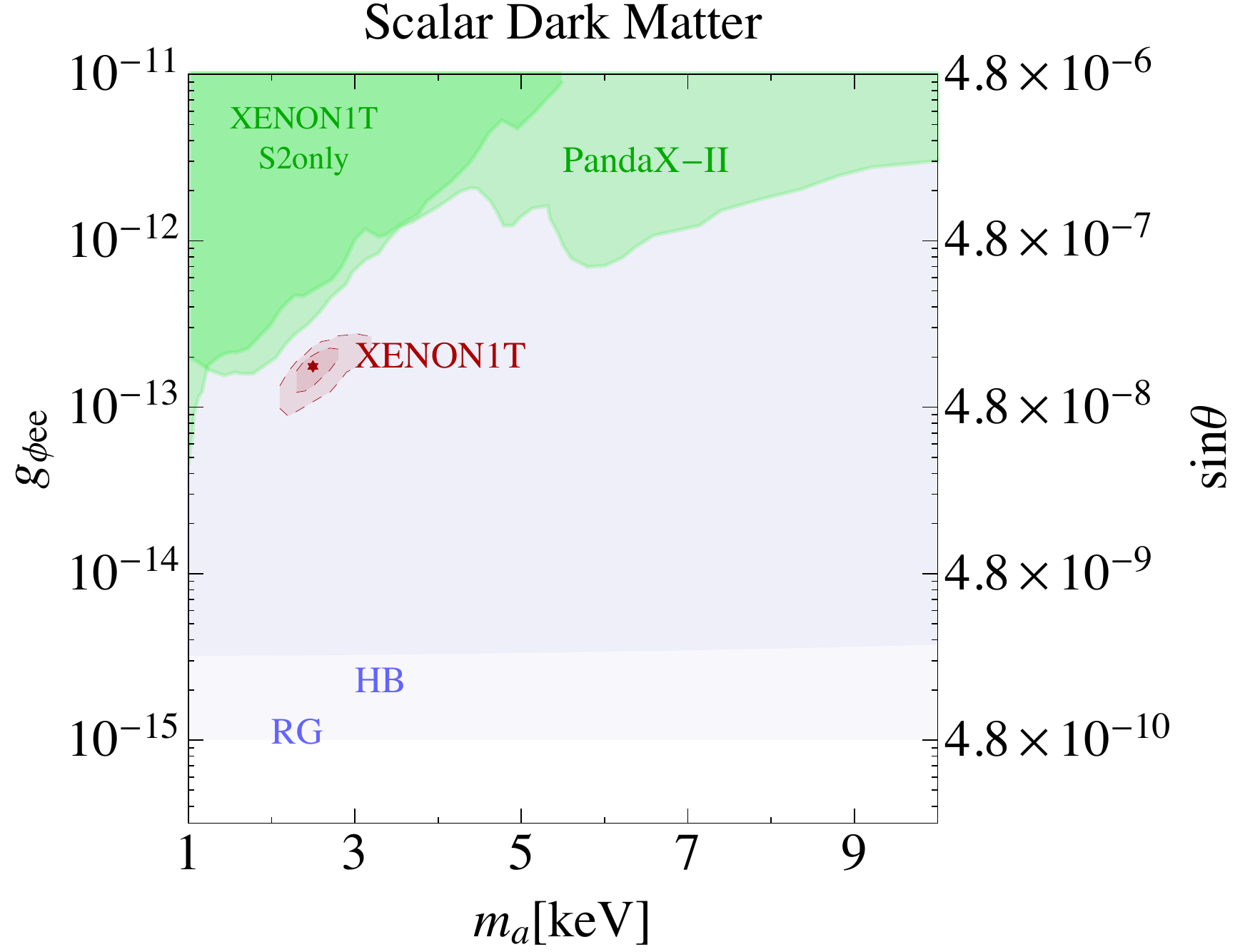}
\hfill
\includegraphics[width=0.42\textwidth]{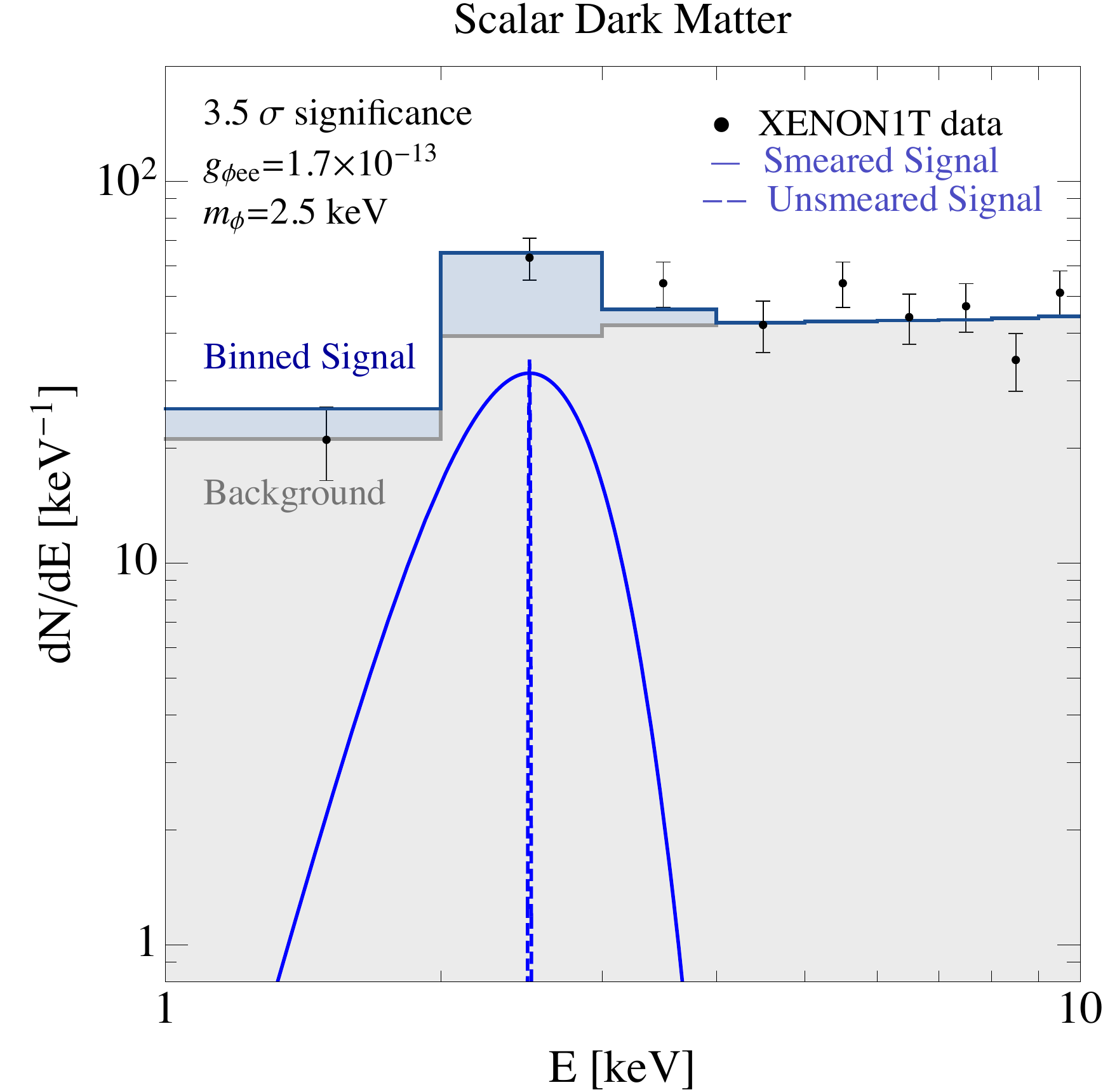}
\caption{{\bf Left:} Allowed parameter space for scalar dark matter in the $(m_a,g_{aee})$ plane. The {\bf red star} is our best fit point in Eq.~\eqref{eq:bestfitALPDM} and and the {\bf dark red} regions are the $1\sigma$ and 2 sigma regions around it. In {\bf blue} we show the bounds from star cooling of red giants and horizontal branch stars~\cite{Hardy:2016kme} and in {\bf green} the present direct detection constraints from Xenon1T and PandaX~\cite{Fu:2017lfc, Aprile:2019xxb}.  {\bf Right:} Signal shape for the best fit point in Eq.~\eqref{eq:bestfitscalar}. {\bf black dots} are the Xenon1T data. The {\bf gray shaded} region is the expected background, the {\bf blue line} is the signal shape and the {\bf blue shaded} region is the resulting signal plus background distribution.}
\label{fig:ScalarDM}
\end{figure}

A  singlet obtaining a VEV would generically mix with the Higgs through the quartic $\lambda_{\phi H} \phi^2 H^{\dagger}H$. The mixing can be written in terms of the ratio of the Higgs and the singlet VEVs, $\sin\theta= v/f$, and the final couplings of the singlet to photons and electrons are generated once the mixing is resolved~\cite{Carmi:2012yp, Carmi:2012zd, Clarke:2013aya}  
\begin{equation}
g_{\phi\gamma\gamma}=\sin\theta\frac{\alpha_{\text{EM}}}{2\pi v}\kappa_{\phi\gamma\gamma}^{\text{SM}}\qquad  ,\qquad g_{\phi ee}=\sin\theta \frac{m_e}{v}\,.\label{eq:mixhiggs1}
\end{equation} 
Here $\kappa_{\phi\gamma\gamma}^{\text{SM}}\simeq11/3+\mathcal{O}(m_\phi/m_e)^2$ is the asymptotic value of the SM loop functions from $W$'s and Standard Model fermions for $m_\phi\ll m_e$, and we fixed the coupling of the Higgs to electrons to be $y_e=m_e/v$, ignoring possible deviations from its predicted SM value. In this simple framework, the ratio between the photon and the electron coupling is fixed to $g_{\phi\gamma\gamma}m_e/g_{\phi ee}=4.2\cdot10^{-3}$, and a large coupling to nucleons is also generated from the couplings of the Higgs to gluons.  

Conversely, if the scalar in Eq.~\eqref{eq:scalarcoupl} is a dilaton, its coupling to the SM are more model dependent and controlled by the infrared (IR) trace anomaly contributions induced by direct UV couplings between the CFT and the SM. In this framework, the dilaton mixing with the Higgs can be arbitrarily suppressed~\cite{Chacko:2012sy}, and the prediction of Eq.~\eqref{eq:mixhiggs1} are changed.  In particular, for a dilaton, one can entertain the possibility of a loop-suppressed photon coupling, which decouples as $m_\phi/m_e$.  Thus, analogously to the ALP case, we consider two possibilities: 
\begin{itemize}
\item The Higgs-mixing scenario, where the ratio of the relative strength of photon and electron couplings is fixed. 
\item The photophobic dilaton scenario, where $g_{\phi \gamma\gamma}$ is suppressed as $m_\phi/m_e$ and the electron coupling dominates the phenomenology.
\end{itemize}

\begin{figure}[t]
\centering
\includegraphics[width=0.46\textwidth]{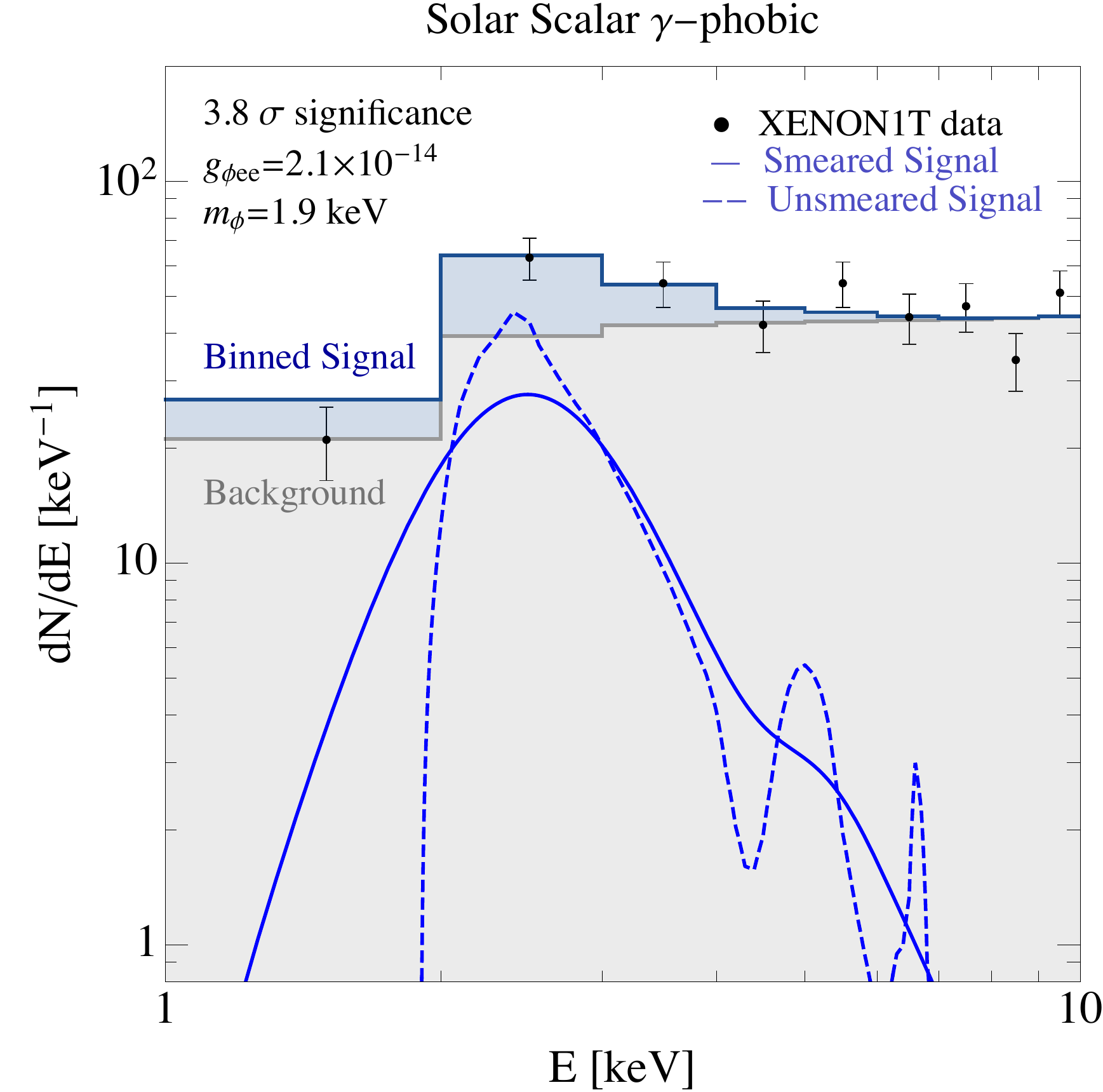}\hfill
\includegraphics[width=0.46\textwidth]{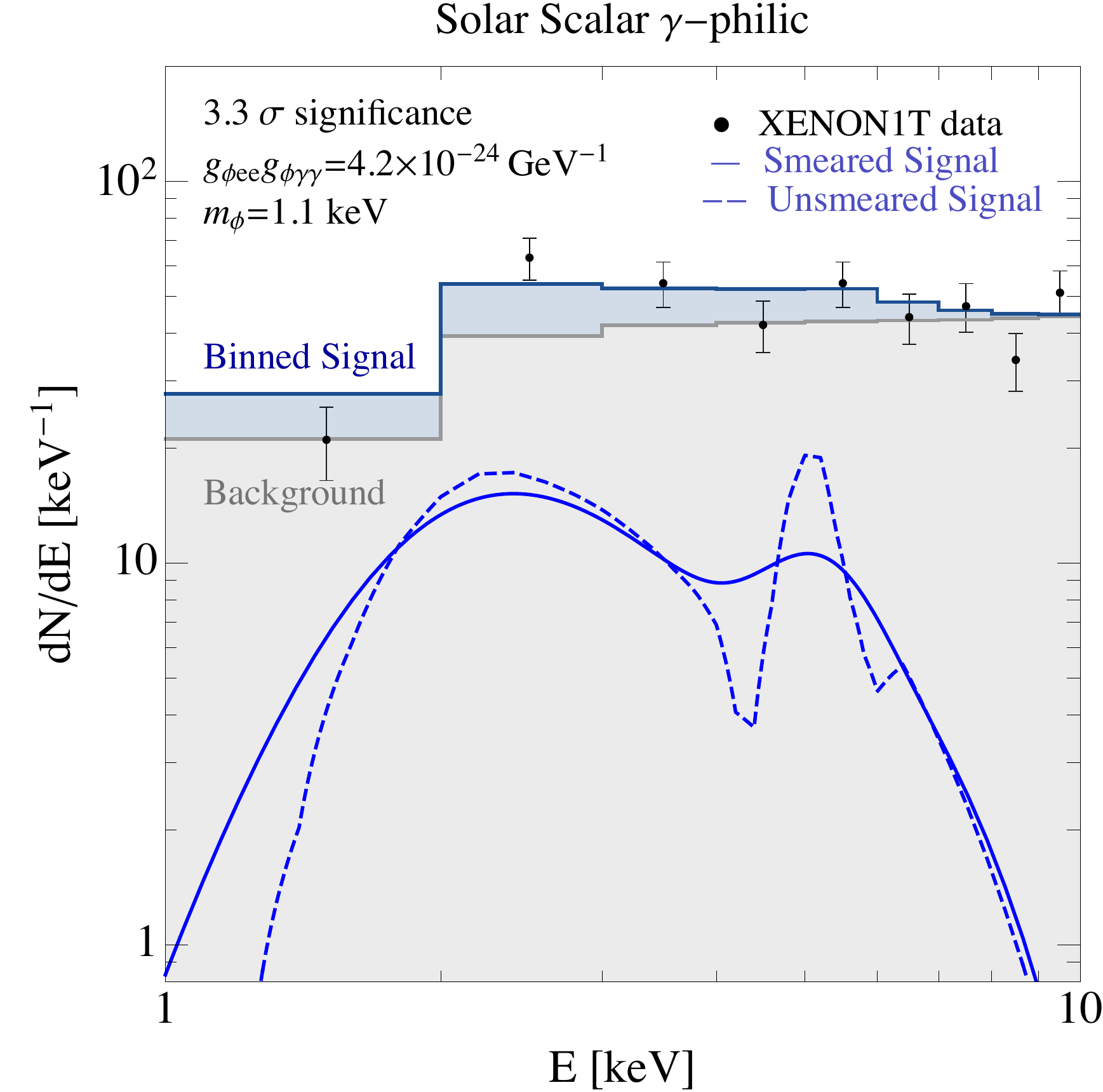}
\caption{Predicted spectrum for the solar production of a  photophobic scalar with a  best-fit value of $m_\phi=2. \units{keV}$ ({\bf left}) and photophilic scalar with a  best fit value $m_\phi=1.1\units{keV}$ ({\bf right}).  The {\bf dashed} and {\bf solid} lines show the signal spectrum before and after detector smearing effects respective.   The measured \xe data is shown as {\bf black dots} while the {\bf gray-shaded} region is the expected binned background and {\bf blue-shaded} region is the predicted binned signal. 
} \label{fig:spectrumS}
\end{figure}

\subsubsection{Scalar Dark Matter}
\label{sec:scalarDM}

As in the ALP case, the absorption spectrum of the scalar is sharply peaked around its mass, as can be seen in the spectrum plotted for the best-fit scalar DM model on the right of Fig.~\ref{fig:ScalarDM},
with values, 
\begin{equation} 
m_a=2.5\text{ keV}\quad ,\quad g_{\phi ee}=1.7\times 10^{-13}\quad ,\quad  2\text{log}(\mathcal{L}(S+B)/\mathcal{L}(B))=15.7\, .
\label{eq:bestfitscalar}
\end{equation}
The number of signal events is given by 
\begin{equation}
R_{\text{AE}}=33\, \left(\frac{\rho_{\text{DM}}}{0.4\text{ GeV}/\text{cm}^3}\right)\left(\frac{m_a}{2.5\text{ keV}}\right)\left(\frac{g_{\phi ee}^2}{1.7\times 10^{-13}}\right)\left(\frac{\mathcal{E}(2.2\text{ keV})}{200\text{ tonne-day}}\right)\ , 
\end{equation}
The predicted coupling to electrons corresponds to a mixing angle with the Higgs of order  $\sin\theta\simeq 2.1\times10^{-7}$. On the left of Fig.~\ref{fig:ScalarDM}, we show in red the $1\sigma$ and $2\sigma$ best-fit regions for the scalar DM case in the $g_{\phi ee}$-$m_\phi$ plane.  On the right y-axis, we map $g_{\phi ee}$ to the mixing angle for the doublet-singlet model, Eq.~\eqref{eq:mixhiggs1}.  Regions excluded by RG cooling constraints~\cite{Raffelt:1994ry,Viaux:2013lha} are shown in light blue, while the exclusion regions due to the \xe S2-only analysis~\cite{Aprile:2019xxb} and PandaX-II analysis~\cite{Fu:2017lfc} are shown in dark and light green, respectively. As one can see from~Fig.~\ref{fig:ScalarDM} the scalar DM cannot explain the \xe excess because of the large suppression of its absorption rate compared to the ALP case.  

\subsubsection{Solar scalar}
\label{sec:solarscalar}

Much like ALPs, light scalars can be produced in the Sun, whether or not they constitute DM.
For a photophobic scalar, the production in the Sun is dominated by electron-nucleus scalar-bremsstrahlung $N+e\to N+e+\phi$. The rate can be obtained through  the rescaling of the regular photon-bremsstrahlung by the ratio of the matrix elements squared. Doing so we find 
\begin{equation}
\frac{\Gamma(N+e\to N+e+\phi)}{\Gamma(N+e\to N+e+\gamma)}=\frac{g_{\phi ee}^2}{4\pi\alpha_{\text{EM}}}\ .
\end{equation} 
The above agrees numerically with the one given in~\cite{Redondo:2013wwa}. 
Similarly, a photophilic scalar is produced via the Primakoff process, with a rate similar to that of the ALP.   The predicted fluxes are shown in Fig.~\ref{fig:sunfluxes}.

We fit both the photophilic and photophobic scalar to the \xe data.   We find the best-fit points 
\begin{align}
& \gamma\text{-phobic}:\quad m_\phi=1.9\, \text{keV}\,, \quad g_{\phi ee}=2.1\times10^{-14}\, ,\quad  2\text{log}(\mathcal{L}(S+B)/\mathcal{L}(B))=18\ ,\label{eq:solarscalar1}\\
& \gamma\text{-philic}:\quad m_\phi=1.1\, \text{keV}\,,\quad g_{\phi\gamma\gamma}g_{\phi ee}=4.2\times10^{-24}\, ,\quad  2\text{log}(\mathcal{L}(S+B)/\mathcal{L}(B))=13.7\ ,\label{eq:solarscalar2}
\end{align}
for which we show with dashed and solid blue lines the predicted spectrum before and after smearing respectively in Fig.~\ref{fig:spectrumS}.   As before, the gray region shows the expected binned background while the  blue fillings show the binned contribution of the signals.  The \xe data are shown in black.

\begin{figure}[t]
\centering
\includegraphics[width=0.47\textwidth]{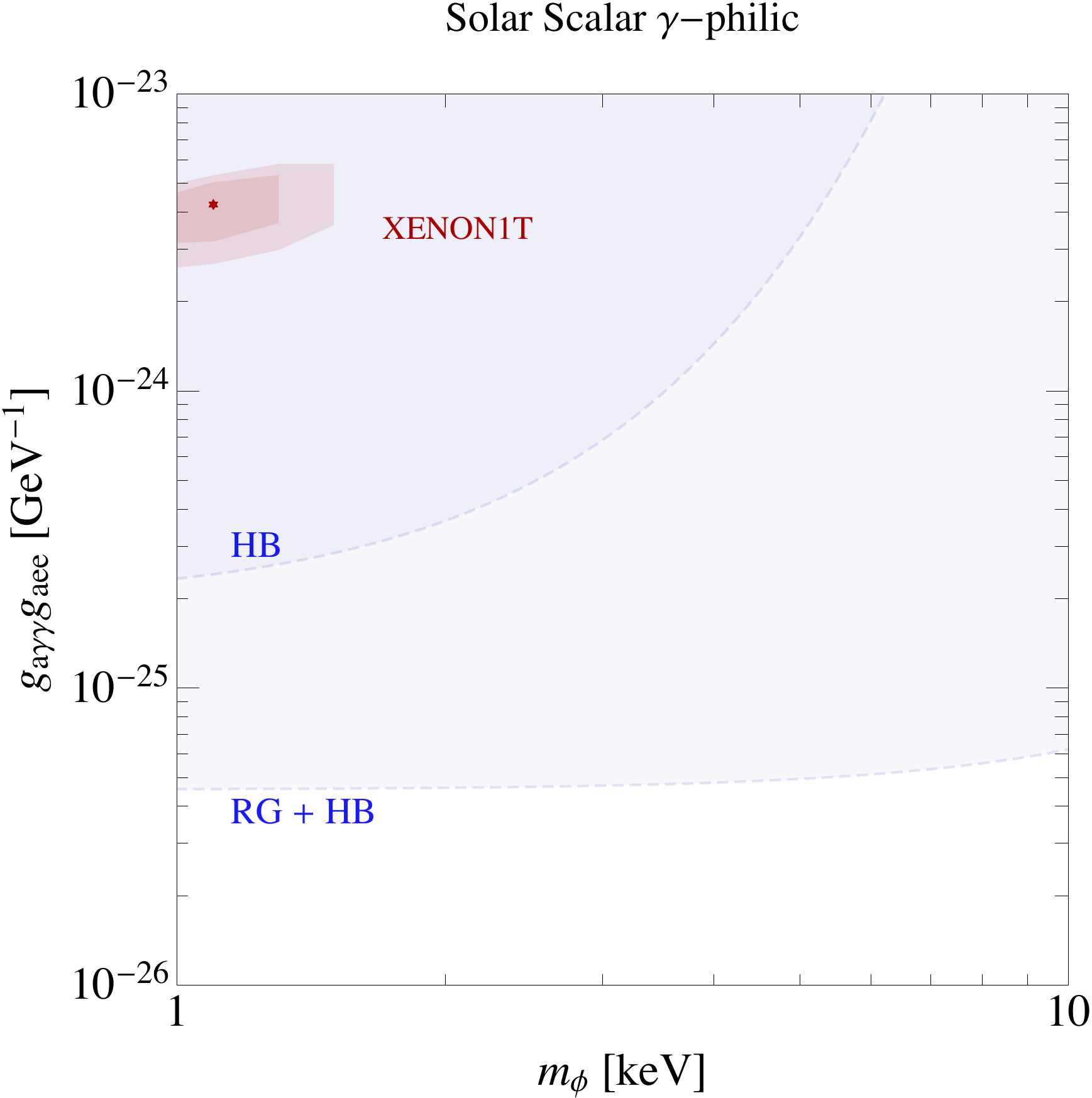}
\hfill
\includegraphics[width=0.47\textwidth]{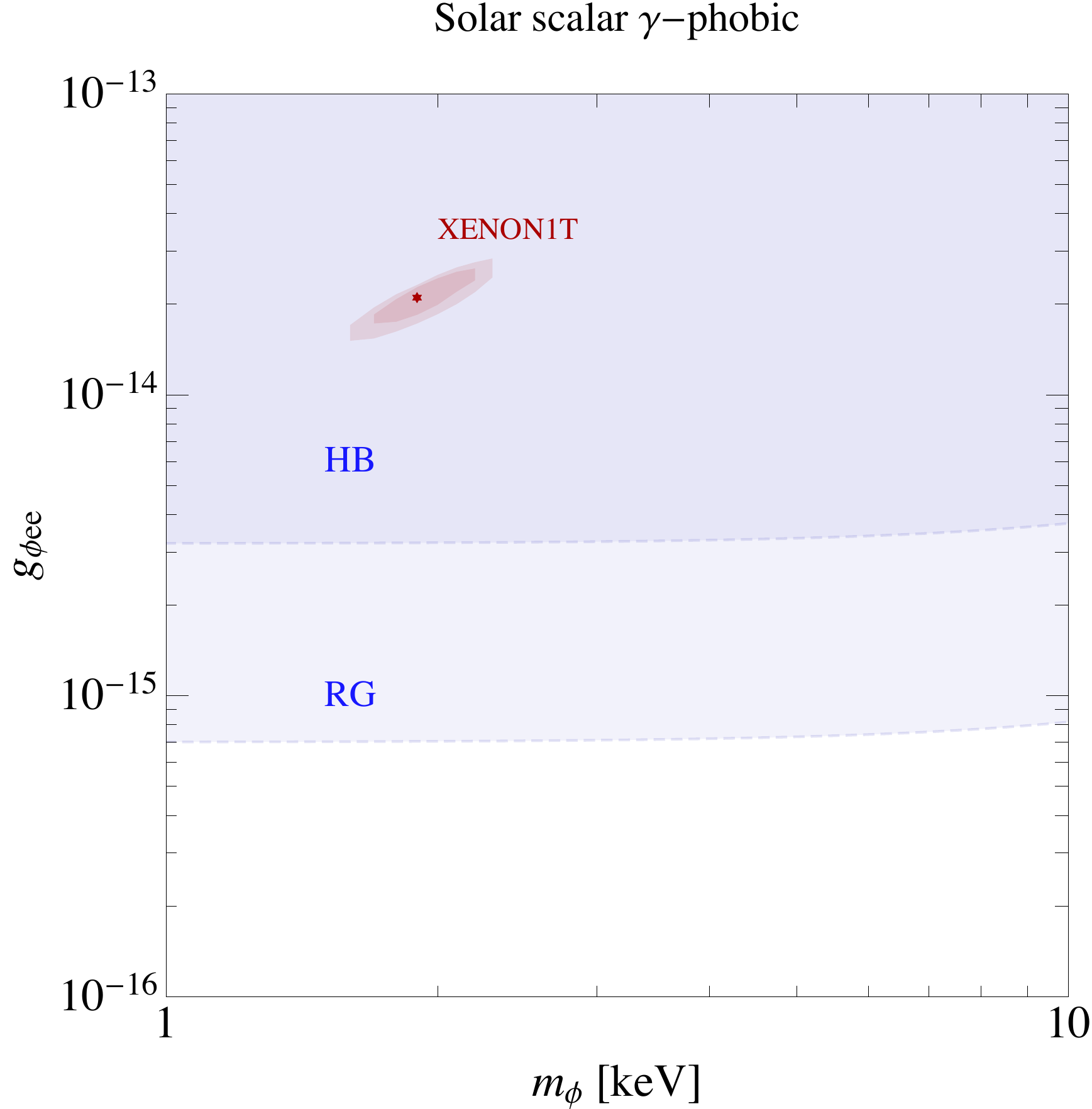}
\caption{{\bf Left:} $1\sigma$ and $2\sigma$ best-fit regions ({\bf red}) for the photophilic solar scalar ({\bf left}) and photophobic solar scalar ({\bf right}) scenarios.    
Red giants (RG)~\cite{Raffelt:1994ry,Viaux:2013lha}, and horizontal branch (HB)~\cite{Raffelt:1996wa,Hardy:2016kme} stellar cooling constraints are shown by the {\bf blue-shaded} regions. The   {\bf red stars} indicate the best fit points in both cases.   In contrast to a photophilic ALP, the scalar must be massive in order to explain the data, due to its sharply rising absorption rate at low energy.}

 \label{fig:SolarScalar}
\end{figure}

In Fig.~\ref{fig:SolarScalar}, we show in red the $1\sigma$ and $2\sigma$ best-fit regions of the solar production of a photophilic (left) and photophobic (right) scalars.    In both cases, only a massive scalar can explain the \xe anomaly.   This is in contrast to the photophilic ALP case for which the massless ALP provided the best fit.  The reason for this can be traced back to the rapidly falling absorption rate at high energies, Eq.~\eqref{eq:scalarabsorption}.  This implies a soft spectrum, which must be cut off at production through kinematic effects from a massive particle.    For the photophilic case, combined stellar cooling constraints are shown in blue while those are shown separately for the photophobic scalar.

\subsection{The Dark Photon}
\label{sec:darkphoton}

As the final absorption scenario of this section, let us consider the dark photon $A'$, a massive gauge boson of a broken (dark) gauge group $U(1)'$.  The dark photon may couple to ordinary matter via its kinetic mixing with the visible photon~\cite{Holdom:1985ag}. Much as in the previous sections, we consider the absorption of a dark photon DM and the production of a dark photon in the Sun as explanations to the \xe anomaly.

The relevant interactions are
\begin{equation}
	\mathcal{L} = -\frac{1}{4}F'^{\mu\nu}F'_{\mu\nu}-\frac{\epsilon}{2}F^{\mu\nu}F'_{\mu\nu} + \frac{1}{2}m^2_{A'}A'^{\mu}A'_{\mu}+e A_\mu \bar{e}\gamma_{\mu} e\ ,
\end{equation}
where $m_{A'}$ is the mass of the dark photon, $F^{\mu\nu}$ and $F'^{\mu\nu}$ are the photon and dark photon field strength respectively, and $\epsilon$ is the kinetic-mixing parameter. After the kinetic terms are diagonalized, the dark photon couples to the electron vector current with a coupling strength $\epsilon e$, and the dark-photon absorption cross-section can be related to the SM photelectric cross section by a simple rescaling,
\begin{equation}
\label{eq:DPPE}
\sigma_{\text{DP}}(E) = \epsilon\cdot\sigma_{\text{PE}}(E)\,.
\end{equation}

Inside a medium, the propagation of electromagnetic fields is determined by the polarization tensor $\Pi^{\mu\nu} = e^2 \langle J^{\mu}_{EM},J^{\nu}_{EM}\rangle$, which can be decomposed into  longitudinal and transverse components as,
\begin{equation}
	\Pi^{\mu\nu} = \Pi_T \sum_{i=1,2}\epsilon_i^{T\mu}\epsilon_i^{T\nu} + \Pi_L \epsilon^{L\mu}\epsilon^{L\nu}\,,
\end{equation}
where $\epsilon^{L,T}$ are the polarization vectors. In general, in-medium effects should be accounted for in order to correctly compute the dark photon absorption rate. We implement these effects following the discussion in~\cite{An:2013yfc,Redondo:2008aa}.
For dark photon DM with mass near 1~keV, we find that the absorption is dominated by the transverse modes, and the inclusion of the longitudinal ones modifies the rate by less than $10\%$. 

For $m_{A'}$ larger than the typical solar plasma frequency $\omega^{pl}_{\odot} \simeq 0.3\text{ keV}$, the production of dark photons in the Sun is dominated by the transverse modes  at energies $\omega\sim\text{keV}$. In such a case the flux at the Earth is found to be~\cite{Redondo:2008aa} 
\begin{equation}
	\frac{d\Phi_T}{d\omega} = \frac{1}{4\pi R^2}\int^{R_{\odot}}_0 4\pi r^2 dr \frac{1}{\pi^2}\frac{\omega\sqrt{\omega^2-m^2_{A'}}}{\exp{\omega/T}-1}\epsilon^2 \Gamma_T\,,
\end{equation}  
where the interaction rate $\Gamma_T$ is dominated by free-free absorption and Compton scattering.
At lower masses, the behavior of the flux from the Sun depends crucially on the nature of the dark photon mass~\cite{An:2013yfc}.  For a non-dynamical Stuckelberg mass, the dark and visible sectors  decouple in the $m_{A'}\to0$ limit for an on-shell $A'$. As a consequence, the rate of production/absorption of the transverse modes falls off as $(m_{A'}/T)^4$, where $T$ is the Sun's temperature. Adding a Stuckelberg mass to the dark photon will then cut off the solar flux around $\omega\simeq m_{A'}$ as shown in Fig.~\ref{fig:sunfluxes}.
Conversely, if the dark photon's mass is generated through the VEV of a dark Higgs, then the ratio between the dark photon mass and the dark Higgs mass is controlled by the ratio of the Higgs quartic and the dark gauge coupling $m_{h'}/m_{A'}\sim \sqrt{\lambda}/e'$. For $m_h\sim m_{A'}$ the production/absorption of a dynamical dark photon therefore goes predominantly through the radial component in the  $m_{A'}\to0$ limit~\cite{Pospelov:2008jk}. This case shares many features with the absorption scenarios discussed so far, and we will not discuss it here for the sake of brevity.

\subsubsection{Dark Photon Dark Matter}\label{sec:darkphotonDM}

\begin{figure}[t]
\centering
\includegraphics[width=0.47\textwidth]{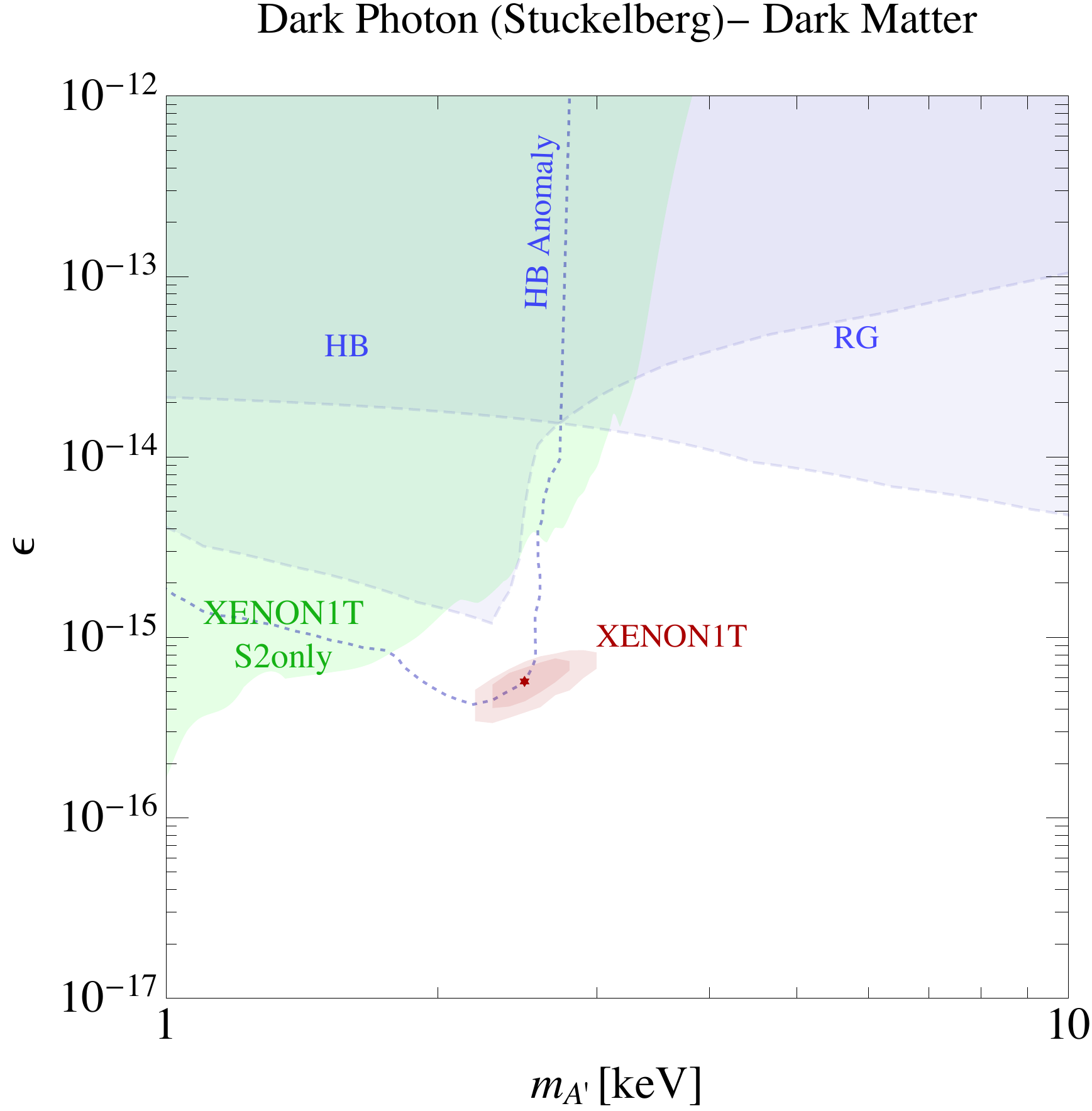}\hfill
\includegraphics[width=0.47\textwidth]{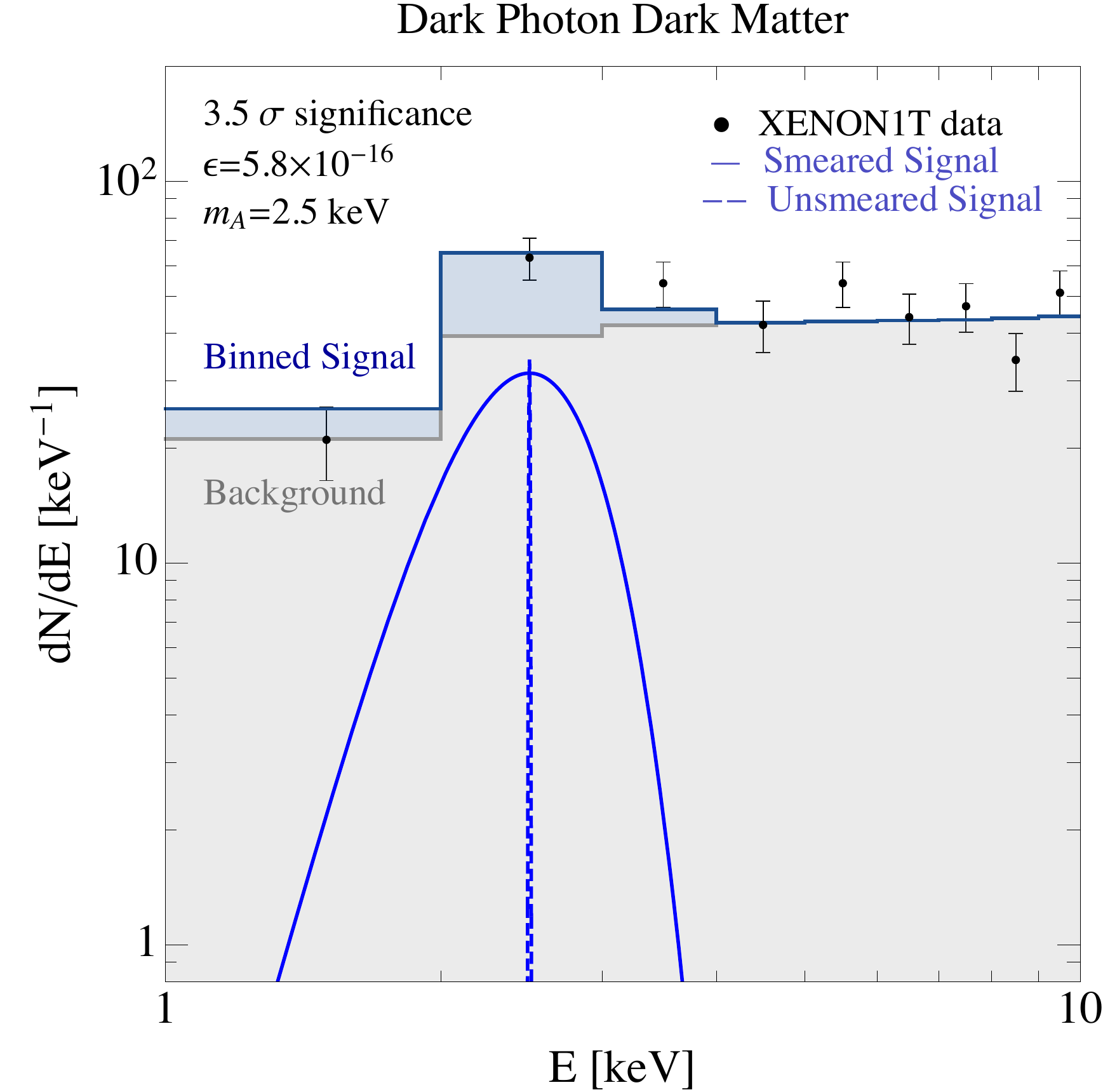}\caption{{\bf Left}: The $1\sigma$ and $2\sigma$ best-fit regions  ({\bf red}) for dark photon DM with a Stuckelberg mass. {\bf Light and darker blue} represent the RG and HB cooling limits, respectively, and the {\bf light green} region is excluded due to the \xe S2-only analysis~\cite{Aprile:2019xxb}. The {\bf dashed blue} line delimits the $2\sigma$ region which can also explain the anomalous cooling of HB stars\cite{Alonso-Alvarez:2020cdv, Giannotti:2015kwo}. 
{\bf Right}: 
An example of the predicted spectrum for dark photon DM using  the best-fit value $m_{A'}=2.5 \units{keV}$.  The {\bf dashed} and {\bf solid} lines show the signal spectrum before and after detector smearing effects, respectively.   The measured \xe data is shown as {\bf black dots}, while the {\bf gray-shaded} region is the expected binned background.  The {\bf blue-shaded} region is the predicted binned signal.  
} \label{fig:spectrumDP}
\end{figure}

If the dark photon plays the role of DM, its predicted absorption  spectrum in the \xe
 detector is very similar to the other bosonic DM cases discussed in the previous subsections. On the right panel of Fig.~\ref{fig:spectrumDP}, we show an example for the best-fit model,
 \begin{equation} 
m_{A'}=2.5\text{ keV}\quad ,\quad \epsilon=5.8\times10^{-16}\quad ,\quad  2\text{log}(\mathcal{L}(S+B)/\mathcal{L}(B))=15.7\,,
\label{eq:bestfitDP}
\end{equation}
for which the number of signal events is given by  
\begin{equation}
R_{\text{PE}}=33\, \left(\frac{\rho_{\text{DM}}}{0.4\text{ GeV}/\text{cm}^3}\right)\left(\frac{m_a}{2.5\text{ keV}}\right)\left( \frac{\epsilon}{5.7\times10^{-16}}\right)\left(\frac{\mathcal{E}(2.2\text{ keV})}{200\text{ ton}/\text{day}}\right)\,.
\end{equation}
 Dashed and solid lines represent the unsmeared and smeared spectrum, respectively. 
 We see that, as with the axion and scalar, the spectral shape  is peaked around the dark photon mass, and detector resolution allow for a reasonable fit to data.   As in previous plots, the expected binned background is shown in the figure in gray, while the binned signal is shown in blue.  The \xe data is presented with black dots.  

On the left plot of Fig.~\ref{fig:spectrumDP}, we show in red the $1\sigma$ and $2\sigma$ best-fit regions for dark photon DM with a Stuckelberg mass. In light and darker blue, we show the RG and HB cooling limits respectively and in light green, the  constraint from the \xe S2-only analysis~\cite{Aprile:2019xxb}.   We learn that the explanation of the \xe anomaly with dark photon DM is viable.

Finally, two remarks are in order.   First, a major advantage of dark photon DM compared to the ALP and scalar cases is that the decay rate of a keV dark photon into SM particles is extremely suppressed~\cite{Pospelov:2008jk}. The only decay channel allowed kinematically is $A'\to 3\gamma$, which is induced by dimension eight operators generated at one loop from the electron coupling. The width of this process is suppressed by $\sim\alpha^5\epsilon^2(m_{A'}/m_e)^8$, and the dark photon explanation to the \xe anomaly  is safely outside any bound from decaying DM. 
Second, the misalignment mechanism, which comfortably explains the scalar- and axion-DM relic densities, fails to generate the observed dark photon abundance unless a non-minimal coupling of the dark field-strength to gravity is taken into account~\cite{Arias:2012az,AlonsoAlvarez:2019cgw}. The contribution from inflationary fluctuations explored in Ref.~\cite{Graham:2015rva} explains the DM relic abundance relating directly the scale of inflation with the dark photon mass
\begin{equation}
H_I=8\times 10^{11}\text{ GeV}\left(\frac{2.5\text{ keV}}{m_{A'}}\right)^{1/4} 
\end{equation}
Lower scales of inflation can be achieved by producing the dark photon with other non-minimal mechanisms~\cite{Bastero-Gil:2018uel,Agrawal:2018vin,Co:2018lka,Dror:2018pdh}. In particular, the mechanism in~\cite{Bastero-Gil:2018uel} can  accommodate the correct  DM abundance for a keV dark photon by postulating a coupling $\phi F'\tilde{F}'$ between the inflaton, $\phi$ and the dark photon. In principle the different inflationary production mechanisms of dark photon DM could be distinguished by looking at the detailed features of the matter power spectrum at short scales.

\subsubsection{Solar Dark Photon}\label{sec:solardarkphoton}

\begin{figure}[t]
\centering
\includegraphics[width=0.49\textwidth]{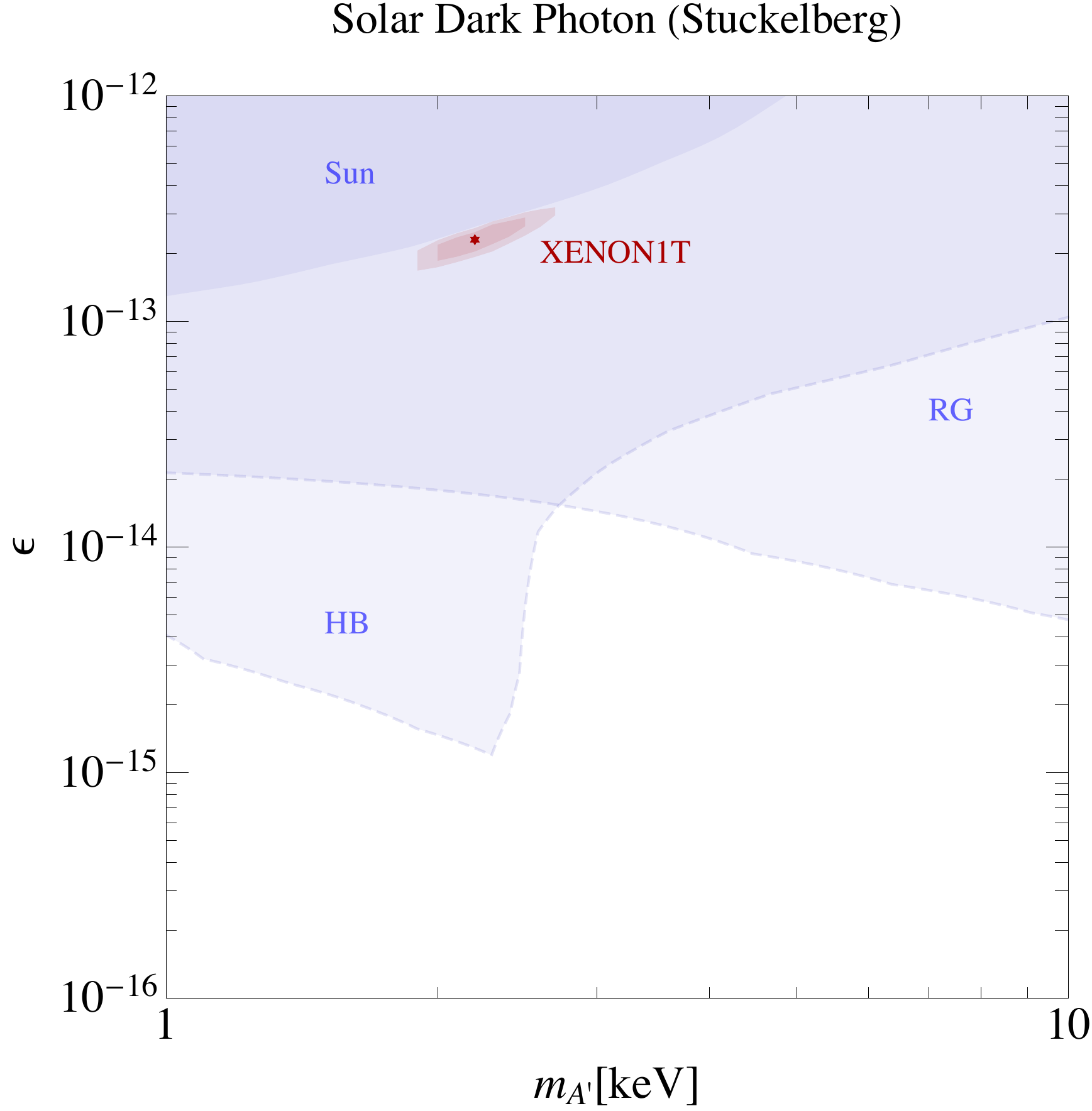}
\hfill
\includegraphics[width=0.49\textwidth]{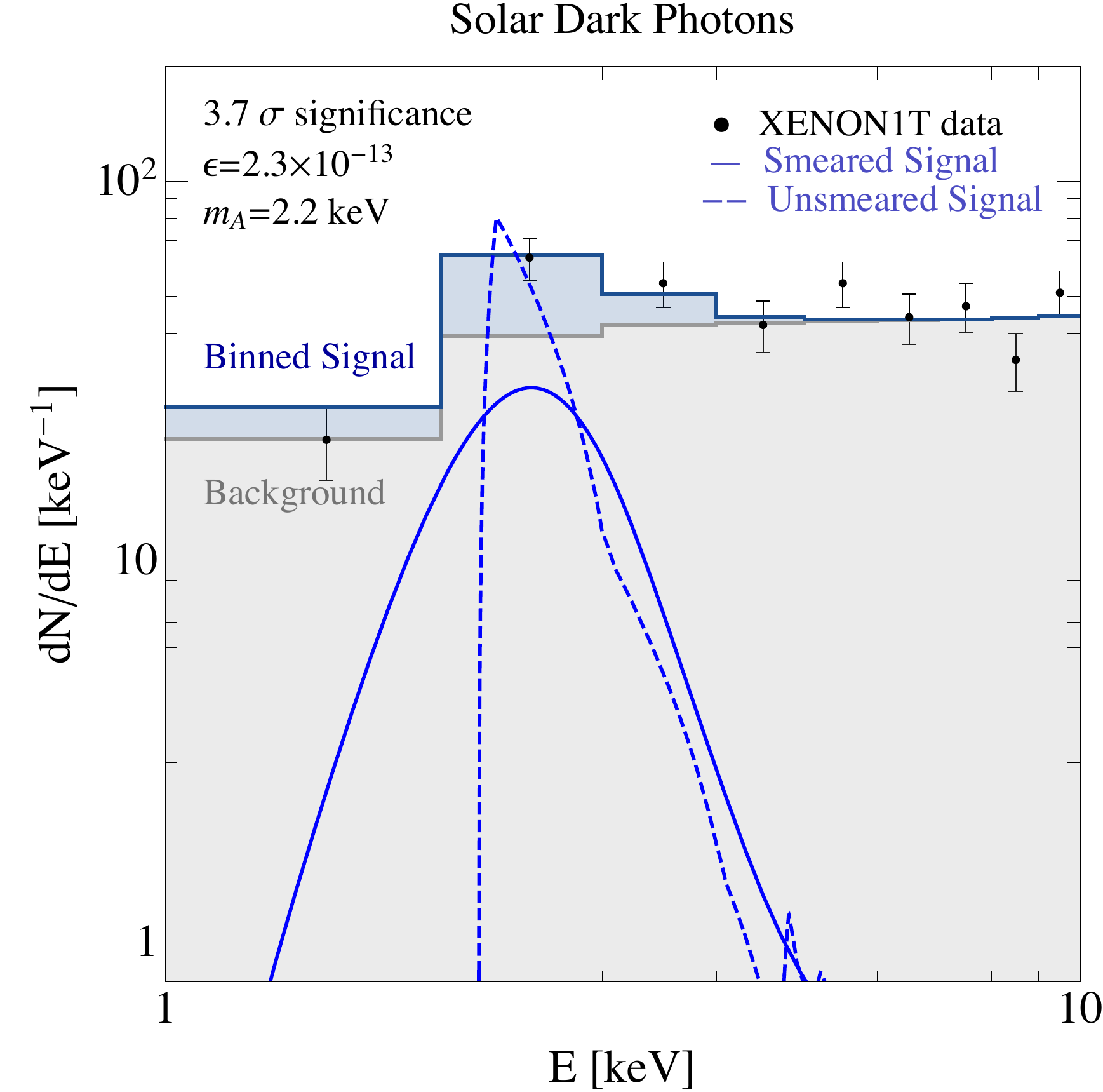}
\caption{{\bf Left}: The 1- and 2-$\sigma$ best-fit regions  ({\bf red}) for a dark photon with a Stuckelberg mass produced in the sun. {\bf Blue regions} represent the Sun~\cite{Redondo:2013lna,Giannotti:2015kwo},RG and HB~\cite{Raffelt:1994ry,Viaux:2013lha} cooling limits. The {\bf red star} indicates the  best fit point.
{\bf Right}: 
An example of the predicted spectrum for the solar dark photon  using  the best-fit value $m_{A'}=2.3 \units{keV}$.  The {\bf dashed} and {\bf solid} lines show the signal spectrum before and after detector smearing effects respective.   The measured \xe data is shown as {\bf black dots} while the {\bf gray-shaded} and {\bf blue-shaded} regions are the expected binned background and signal respectively. 
} \label{fig:solarDP}
\end{figure}

For a Stuckelberg dark photon produced in the Sun, the best fit point is 
\begin{equation} 
m_{A'}=2.2\text{ keV}\quad ,\quad \epsilon=2.3\times 10^{-13} ,\quad  2\text{log}(\mathcal{L}(S+B)/\mathcal{L}(B))=17.3\ .\label{eq:bestfitDPsolar}
\end{equation}
As for the case of the scalar, the presence of a mass cuts off the low-energy  flux to reduce the signal yield in the lower \xe bins. 
The unsmeared and smeared spectrum, together with the binned background, signal, and data is shown in Fig.~\ref{fig:solarDP} (right).  In the left plot, we show the best fit region for the model, together with the HB and RG stellar cooling bounds.   We learn that as for the scalar and ALP, the best-fit regime is robustly excluded by the astrophysical bounds.

\section{Chameleon-like ALPs: Circumventing the Stellar Cooling Bounds}\label{sec:Chameleon}

As discussed in the previous section, particles produced in the Sun are excluded as an explanation for the \xe anomaly due to stringent stellar cooling constraints.  These constraints arise from the energy loss induced by the emission of light bosons in the star environment. In principle, the constraints can  be evaded if the properties of these particles depend on the environment, thereby allowing for a suppressed production in stars.  Such Chameleon-like particles have been studied extensively in a broader context, for example in order to evade fifth-force constraints or play the role of dark energy (see e.g.~\cite{Khoury:2003aq, Khoury:2013yya}), but also for the particular case of ALPs~\cite{Masso:2005ym,Masso:2006gc,Jaeckel:2006xm,Ganguly:2006ki,Kim:2007wj,Brax:2007ak,Redondo:2008tq}.

Here we focus on the specific case of chameleon-like ALPs (cALPs).   While most previous work has focused on suppressing the axion-photon couplings in stars, we choose to study the suppression of the axion-electron coupling, which is sufficient to open up the parameter space for solar ALP models that predict either only the latter or both couplings (see Fig.~\ref{fig:SolarAxion}).    Below we entertain a simple novel model of this kind, leaving a more general framework as well as possible generalizations for future work. 

\begin{table}[t]
\renewcommand{\arraystretch}{1.2}
\centering
\begin{tabular}{lcccc}
Star & $g_{aee}$ bound & $\rho_{\text{core}} (\text{MeV}^4)$& $T_{\text{core}} (\text{keV})$ & Ref. \\ \hline
RG  & $4.3\times10^{-13}$ & $4.3$ & 8.6 & \cite{Raffelt:1994ry,Viaux:2013lha,Straniero:2018fbv}  \\
WD & $2.8\times10^{-13}$ &  $7.7$ & 0.8 & \cite{Raffelt:1985nj,Bertolami:2014wua,Giannotti:2017hny,Corsico:2019nmr}  \\
HB & $9.5\times 10^{-13}$ & $4.3\times 10^{-2}$ & 8.6 &\cite{Raffelt:1985nk}\\
Sun & $ 2.4\times 10^{-11}$ & $6.7\times10^{-4}$ &1.3 &\cite{Gondolo:2008dd,Redondo:2013wwa} \\
\end{tabular}
\caption{\label{tab:astroinput} Summary of the bounds on the electron coupling $g_{aee}$ from star cooling with the rough value of the density at the core.}\label{tab:electronbounds}
\end{table}

For the axion-electron coupling, $g_{aee}$, four stellar cooling bounds may need to be addressed: RG, WD, HB stars, and Sun cooling. The resulting bounds on the ALP electron couplings are summarized in Table~\ref{tab:electronbounds}. Among the four, the solar cooling bound is the least constraining and does not exclude the ALP explanation of the \xe anomaly (see for instance Fig.~\ref{fig:SolarAxion}). The HB bound is in marginal tension with the \xe explanation if one accounts for the potentially large systematical uncertainties. 

For this reason, we focus here mostly on evading the RG and WD bounds. The energy losses in RG and WD are dominated by the production of light bosons in the highly degenerate core, where the central density is of order $\rho_{\rm WD,RG} \sim \units{MeV}^4$, roughly four orders of magnitude larger than the core density of the Sun (see Table~\ref{tab:electronbounds}). Therefore, a model that suppresses production only in high density stars while keeping it unaltered in low density ones may evade RG and WD constraints and, at the same time, leave the ALP production in the Sun unchanged. To illustrate this point, we now discuss a simple model for which production in high-density objects is suppressed.  A more thorough study of the constraints, as well as a UV-completion of this model, is left for future work. 

\begin{figure}[t]
\centering
\includegraphics[width=0.47\textwidth]{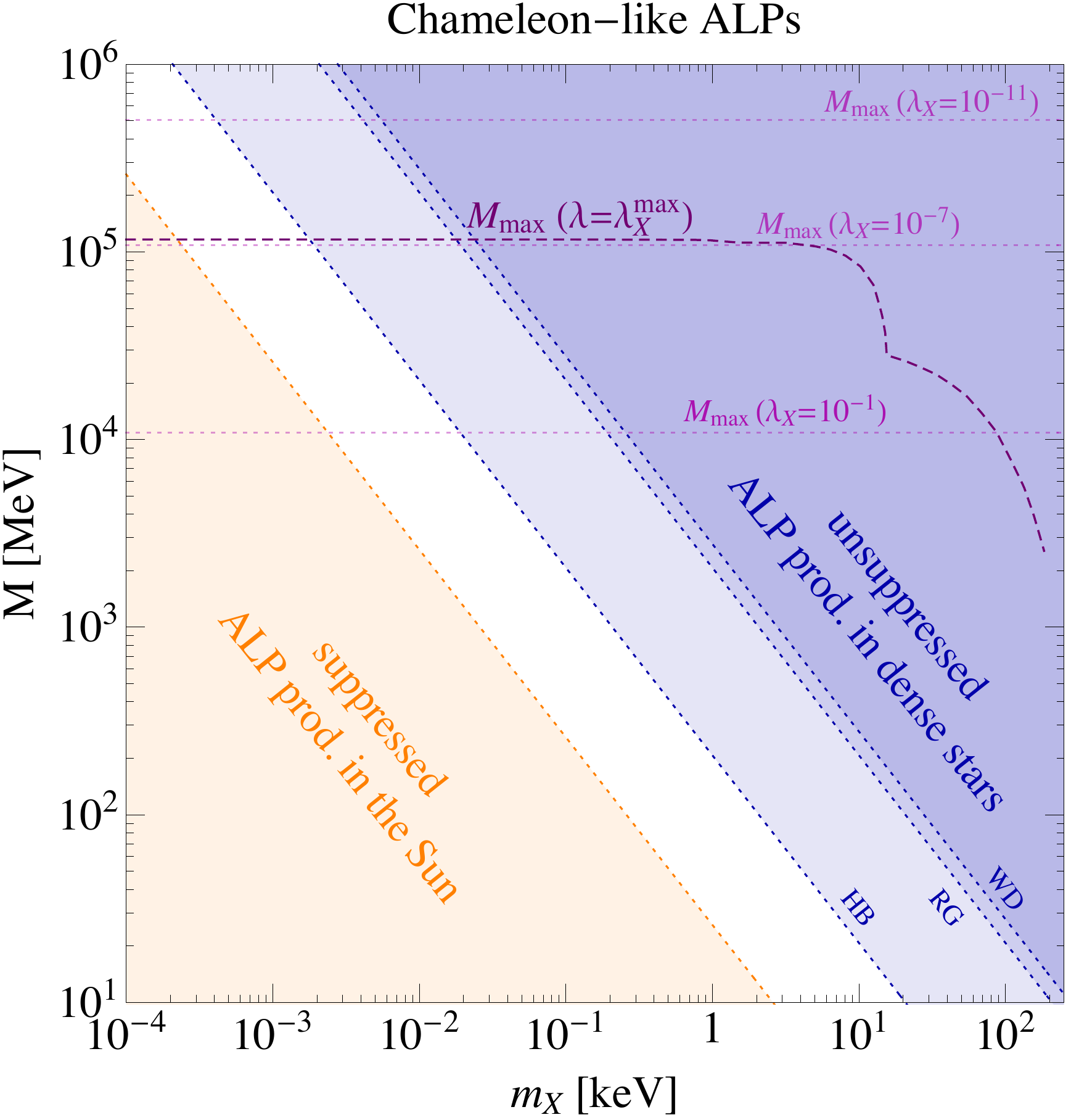}\hfill
\includegraphics[width=0.5\textwidth]{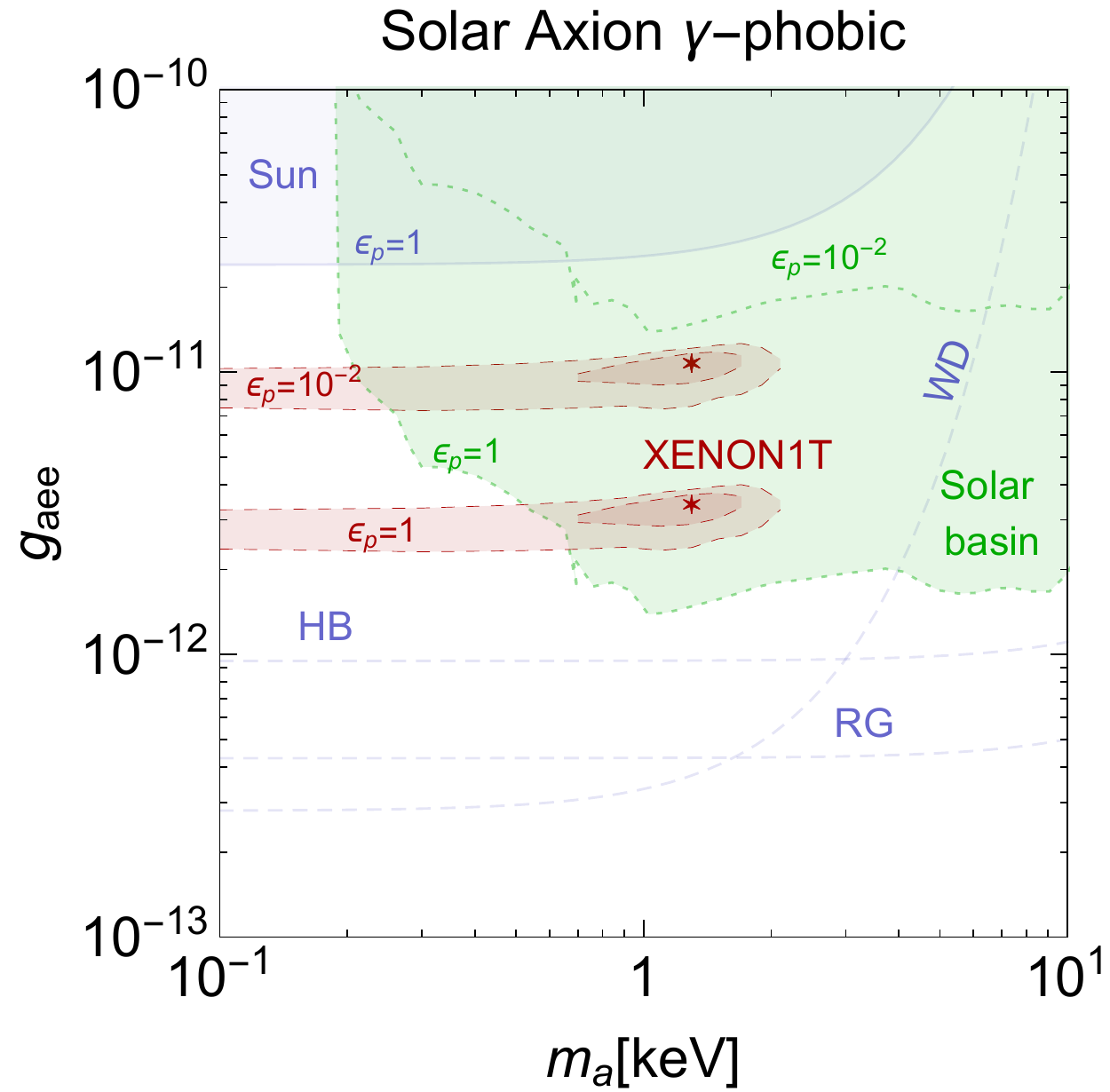}
\caption{{\bf Left:} Allowed parameter space for chameleon-like ALPs in the $(m_X, M)$ plane. In the {\bf white region} defined in Eq.~\eqref{eq:cutoff1} RG, WD and HB are shut off while the Sun production is unchanged. In the {\bf orange} and {\bf blue} either one or the other requirement is not satisfied. The maximal cutoff scale $M_{\text{max}}$ depends on the chameleon quartic coupling $\lambda_X$ as in Eq.~\eqref{cutoffmax}. The {\bf dotted  magenta} contours show different values of $M_{\text{max}}$ for different choices of $\lambda_X$. The smallness of the chameleon quartic can be taken as a measure of the fine tuning of the model. The {\bf dashed dark magenta} line shows $M_{\text{max}}$  evaluated at the maximal quartic $\lambda^{\text{max}}_X$  allowed by star cooling bounds (we fix $\lambda_S=1$) once the ALP coupling to electrons $g_{aee}$ is fixed at its best fit point. A lower quartic $\lambda_X$ will allow for a higher cutoff scale. {\bf Right:} Parameter space of the chameleon ALP produced in the Sun. The star cooling bounds from HB, WD and HB stars summarized in Table~\ref{tab:astroinput} are circumvented by the chameleon mechanism for all the parameter space shown in the left panel. The {\bf dark red} regions are the $1\sigma$ and $2\sigma$ band around the best fit point in Eq.~\eqref{eq:solarscalar1} for unsuppressed Sun flux ($S=1$) and for a suppressed Sun flux ($S=10^{-2}$). In {\bf green} we show the bounds from Sun basins~\cite{VanTilburg:2020jvl} for $S=1$ and $S=10^{-2}$. As we can see, suppressing the sun flux extend the parameter space of the model as discussed in the text. The shaded blue band is excluded by Sun cooling.}\label{fig:chameleon}
\end{figure} 

Consider a complex Standard Model (SM) singlet, $S$, charged under a Peccei-Quinn (PQ) symmetry~\cite{Peccei:1977hh}  and a real SM-singlet $X$. The two fields are odd under the same ${\mathbb Z}_2$, and $X$ couples to density. Below a given cutoff scale, $M$, we assume that the following ${\mathbb Z}_2$-invariant interactions are generated 
\begin{equation}
\label{eq:cALPModel}
\mathcal{L} \supset c_{ee} \frac{X S}{M^2} m_e e_L e_R + \frac{1}{2}\left(\frac{\rho}{M^2} -m_X^2\right) X^2 + \frac{1}{4}\lambda_X X^4 + V(S) + \text{c.c.}\,.
\end{equation}
The interaction term with the electrons can be induced in a Froggatt-Nielsen construction~\cite{Froggatt:1978nt}, where the SM electrons carry charges under the same $U(1)_{\text{PQ}}$ that rotates the complex singlet $S$. Ensuring that under that symmetry $[e_L]+[e_R]+1=0$, allows the operator above while forbidding unwanted others (we normalize the singlet charge to be $[S]$=1). The cut-off scale in such a construction would correspond to the scale of the vector like-fermions required to generate this interaction~\cite{Leurer:1992wg,Calibbi:2012yj}. For simplicity, we consider the theory below the Higgs mass scale, ignoring further complications that might arise above it. 

The potential $V(S)$ is such that  $S$ develops a VEV, $S = \frac{1}{\sqrt{2}}(f_a+s) e^{i a / f_a}$, where $s$ is the massive singlet with mass $m_s=\sqrt{\lambda_s} f_a$ and $a$ is the ALP, which is massless up to the addition of operators breaking the $U(1)_{PQ}$ explicitly. For $m_S\gg m_X$, we can neglect the $s$ dynamics and write the effective coupling of the ALP to the electrons  
\begin{equation}
g_{aee}^2=c_{ee}^2\frac{m_e^2}{M^2}\left(\frac{\rho-M^2m_X^2}{\lambda_X M^4}\right)\Theta(-\rho+M^2m_X^2)\ ,\label{eq:couplingtoel}
\end{equation}
where $\rho$ is the matter density and $\Theta(x)=0$ if $x<0$ and 1 otherwise. The second term in Eq.~\eqref{eq:cALPModel}  expresses nothing more than the idea discussed in~\cite{Hinterbichler:2010es}: at low densities, $X$ has a negative mass, obtaining a VEV.   Conversely, at high densities, its squared mass is positive, and the ${\mathbb Z}_2$ symmetry is restored.  As shown in Eq.~\eqref{eq:couplingtoel}, for $\rho\gtrsim M^2 m_X^2$ one finds $\langle X\rangle =0$, and the coupling of the ALP, $a$, to electrons vanishes, shutting down its production in stars.  

Several conditions limit the parameter space of the example above:
\begin{itemize}
\item First, in accordance with the discussion above 
\begin{equation}
\rho_{\odot, {\rm core}} \lesssim m_X^2 M^2 \lesssim \rho_{\rm WD, RG, HB}\label{eq:cutoff1}
\end{equation}
if we want to avoid WD, RG, or HB constraints while keeping the Sun flux unsuppressed. The allowed parameter space in the $(m_X, M)$ plane is shown in the white band of Fig.~\ref{fig:chameleon}.  

\item Second, the quartic $\lambda_{SX} X^2 \vert S \vert^2$ was omitted from Eq.~\eqref{eq:cALPModel} even though it is allowed by all symmetries. 
When $S$ obtains a VEV, such a quartic induces a new mass term for X that could destroy the density-dependent VEV of $X$. To avoid this,  we require $\lambda_{SX}f_a^2\lesssim m_X^2$. Independently of its bare value, this quartic will be generated at one loop via the electrons.  Putting all together we get an upper bound on the VEV of S,
\begin{equation}
f_a\lesssim  24.6\text{ MeV}\left(\frac{1}{c_{ee}}\right)\left(\frac{\rho_{\text{core}}}{1 \text{ MeV}^4}\right)\ .
\end{equation}
\item Third we want to fit still the \xe hint with the cALP. Using as a benchmark the solar ALP best-fit model in Eq.~\eqref{eq:solarALP1}, we get 
\begin{equation}
c_{ee}= 5\times 10^{-12} \lambda_X^{1/2}\left(\frac{g_{aee}}{2.6\times 10^{-12}}\right)\left(\frac{M}{\text{MeV}}\right)^3\left(\frac{\text{ MeV}}{\rho_{\text{core}}}\right)^{1/2}\,.
\end{equation}
Requiring $c_{ee}\lesssim1$ to comply with perturbativity, we get an upper bound on the cutoff scale $M$
\begin{equation}
M\lesssim M_{\text{max}}\equiv 6\text{ GeV}\left(\frac{\rho_{\text{core}}}{1\text{ MeV}}\right)^{1/6}\left(\frac{1}{\lambda_X}\right)^{1/6}\,. \label{cutoffmax}
\end{equation}
\item Finally, we need to avoid the phenomenological constraints on $X$. In the limit $m_s\gtrsim m_X$ the coupling of the chameleon field $X$  to electrons $g_{Xee}=\langle S\rangle m_e c_{ee}/M^2$ is enhanced compared to the one of the ALP and is bounded from below by 
\begin{equation}
g_{Xee}\gtrsim g_{Xee}^{\text{min}}\equiv g_{aee} \left(\frac{\lambda_X}{\lambda_S}\right)^{1/2}\ .
\end{equation}

A conservative bound on the parameter space can be obtained by requiring $g_{Xee}^{\text{min}}$ to satisfy the stellar cooling constraints~\cite{Hardy:2016kme}. Setting $g_{aee}$ to the \xe best fit and setting $\lambda_S=1$,  we get the maximal value of $\lambda_X^{\text{max}}$ allowed by stellar cooling constraints. In the mass range $10^{-4}\text{ keV}\lesssim m_X\lesssim 10\text{ keV}$ the RG bounds are the most stringent, and we find, 
\begin{equation}
\lambda_X\lesssim\lambda_X^{\text{max}}\equiv 7\times 10^{-8}\left(\frac{2.6\times 10^{-12}}{g_{aee}}\right)^2\left(\frac{g_{Xee}}{6.7\times 10^{-16}}\right)^2\left(\frac{\lambda_S}{1}\right)\,.
\end{equation}

The above reveals a hierarchy between the quartic of the PQ-breaking field $\lambda_S$, and that of the chameleon, $\lambda_X$, needed in order to make this model phenomenologically viable. This hierarchy might be difficult to realize quantum mechanically. For instance, three loop contributions to the singlet and chameleon quartics induced by their electron couplings, will act to make them both of the same order. Higher chameleon masses weaken the phenomenological bounds, allowing for a milder hierarchy between the couplings, but at the price of lowering the cut-off scale $M$ as in dictated by Eq.~\eqref{eq:cutoff1}. 
\end{itemize}

In summary, cALPs could avoid stellar cooling bounds. As shown in Fig.~\ref{fig:chameleon} right, the stellar cooling from dense stars can be circumvented if a new light scalar $X$ controls the coupling of the ALP to matter. If chameleon-like scalar $X$ lies in the mass vs cut-off range shown in Fig.~\ref{fig:chameleon} left, its potential is modified by density dependent effects. In the simplest construction, the chameleon-like scalar can be light and the cut-off of the theory can be arranged to be sufficiently high if a hierarchy between the quartic of the PQ radial mode and the quartic of the chameleon is arranged as shown in Fig~\ref{fig:chameleon} left.  

Our cALP construction is still challenged by the Sun basins constraint pointed out in~\cite{VanTilburg:2020jvl}. A possibility to relax this constraint, which we do not pursue here, is to suppress the solar production in order to relax stellar cooling bounds with respect to direct detection (see Ref.~\cite{Jaeckel:2006xm} for a first discussion of such a possibility). Indeed, for a given suppression factor, $\epsilon_p\ll 1$, in the solar production of ALPs, the solar flux scales as $\epsilon_p g_{aee}^2$, while the the solar detection rate scales as $\epsilon_p g_{aee}^4$. Increasing $g_{aee}$, while keeping the detection rate fixed, implies a relative suppression in the solar cooling bound, which scales as $\epsilon_p^{1/2}$.  Achieving this suppression requires extra fine-tuning in the model presented here but could play an important role in generalizing cALPs to the case of light scalars and dark photons.


\section{Dark Matter-Electron Scattering}\label{sec:scattering} 
If DM interacts with electrons, it can scatter off the electrons in the target material and produce an electron recoil signal~\cite{Essig:2011nj}.  
Due to the distinctive kinematics of this process, the electron recoil signal for ``standard'' DM-electron scattering peaks at recoil energies well below the keV energies needed to explain the XENON1T data; this standard process is thus in conflict with lower threshold direct-detection searches.  However, we will investigate here whether other scenarios can explain the XENON1T data: exothermic scattering off electrons as well as DM-electron interactions that increase as a function of the momentum transfer (up to some cutoff scale).  We will find that exothermic scattering off electrons work well, and momentum-dependent interactions also provide a potential explanation of  the \xe excess.

\subsection{Standard DM-Electron Scattering}\label{sec:elscattering}

We begin by reviewing the standard DM-electron scattering kinematics and formalism discussed in~\cite{Essig:2011nj,Essig:2015cda}, before discussing momentum-dependent and exothermic interactions.  
Consider a DM particle with mass $m_{\chi}$ and initial velocity $\textbf{v}$, which scatters off a bound electron, transferring a momentum $\textbf{q}$ to the electron. Energy conservation of the DM-atom system gives,
\begin{equation}
\Delta E_e + \frac{|m_{\chi}\textbf{v}-\textbf{q}|^2}{2 m_{\chi}} + \frac{q^2}{2 m_N} = \frac{1}{2} m_{\chi} v^2\,,  
\end{equation}    
where $\Delta E_e$ is the energy transferred to the electron and $m_N$ is the mass of the nucleus. 
This can be written as 
\begin{equation}\label{eq:DeltaEusual}
\Delta E_e = \textbf{q}\cdot \textbf{v} - \frac{q^2}{2\mu_{\chi N}}\,.
\end{equation}    
As the initial electron is in a bound state, it can have arbitrary momentum, and hence the momentum transfer $\textbf{q}$ could take any value. The maximum energy that can be deposited is then found by maximizing the above equation with respect to $q$, and we get 
\begin{equation}
\Delta E_e \lesssim \frac{1}{2} \mu_{\chi N} v^2\,.
\end{equation}   
For $m_\chi \ll m_N$, $\mu_{\chi N} \simeq m_{\chi}$, and almost the entire kinetic energy of the incoming DM particle can be transferred to the electron.  Since the typical DM halo velocity is $v \sim 10^{-3}$, a DM particle with mass of a few GeV can in principle produce a $\mathcal{O}({\rm keV})$ electron recoil. However, for DM with masses above the MeV scale, the \emph{typical} momentum-transfer scale is set by the electron's momentum, given by $q_{\rm typ}\sim Z_{\rm eff} \alpha m_e \sim Z_{\rm eff} \times 4$~keV, where $Z_{\rm eff}$ 
is the effective charge seen by the electron.  From Eq.~(\ref{eq:DeltaEusual}) (neglecting the second term, which is usually small), $\Delta E_e \sim 10^{-3} q_{\rm typ} \sim Z_{\rm eff} \times \textrm{few eV}$.  While higher momentum transfers are possible, they are dramatically suppressed, since it is unlikely for the electron to have a momentum that is much higher than the typical momentum.   

We can see this behavior in more detail by calculating the atomic form factor $f_{1 \rightarrow 2}(\textbf{q})$, which captures the transition 
from state 1 to state 2,
\begin{equation}
f_{1 \rightarrow 2}(\textbf{q}) = \int d^3 x\ \psi_2^*(\textbf{x}) \psi_{1}(\textbf{x}) e^{i\textbf{q}\cdot \textbf{x}}\,,
\end{equation}   
where $\psi_{1}(\textbf{x})$ $(\psi_{2}(\textbf{x}))$ is the initial bound-state (final-state) electron wavefunction. 
There are various methods to calculate the wave functions. We here consider three different approaches for calculating the form factors: 
\begin{itemize}
\item First, we follow~\cite{Essig:2012yx}, taking the initial bound-state wave functions from~\cite{BUNGE1993113} and numerically solving the Schr\"{o}dinger equation with a central potential that reproduces the bound state wavefunctions for the outgoing wave functions. We will refer to this scheme of form factors as the `Non-relativistic' form factors as these do not take into account the relativistic corrections important at high momenta. 
\item Another simple approximation for calculating the form factors without taking into account the relativistic corrections is to consider the outgoing wavefunctions as plane waves. We also consider this approach here and call the form factors so obtained as the `Plane Wave' form factors. In this approach, we also do not subtract the identity operator from the operator $e^{i\textbf{q}\cdot \textbf{x}}$; our  `Plane Wave' form factors will therefore not be correctly behaved for $q\lesssim\mathcal{O}$(keV), since the outgoing wave functions are not orthogonal to the bound state wave functions. This issue ends up not affecting our results by much, since DM-electron scattering does not typically sample the atomic form factor at $q\lesssim\mathcal{O}$(keV). Also, within this scheme, the form factors are multiplied by a Fermi factor (see below). 
\item Finally, it is important to include the relativistic corrections for high $q$. We use the available atomic form factors with relativistic corrections computed in~\cite{Roberts:2016xfw}, calling them the `Relativistic' form factors. These form factors are given for $q\ge 100$~keV.
\end{itemize}

\begin{figure}[t]
\centering
\includegraphics[width=0.46\textwidth]{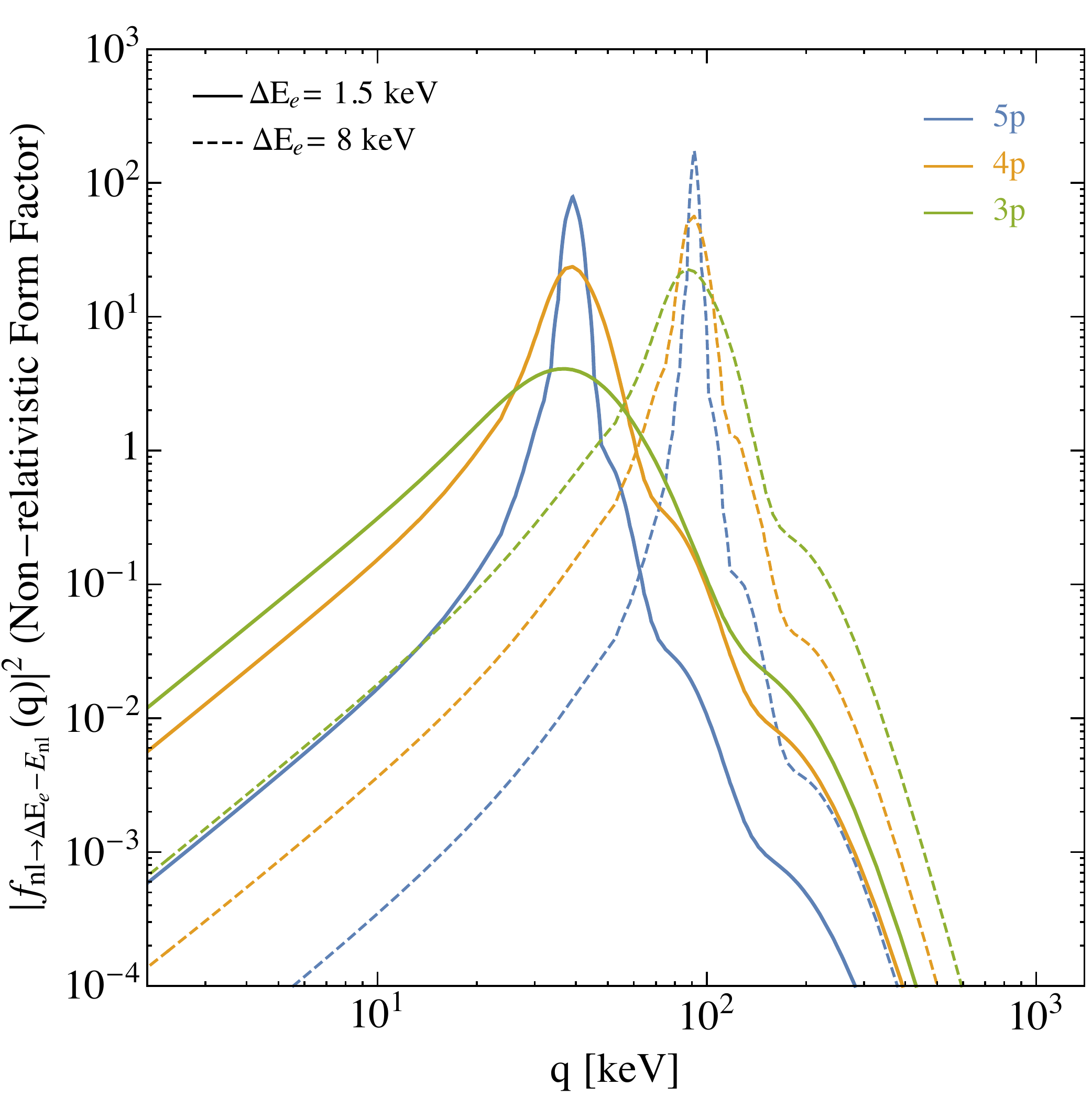}\hfill
\includegraphics[width=0.46\textwidth]{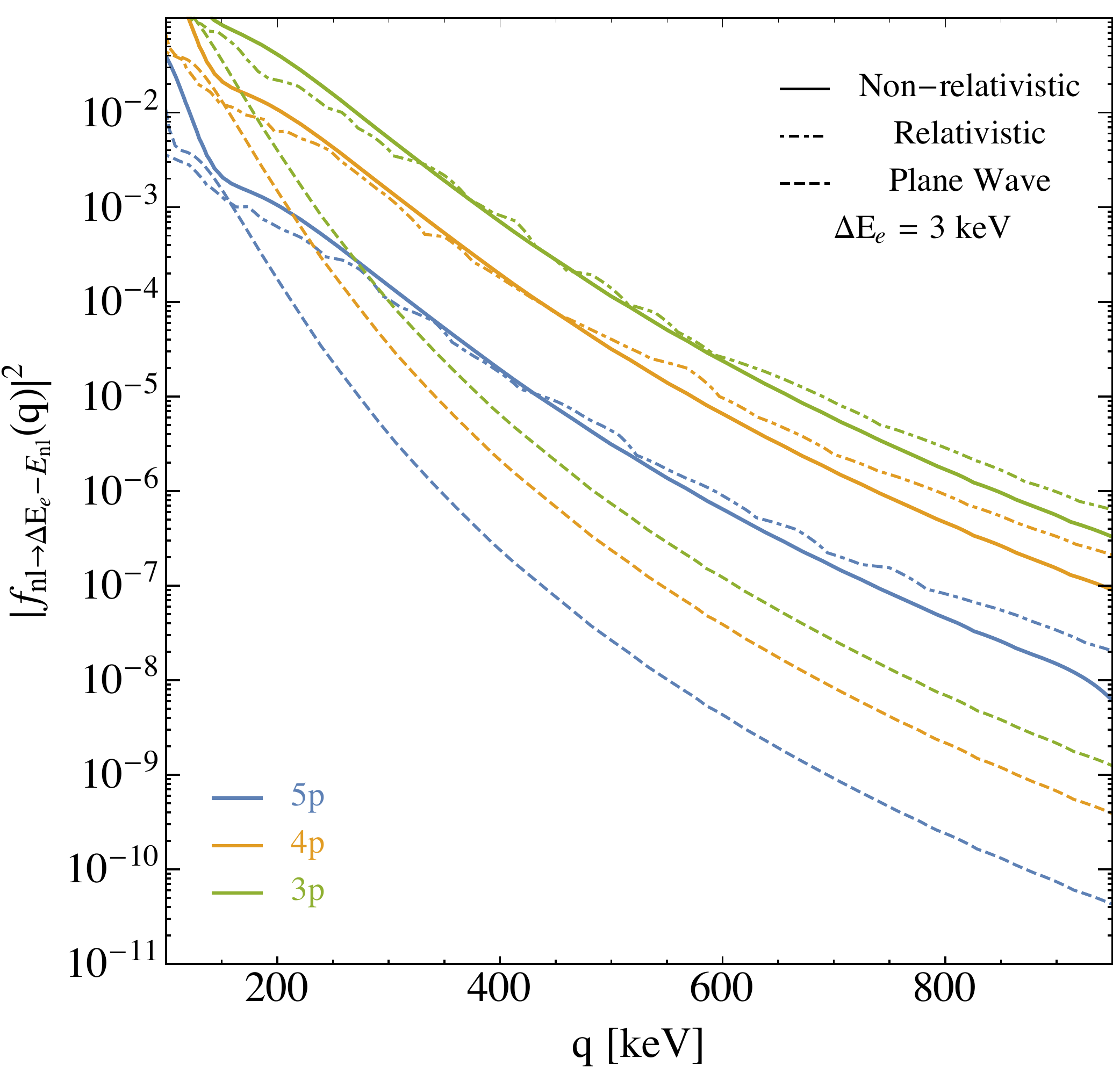}\caption{\textbf{Left:} Non-relativistic form factors for DM-electron scattering in xenon for the indicated electron shells for $\Delta E_e=1.5$~keV (solid lines) and $\Delta E_e=8$~keV (dashed lines),  where $\Delta E_e$ is the entire deposited energy. \textbf{Right:} Non-relativistic (solid lines), Plane Wave without Fermi factor (dashed lines) and Relativistic form factors (dot-dashed lines) for $\Delta E_e=3$~keV.} 
\label{fig:form-factor}
\end{figure}

In the left panel of Fig.~\ref{fig:form-factor}, we plot the non-relativistic form factors $|f_{nl \rightarrow \Delta E_e - E_{nl}}(\textbf{q})|^2$ for different initial electron shells \{n,l\} of the xenon atom and for two different values of $\Delta E_e$. This corresponds to different final outgoing electron energies $\Delta E_e -E_{nl}$, where $E_{nl}$ is the binding energy of the shell  \{n,l\}. We see that the form factor drops sharply for $q \gtrsim \alpha m_e$. For every shell, the peak also shifts to higher $q$ for higher $\Delta E_e$. 
In order for an electron in any shell to give $\Delta E_e\gtrsim 1$~keV, we need $q\gtrsim {\rm MeV}$ (see Eq.~(\ref{eq:DeltaEusual})).  This is possible, but highly suppressed. 

In the right panel of Fig.~\ref{fig:form-factor}, we compare the three different form factor schemes considered in this paper at a fixed value of $\Delta E_e=3$~keV. We see that for $q \gtrsim 500$~keV, the relativistic corrections start becoming important. We also see that for $q \gtrsim 200$~keV, the plane wave calculation underestimates the form factor, justifying the inclusion of the Fermi factor in the calculation of the scattering rates (see below). 

\begin{figure}[t]
\centering
\includegraphics[width=0.46\textwidth]{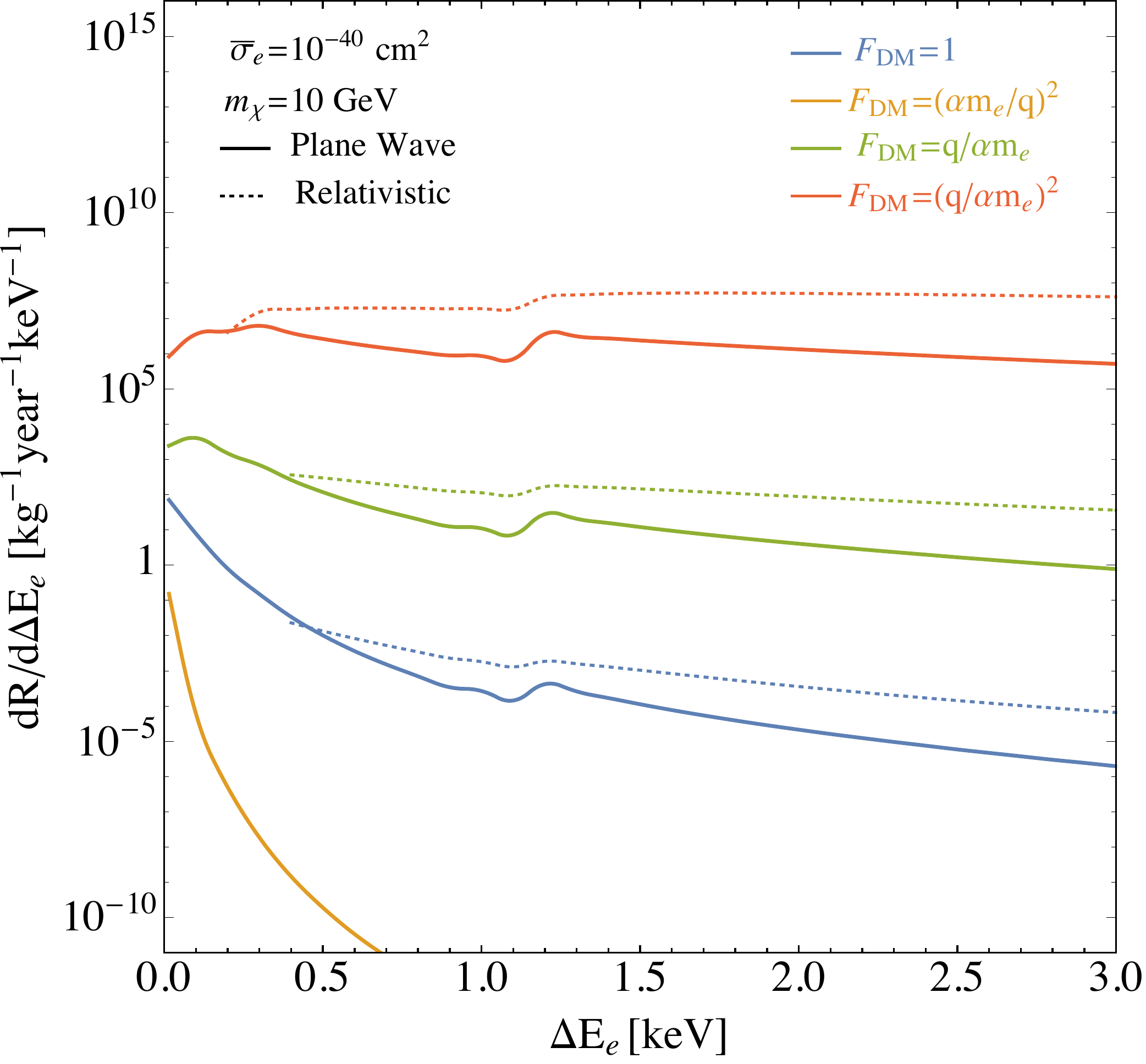}
\caption{Electron recoil spectra for standard DM-electron scattering for $m_\chi = 10$~GeV and $\overline\sigma_e=10^{-40}$~cm$^2$ for four different DM form factors. Solid lines show results calculated with plane waves as outgoing electron wave functions while the dotted lines include relativistic effects using results from~\cite{Roberts:2016xfw}.  See text for details.}
\label{fig:spectra}
\end{figure}

We can now write the cross section for the scattering rate as~\cite{Essig:2011nj,Essig:2015cda}
\begin{equation}\label{eq:sigma-scatter}
\sigma v_{1 \rightarrow 2} = \frac{\overline{\sigma}_e}{\mu_{\chi e}^2} \int \frac{d^3 q}{4 \pi} \delta (\Delta E_e + \frac{q^2}{2 \mu_{\chi N}} - \textbf{q}\cdot\textbf{v}) |F_{\text{DM}}(q)|^2 |f_{1 \rightarrow 2}(\textbf{q})|^2\,,
\end{equation}
where $|F_{\text{DM}}(q)|^2$ is the DM-electron interaction form factor and $\overline{\sigma}_e$ is the reference DM-electron cross section 
defined as 
\begin{eqnarray}
\overline{|\mathcal{M}_{\rm free}(\textbf{q})|^2} & \equiv & \overline{|\mathcal{M}_{\rm free}(\alpha m_e)|^2} \times |F_{\rm DM}(q)|^2 \, \\
\overline{\sigma}_e & \equiv & \frac{\mu_{\chi e}^2{|\overline{|\mathcal{M}_{\rm free}(\alpha m_e)|^2}|}}{16 \pi m_\chi^2 m_e^2}\,, 
\end{eqnarray} 
where $\overline{|\mathcal{M}_{\rm free}(\alpha m_e)|^2}$ is the absolute value squared of the matrix element describing the elastic 
scattering between DM and a free electron. While using the plane wave form factors, we also include a Fermi factor in the scattering rate given by, 
\begin{equation}
F_{\rm fermi}(\Delta E_e,Z_{\rm eff})=\frac{2\pi \zeta}{1-e^{-2\pi \zeta}}\,,
\end{equation}
with $\zeta=Z_{\rm eff} \alpha \sqrt{\frac{m_e}{2 E_e}}$.  
We take $Z_{\rm eff}=$~\{12.4, 14.2, 21.9, 25.0, 26.2, 39.9, 35.7,  35.6, 49.8, 39.8, 52.9\} for the shells \{5p, 5s, 4d, 4p, 4s, 3d, 3p, 3s, 2p, 2s, 1s\}~\cite{Clementi1963,Clementi1967}. 
The differential scattering rate will then be given by, 
\begin{eqnarray}
\frac{dR}{d \Delta E_e} & = &\frac{\overline{\sigma}_e}{8\mu_{\chi e}^2} \sum_{n,l} (\Delta E_e - E_{nl})^{-1} \frac{\rho_{\chi}}{m_{\chi}}  \\ 
& & \times \int q dq |F_{\text{DM}}(q)|^2 |f_{nl \rightarrow (\Delta E_e - E_{nl})}(\textbf{q})|^2 \eta(v_{\text{min}}(q,\Delta E_e)), \nonumber 
\end{eqnarray}
where we sum over all the occupied initial shells $\{n,l\}$ with respective binding energies $E_{nl}$. The $\eta(v_{\text{min}})$ is defined by,
\begin{equation}
\eta (v_{\text{min}}) = \int \frac{d^3 v}{v} g_{\chi}(v) \Theta(v-v_{\text{min}})\,,
\end{equation}
where $v_{\text{min}}$ is given by,
\begin{equation}\label{eq:vmin}
v_{\text{min}} = \frac{\Delta E_e}{q} + \frac{q}{2 m_{\chi}} \,,
\end{equation}
and 
\begin{equation}
g_\chi(\textbf{v}_\chi) \propto e^{-\frac{|\textbf{v}_\chi+\textbf{v}_{\rm E}|^2}{v_0^2}} \Theta(v_{\rm esc}-|\textbf{v}_\chi+\textbf{v}_{\rm E}|)\,,
\end{equation}
(normalized as $\int d^3 v \ g_\chi(\textbf{v})=1$) 
where $\textbf{v}_\chi$ is the DM velocity in the Earth frame, and $\textbf{v}_{\rm E}$ is the Earth's velocity in the galactic rest frame.  
We take a peak velocity of $v_0=220$~km/s,  an average Earth velocity of $v_{\rm E}=240$~km/s, and a galactic escape velocity of 
$v_{\rm esc} = 544$~km/s.  We set $\rho_\chi = 0.4$~GeV/cm$^3$. 

The DM form factor depends on the precise DM-electron interaction, but we will consider
\begin{eqnarray}
F_{\rm DM} & = & 1  \qquad \qquad\qquad \textrm{``heavy'' mediator}\\
F_{\rm DM} & = & \left(\frac{\alpha m_e}{q}\right)^2 \qquad \textrm{ ``light'' mediator}\\
F_{\rm DM} & = & \left(\frac{q}{\alpha m_e}\right)  \qquad \textrm{\,\;\;  $q$-dependent  ``heavy" mediator}\,,\\
F_{\rm DM} & = & \left(\frac{q}{\alpha m_e}\right)^2  \qquad \textrm{\,\;\;  $q^2$-dependent  ``heavy" mediator}\,,
\end{eqnarray}  
where ``heavy'' and ``light'' refer to the mass of the mediator, which is respectively above or below the typical momentum transfer. The resulting differential DM-electron scattering rates for $m_\chi=10$~GeV are shown in Fig.~\ref{fig:spectra} for $\overline{\sigma}_e$=$10^{-40} \rm{cm}^2$.  

\begin{figure}[t]
\centering
\includegraphics[width=0.46\textwidth]{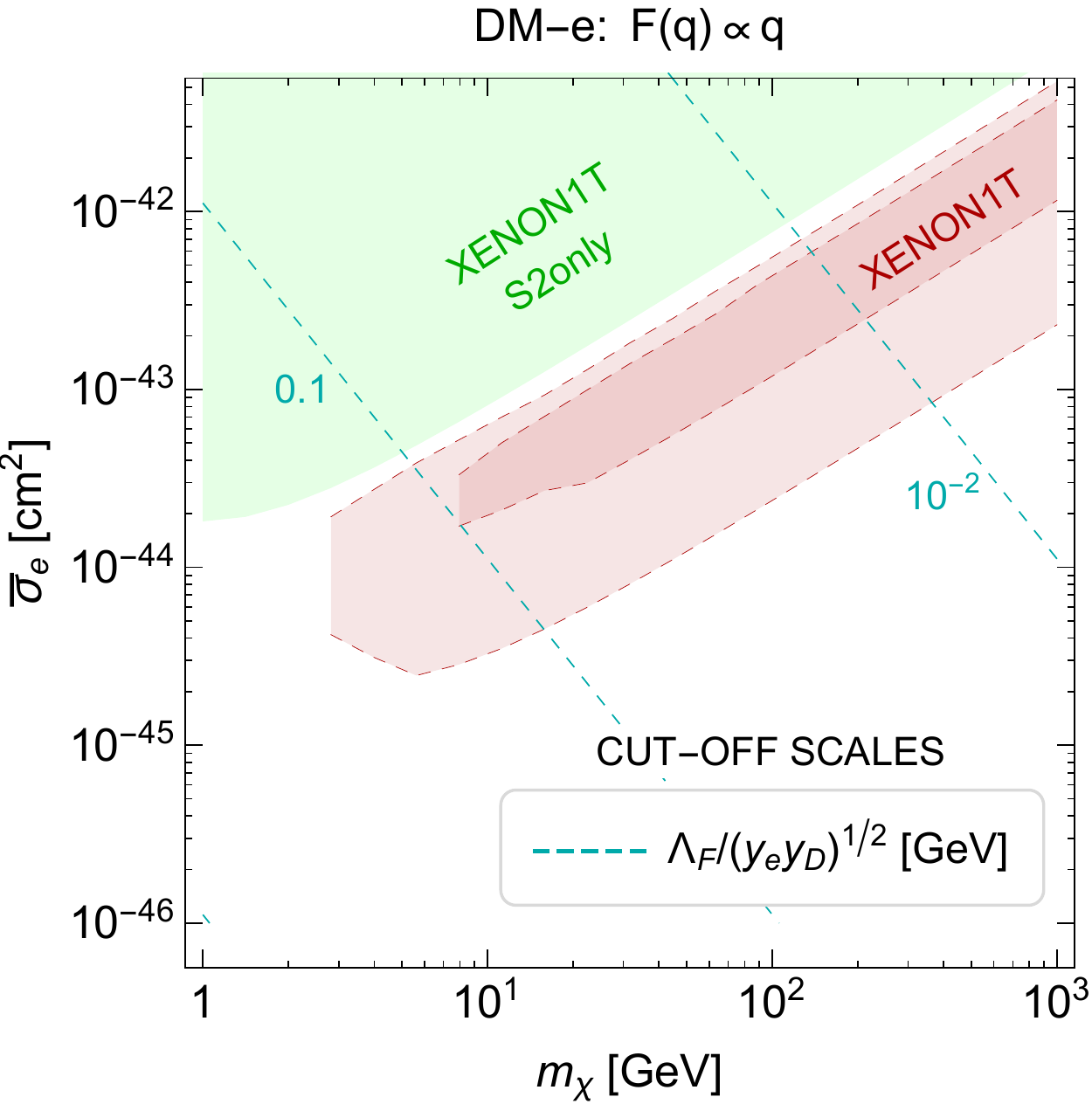}
\hfill
\includegraphics[width=0.46\textwidth]{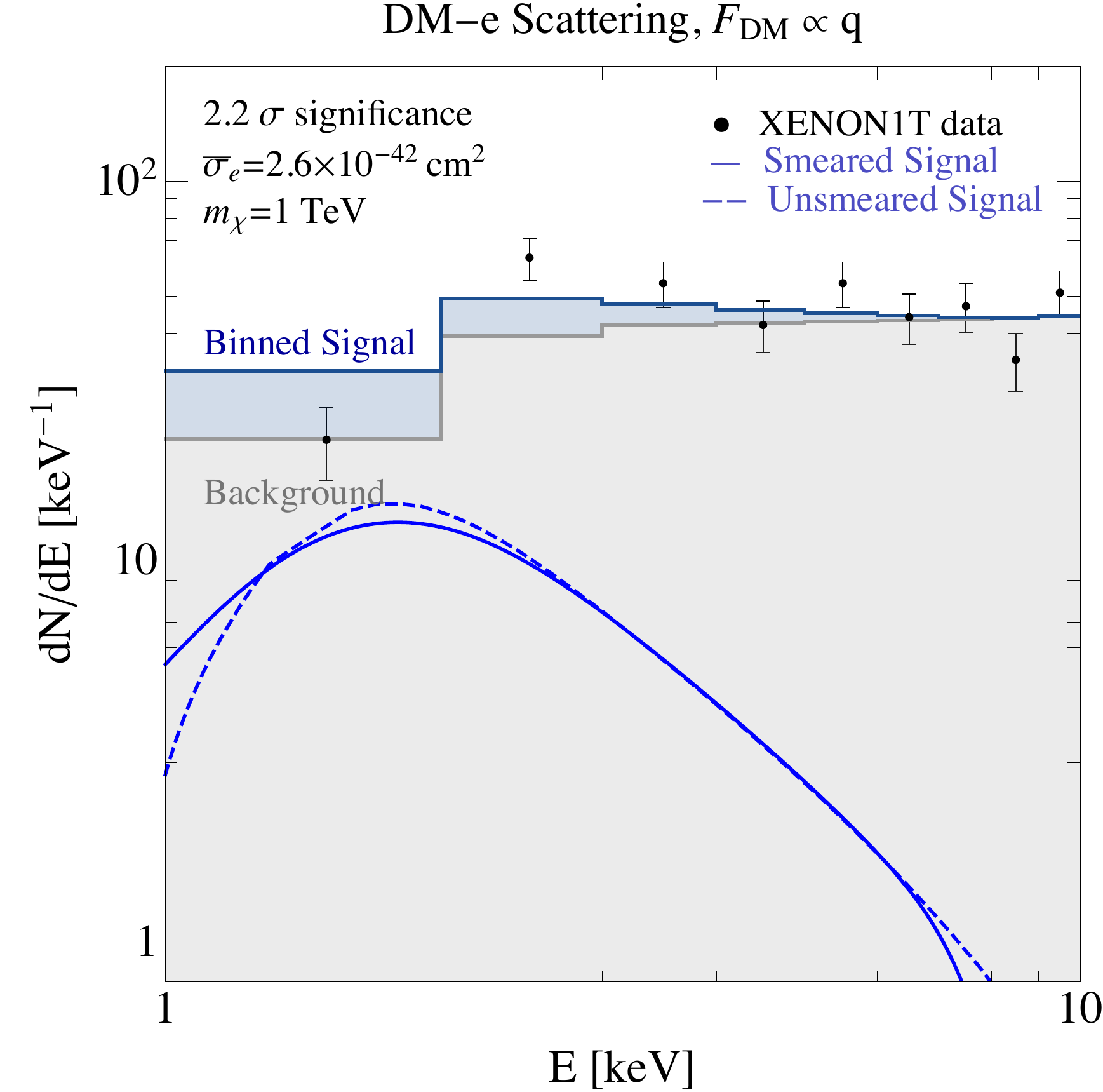}
\caption{
 {\bf Left:} Allowed parameter space for DM-electron scattering through a $q$-dependent heavy mediator, with $F_{\rm DM} = q/\alpha m_e$. The {\bf dark red} regions are the $1\sigma$ and $2\sigma$ regions fitting the \xe excess. The {\bf green} shaded region shows the current bound from the XENON1T only analysis based on S2 only. {\bf Dashed cyan}  lines show the scales at which the operator in Eq.~\eqref{eq:qoperator} is generated to obtain the corresponding cross section. {\bf Right:} Signal shape for the best fit point in Eq.~\eqref{eq:bestfitqdependent}. The {\bf black dots} are the \xe data, the {\bf gray shaded} region is the expected background, the {\bf blue line} is the signal shape after smearing, and the {\bf blue shaded} region is the resulting signal plus background distribution. 
\label{fig:DMqdependent}
}
\end{figure}

For standard DM-electron scattering, we calculate the rates using plane wave atomic form factors and also using the relativistic form factors. We show both spectra in Fig.~\ref{fig:spectra}. We see that the relativistic corrections (dotted lines) predict a larger signal rate in the region relevant for explaining the \xe excess than that predicted with plane wave form factors (solid lines).

We now briefly describe the different DM  form factors, focusing first on their ability to fit the XENON1T excess without being in conflict with other direct detection experiments and then commenting on possible complementary probes related to the new physics scale encoded in the cutoff of the operators generating the DM-electron interaction. 

\paragraph{``Heavy'' a and light mediator.} Due to the steep rise   at low energy,  the spectra for $F_{\rm DM}\propto 1$ and especially $F_{\rm DM}\propto 1/q^2$ are unable to explain the XENON1T signal without being in dramatic conflict with lower-threshold direct-detection searches from, e.g., XENON1T (S2-only analysis)~\cite{Aprile:2019xxb} (for heavy mediators) and SENSEI~\cite{Barak:2020fql} (light mediators). 

\paragraph{$q$-dependent  ``heavy" mediator.}  A $q$-dependent form factor $F_{\rm DM} = q/\alpha m_e$   does provide a reasonable fit to the \xe excess. The best-fit point is given by 
\begin{equation} 
m_{\chi} \gtrsim 90 \text{ GeV}\,, \quad \overline\sigma_e=2.6\times 10^{-45}~\text{ cm}^2\times\left(\frac{m_{\chi}}{1\text{ GeV}}\right)\,, 
\quad 2\log(\mathcal{L}_{S+B}/\mathcal{L}_{B})=7.3 \, , 
\label{eq:bestfitqdependent}
\end{equation}
and the resulting spectrum is shown in Fig.~\ref{fig:DMqdependent} right. In Fig.~\ref{fig:DMqdependent} left we show the  $1\sigma$ and $2\sigma$ regions in the $\overline\sigma_e$ versus $m_\chi$ parameter space. We include also a rough estimate of the signal yield of the S2-only analysis~\cite{Aprile:2019xxb}. We consider two bins: $(0.2,0.5)\text{ keV}$ and $(0.5,1)\text{ keV}$. We avoid considering the S2-only analysis above $1\text{ keV}$ to ensure that the dataset is completely independent from the one used to fit the signal. We impose a conservative bound by requiring a signal yield of less than 22 events in the $(0.2,0.5)\text{ keV}$ bin, as well as less than 5 events in the $(0.5,1)\text{ keV}$ bin. We see  that the best-fit regions for the $q$-dependent heavy mediators are not constrained from the lower-threshold S2-only analysis.

A  $q$-dependent form factor is predicted, for example, by the  dimension six operator, 
\begin{equation}
\qquad \frac{y_e y_Di\bar\chi\gamma_5\chi \bar e e}{\Lambda_F^2}\ ,\label{eq:qoperator} 
\end{equation} 
where $\chi$ is the fermionic DM and the scalar-pseudoscalar interaction is induced by a heavy scalar which admits a spin dependent interaction with the DM. 
In the same plot we present contours of the cutoff scale divided by the square-root of the mediator-electron ($y_e$) and mediator-DM ($y_D$) couplings. Even by assuming $y_D$ to be at its  perturbativity bound and taking the coupling of the mediator to electrons or be order one, the required cutoff $\Lambda_F$ of the effective operator in Eq.~\eqref{eq:qoperator} implies new physics below the GeV scale.     This is likely to be excluded by collider bounds from electron-positron machines~\cite{Fox:2011fx}.  A possible way to raise the cut-off scale would be to consider a scalar DM $\phi$ interacting with electron through the dimension five operator, 
\begin{equation}
\frac{y_e y_D\phi^{\ast}\phi \bar e \gamma_5 e}{\Lambda_S}\ ,
\end{equation} 
which leads to a $q$-dependent cross section, where the scalar DM interacts with the electron spin. The reduced dimensionality of this operator could help pushing the cut-off up to hundreds of GeVs thereby allowing a  UV completion consistent with collider constraints and electron EDMs. However, a correct treatment of the spin-dependent interactions for the bounded electrons inside the xenon atom is necessary to correctly compute the DM-e cross section. We leave this interesting issue for future investigations.  

\begin{figure}[t]
\centering
\includegraphics[width=0.46\textwidth]{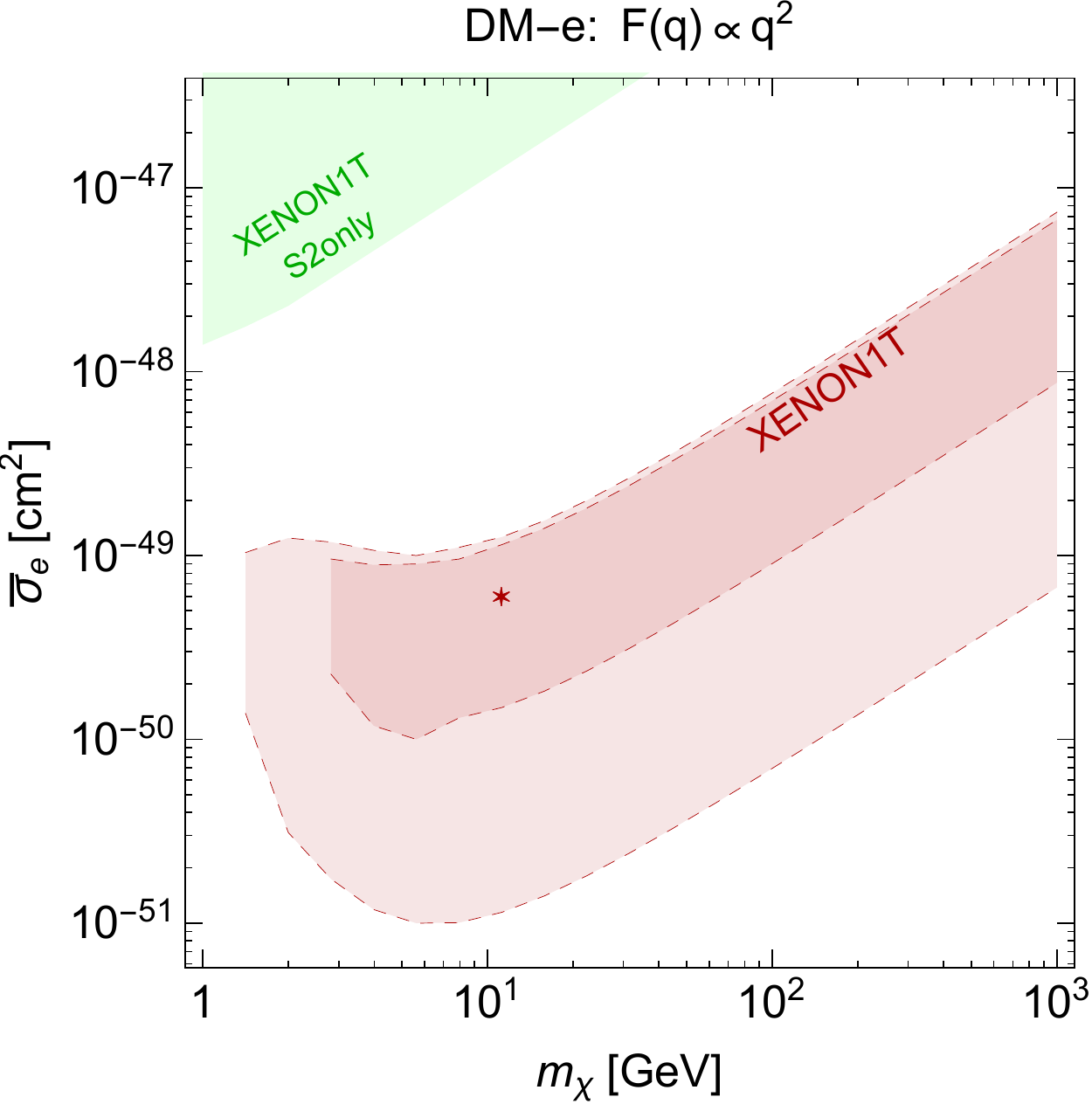}
\hfill
\includegraphics[width=0.46\textwidth]{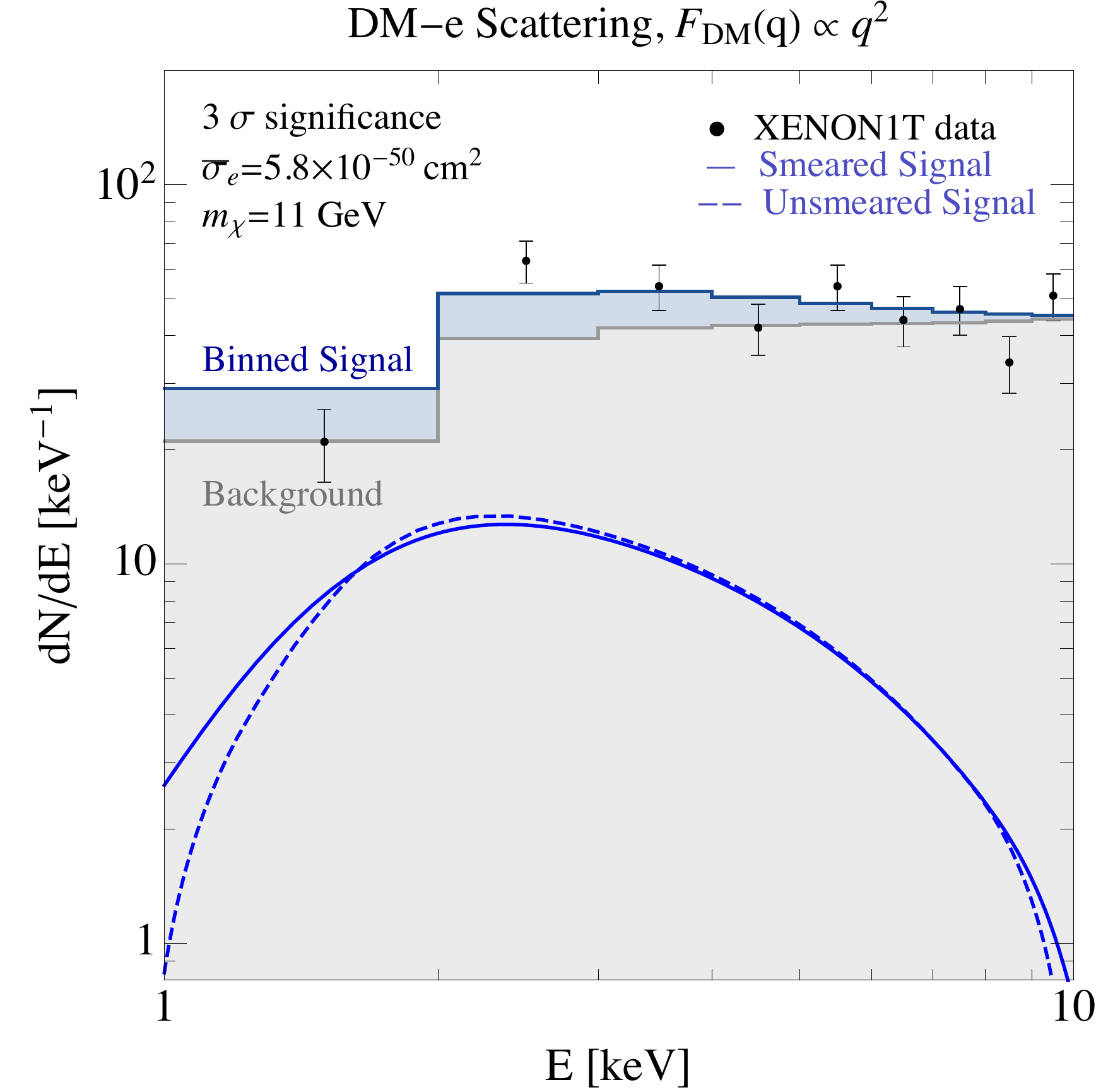}
\caption{ {\bf Left:} Allowed parameter space for DM-electron scattering through a $q$-dependent heavy mediator, with $F_{\rm DM} = \left(q/\alpha m_e\right)^2$.  The {\bf dark red} regions are the $1\sigma$ and $2\sigma$ bands of our fit to the \xe data. The  {\bf green} shaded region shows the current XENON1T bound from the S2 only analysis.  {\bf Right:} Signal shape for the best fit point in Eq.~\eqref{eq:bestfitq2dependent}. The {\bf black dots} are the \xe data, the {\bf gray shaded} region is the expected background, the {\bf blue line} is the signal shape after smearing, and the {\bf blue shaded} region is the resulting signal plus background distribution.
\label{fig:DMq2dependent}
}
\end{figure}

\paragraph{$q^2$-dependent  ``heavy" mediator.}  Last,  we show in Fig.~\ref{fig:DMq2dependent} the 
 allowed parameter  and best fit regions  to the \xe data for the DM form factor $F_{\rm DM} = \left(q/\alpha m_e\right)^2$  together with the bound from the XENON1T S2-only analysis~\cite{Aprile:2019xxb}. The best fit point is given by 
\begin{equation} 
m_{\chi} = 11 \text{ GeV}\,, \quad \overline\sigma_e=6.0\times 10^{-50}~\text{ cm}^2\,, 
\quad 2\log(\mathcal{L}_{S+B}/\mathcal{L}_{B})=12.3 \,, 
\label{eq:bestfitq2dependent}
\end{equation}
and the resulting spectrum is shown in Fig.~\ref{fig:DMq2dependent} right. As it is evident by comparing this result with the previous one for a $q$-dependent form factor, a stronger momentum dependence improves  the \xe fit substantially. 

A $q^2$-dependent form factor could be generated by operators such as  
\begin{equation}
\frac{y_e y_D\partial_\mu(\phi^{\ast}\phi)\partial^\mu (\bar e e)}{\Lambda_S^3}\ ,\label{eq:scalarq2}
\end{equation} 
and
\begin{equation}
\frac{y_e y_D\bar\chi\gamma_5\chi \bar e\gamma_5 e}{\Lambda_F^2}\label{eq:fermionq2}\ ,
\end{equation} 
where again $\phi$ is a scalar DM while $\chi$ is a fermionic DM. The operator in Eq.~\eqref{eq:scalarq2} could be obtained from the dimension six ``derivative'' Higgs portal $\partial_\mu(\phi^{\ast}\phi)\partial_\mu(H^{\dagger}H)/\Lambda^2$ after the Higgs is integrated out.  This will lead to an extra suppression of the wilson coefficient proportional to $m_e/m_h$~\cite{Balkin:2018tma} if $\Lambda\sim\Lambda_S\sim m_h$ . A very low cut-off scale, $\Lambda_S$, is then required to get a cross section in the ballpark of the one required by our fit of the \xe data, making this example not viable phenomenologically. The second operator in Eq.~\eqref{eq:fermionq2} can be obtained by integrating out a heavy axion coupled to fermionic DM and electrons. The expected cut-off for the range of cross section and DM masses of interest is always lower than 1 GeV and hence in tension with colliders constraints. As mentioned earlier, a more in depth analysis should be performed in order to correctly account for the spin dependence on the electronic side.

We now turn our attention to exothermic DM, which can provide an even better fit to the XENON1T excess, and also has several interesting features that deserve 
further study. 



\subsection{Exothermic Dark Matter and Electron Recoils}\label{sec:EXO}

DM could consist of two or more approximately degenerate particles,  see e.g.~\cite{TuckerSmith:2001hy,Finkbeiner:2007kk,ArkaniHamed:2008qn,Finkbeiner:2009mi,Batell:2009vb,Essig:2010ye,Graham:2010ca,Lang:2010cd}. 
We consider two states, $\chi_1$ and $\chi_2$, with masses $m_{\chi_1}$ and $m_{\chi_2}=m_{\chi_1}+\delta$, respectively, with $|\delta|\ll m_{\chi_1}, m_{\chi_2}$.  For example, $\chi_1$ and $\chi_2$ could be two Majorana fermions that originated from a Dirac fermion that is charged under a new $U(1)$ gauge symmetry; if there are mass terms for the Dirac fermion that break the $U(1)$ symmetry, it is possible to split them into the two Majorana fermions, with the gauge boson coupling off-diagonally to $\chi_1$ and $\chi_2$.  Similarly, one can consider two real scalars that originated from a complex scalar.  
In what follows, we will always take $\chi_1$ to be the incoming state,  which then scatters off ordinary matter and converts to $\chi_2$ (which in our notation is always the outgoing state).  The scenario where $\chi_2$ is heavier than $\chi_1$ is often called ``inelastic'' DM ~\cite{TuckerSmith:2001hy} ($\delta>0$), while the scenario where $\chi_1$ is heavier than $\chi_2$ is often called ``exothermic''  DM ($\delta<0$); in the context of direct-detection experiments, the latter was previously discussed for DM-nuclear scattering in~\cite{Essig:2010ye,Graham:2010ca} and for DM-electron scattering in~\cite{Bernal:2017mqb}.  

The relic abundance of the two states depends on the precise model. In the minimal scenario above and  for $|\delta|$ sufficiently small (typically $\lesssim 2m_e$), the lifetime of the heavier state for decays via the (off-shell) mediator into the lighter state plus two neutrinos, or for decays into the lighter state plus three photons, is easily much longer than the age of the universe~\cite{Finkbeiner:2009mi,Batell:2009vb}.  
However, the fractional abundance of the heavier state after freeze-out in the early universe will depend sensitively on the precise DM-mediator interaction strength and the DM and mediator masses~\cite{Finkbeiner:2009mi,Batell:2009vb}.  For sub-GeV DM, the abundance of the heavier state will typically be small. 
However, even a small fractional abundance of the heavier state can leave dramatic signals in direct-detection experiments, since, as we will see, the mass splitting $|\delta|$ can be entirely converted into kinetic energy of the electron when scattering off of it  in a target material.  The exothermic scenario allows all relic particles in the halo to scatter, while the inelastic up-scatter of the lighter to the heavier state will be highly suppressed for $|\delta|\gg 10$'s of eV.  

We focus here on exothermic scattering, since it is able to explain the \xe excess.  
In \S\ref{subsubsec:DM-exo-kin}, we discuss the kinematics and also provide best-fit regions to the \xe excess that are independent of the precise relic abundance of the heavier state, before considering concrete models in \S\ref{subsubsec:relicEXO}. 

\subsubsection{Exothermic Dark Matter-Electron Scattering: kinematics and best-fit regions}\label{subsubsec:DM-exo-kin}
We assume that the incoming DM particle, $\chi_1$, transfers momentum $\textbf{q}$ to the target electron and converts to the lighter (outgoing) state, $\chi_2$. In contrast to Eq.~(\ref{eq:DeltaEusual}), the energy-conservation equation now reads 
\begin{equation}
\Delta E_e + \frac{|m_{\chi_1}\textbf{v}-\textbf{q}|^2}{2 m_{\chi_2}} + \frac{q^2}{2 m_N} + m_{\chi_2} = \frac{1}{2} m_{\chi_1} v^2 + m_{\chi_1},  
\end{equation} 
where $\Delta E_e$ is again the energy transferred to the electron.  Assuming 
a small mass-splitting compared to the mass scale of the DM i.e. $|\delta| \ll m_{\chi_1} \sim m_{\chi_2}$, we can simplify this as,
\begin{equation}
\Delta E_e = \textbf{q}\cdot\textbf{v} - \frac{q^2}{2 m_{\chi_2,N}} - \delta.
\end{equation}
In contrast to the ``standard'' DM-electron scattering discussed above ($\delta=0$) (and in contrast also with exothermic nuclear scattering, see below), $\Delta E_e$ can be well above the ``typical'' energy transfers of $\Delta E_e \sim 10^{-3} q_{\rm typ} \sim Z_{\rm eff} \times \textrm{few eV}$ applicable for $\delta=0$.  In particular, for $\delta \sim \mathcal{O}(-\text{keV})$, the electron recoil spectrum will be peaked at $\mathcal{O}(\text{keV})$, and can explain the \xe excess.  
Below, since $m_{\chi_1}\sim m_{\chi_2}$, we will often simply denote the DM mass as $m_\chi$. Also, for the calculation of exothermic DM-electron scattering, we consider non-relativistic atomic form factors. 

\begin{figure}[t]
\centering
\includegraphics[width=0.46\textwidth]{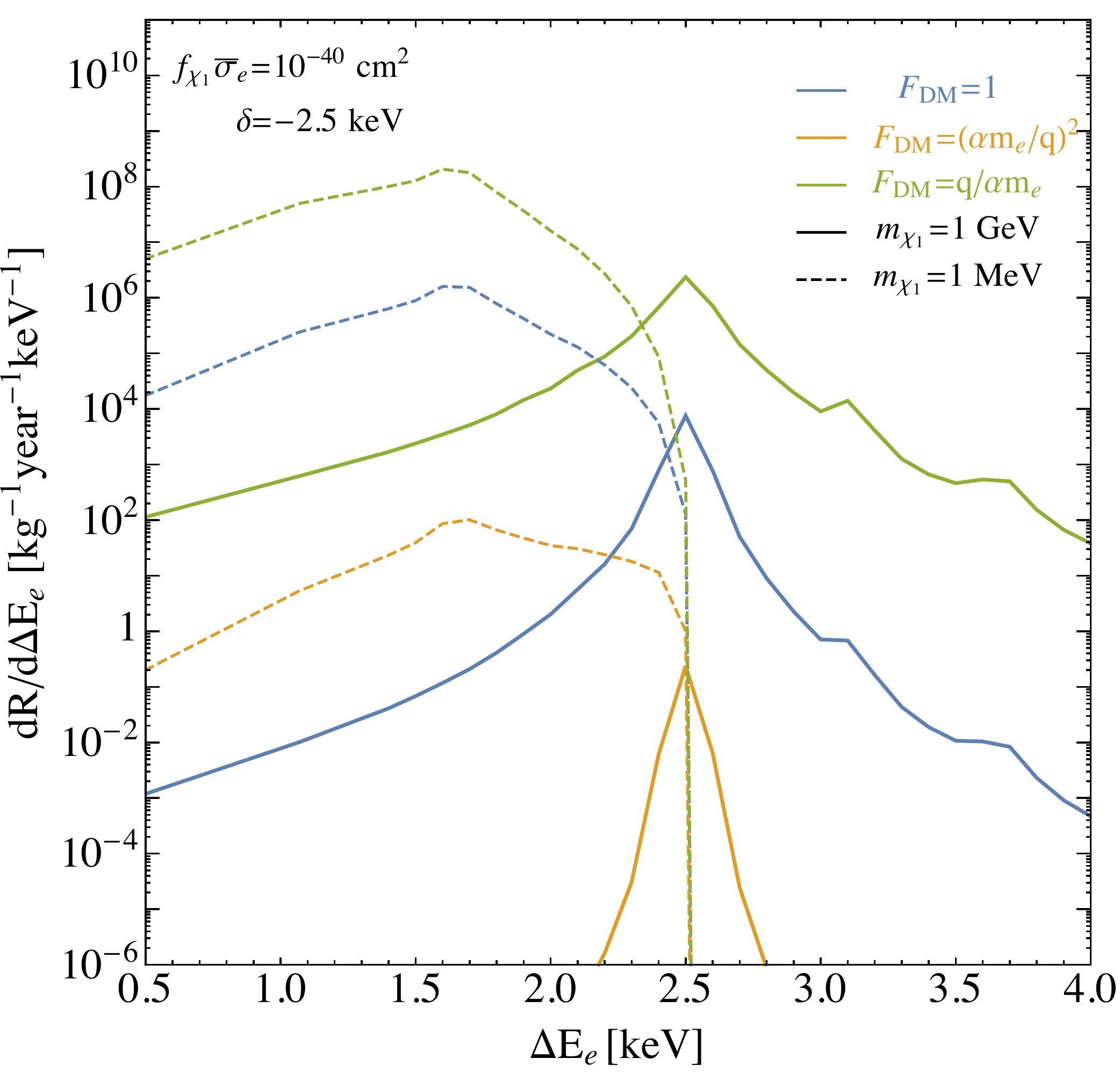}\hfill
\includegraphics[width=0.46\textwidth]{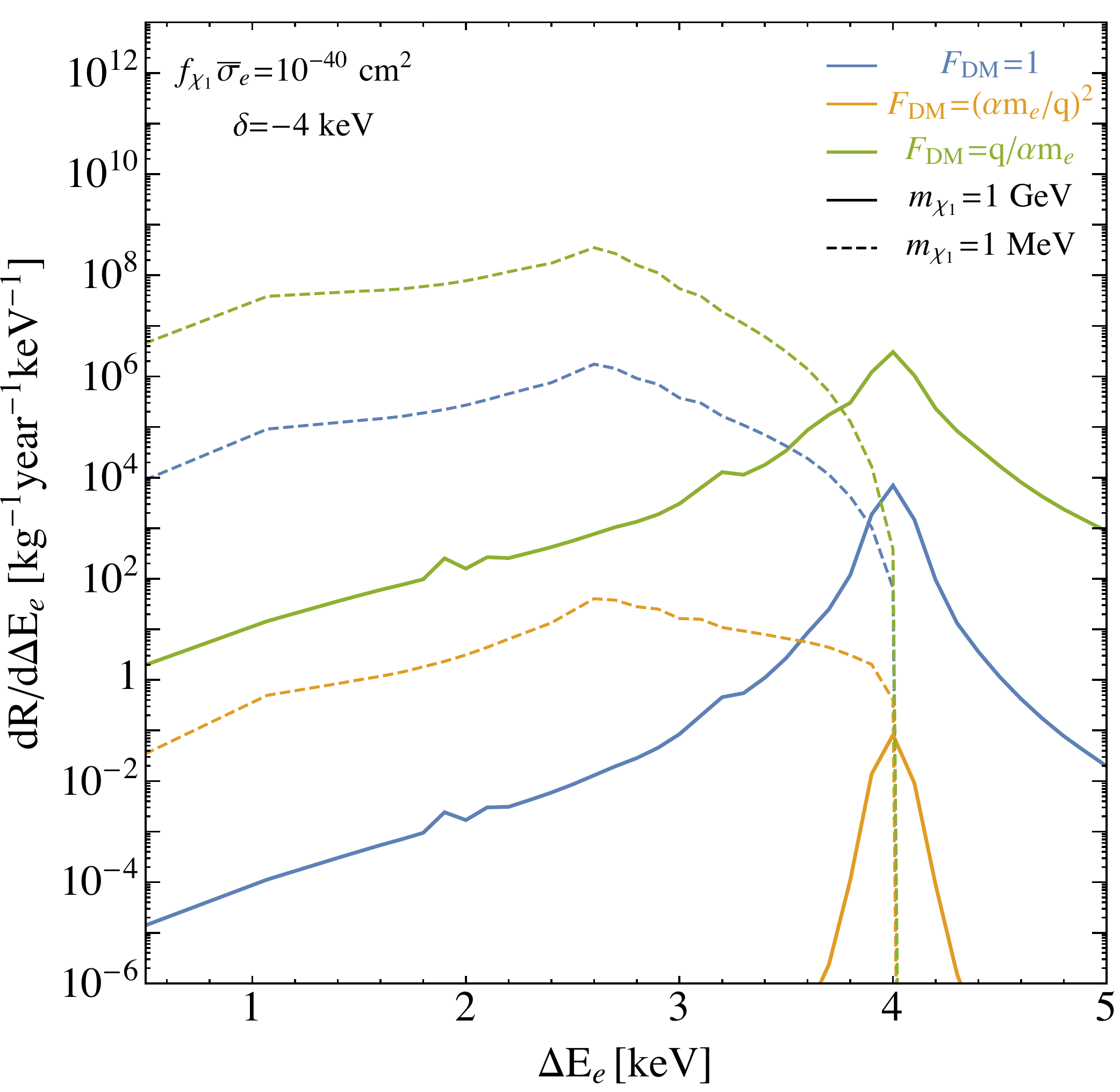}
\caption{Differential recoil spectra for ``exothermic'' DM, in which a heavy incoming DM state, $\chi_1$, scatters off an electron and converts to a lighter (outgoing) DM state, $\chi_2$, which is lighter by $\delta$=$-2.5$~keV (\textbf{left}) and $\delta$=$-4$~keV (\textbf{right}).  We show $m_{\chi_1}$=1~GeV (\textbf{solid}) and $m_{\chi_1}=1$~MeV (\textbf{dashed}). We consider three DM form factors, $F_{\text{DM}}$=1 (\textbf{blue}), $F_{\text{DM}}=(\alpha m_e/q)^2$ (\textbf{orange}), and $F_{\text{DM}}=(q/\alpha m_e)$ (\textbf{green}).
 }  
\label{fig:spectra-twodelta}
\end{figure} 

The differential scattering rate is given by
\begin{eqnarray}\label{eq:scatrate}
\frac{dR}{d \Delta E_e} & = & \frac{\overline{\sigma}_e}{8\mu_{\chi e}^2} \sum_{n,l} (\Delta E_e - E_{nl})^{-1} \frac{\rho_{\chi_1}}{m_{\chi_1}} \\
& & \times \int q dq |F_{\text{DM}}(q)|^2 |f_{nl \rightarrow  \Delta E_e - E_{nl}}(\textbf{q})|^2 \eta(v_{\text{min}}(q,\Delta E_e)), \nonumber 
\end{eqnarray}
where the minimum velocity to scatter is given by 
\begin{equation}\label{eq:vmin-delta}
v_{\rm min} = \left| \frac{\Delta E_e + \delta}{q} + \frac{q}{2\mu_{\chi_2,N}} \right|\,.
\end{equation}
As there is an upper bound of $v_{\text{max}}$=$v_{\text{esc}}+v_{\text{E}}$ on the DM halo velocity, we get upper and lower bounds on the allowed values of $q$ for a given $m_{\chi}$ and a fixed $\Delta E_e$, 
\begin{equation}\label{eq:qmin-delta}
q_{\rm min} = {\rm sign}(\Delta E_e + \delta) m_{\chi} v_{\text{max}} \left(1-\sqrt{1-\frac{(\Delta E_e + \delta)}{\frac{1}{2} m_{\chi} v_{\text{max}}^2}}\right)\,,
\end{equation}
\begin{equation}\label{eq:qmax-delta}
q_{\rm max} =  m_{\chi} v_{\text{max}} \left(1+\sqrt{1-\frac{(\Delta E_e + \delta)}{\frac{1}{2} m_{\chi} v_{\text{max}}^2}} \right)\,.
\end{equation}
For DM scattering off electrons through a light mediator [$F_{\rm DM}=(\alpha m_e/q)^2$],  the scattering rate diverges at low $q$; this was not a problem for the case $\delta=0$, since requiring a sizable value for $\Delta E$ also forces $q$ to be sizable.  However, as we will discuss, this is not true anymore for exothermic scattering: a sizable $\Delta E$ can be obtained even for very small $q$.  To remove the resulting divergence in this case, we will consider that the light mediator is a light dark photon, which couples to electric charge.  At low $q$, Thomas-Fermi screening will then naturally regulate the divergence.  We implement Thomas-Fermi screening as discussed in~\cite{Emken:2019tni,Emken:2019hgy}, using a Thomas-Fermi radius (called $a'$ in~\cite{Emken:2019hgy}) from~\cite{Tsai-screening-RevModPhys.46.815}.

Consider first DM masses of $\mathcal{O}$(GeV). From Eq.~(\ref{eq:qmax-delta}), we see that the value of $q_{\rm max}$ is near $m_{\chi} v_{\rm max} \sim \mathcal{O}$(MeV), which is much higher than $q_{\rm typ}$. Thus, the recoil spectrum in $\Delta E_e$ depends on the behavior of $q_{\rm min}$ as a function of $\Delta E_e$. From Eq.~(\ref{eq:qmin-delta}), we see that, for $\delta=0$ and $\Delta E_e \sim \mathcal{O}$(keV), $q_{\rm min} \gg q_{\rm typ}$ and therefore the integral over $q$ misses the peak of the form-factor, leading to strongly suppressed scattering rates (as discussed above). However, for $\delta \sim \mathcal{O}$($-$keV), we see that $q_{\rm min}=0$ when $\Delta E_e=|\delta|$. For $\Delta E_e$ smaller or larger than $|\delta|$, $q_{\rm min}$ increases and the available phase space decreases again, thus giving a suppression in the rate. Hence, for $m_{\chi} \sim \mathcal{O}$(GeV) and $\delta \sim \mathcal{O}$($-$keV), we get a sharp peak in the spectrum at $\Delta E_e \sim |\delta|$. 

In Fig.~\ref{fig:spectra-twodelta}, the solid lines show spectra for $m_\chi=1$~GeV for $\delta=-2.5$~keV (left) and $\delta=-4$~keV (right) for $f_{\chi_1}\overline{\sigma}_e=10^{-40}~\rm{cm}^2$ and for three different form factors.  Here the fractional abundance of the 
incoming DM particle, $\chi_1$, is $f_{\chi_1}=\frac{n_{\chi_1}}{n_{\chi_1}+n_{\chi_2}}$, with $n_{\chi_1}$ ($n_{\chi_2}$) being the number density of $\chi_1$ ($\chi_2$).   
We see that the spectrum is sharply peaked at $|\delta|$ and is reminiscent of a DM absorption signal, which provides an adequate fit to the \xe excess. 

We next consider $m_{\chi} \sim \mathcal{O}$(MeV), showing the resulting spectra in Fig.~\ref{fig:spectra-twodelta}, where the dashed lines show spectra for $m_\chi=1$~MeV for $\delta=-2.5$~keV (left) and $\delta=-4$~keV (right), both for incoming DM mass fraction $f_{\chi_1}\overline{\sigma}_e=10^{-40}~\rm{cm}^2$ and for three different form factors. The spectrum has a wide peak, wider than for heavier exothermic DM and wider than a DM absorption signal, which (for the larger value of $|\delta|$) provides a very good fit to the \xe data.  

We can understand the shape of the spectra for $m_{\chi} \sim \mathcal{O}$(MeV) as follows.  If $\delta \sim \mathcal{O}$($-$keV), $q_{\rm min}=0$ at $\Delta E_e=|\delta|$. However, $q_{\rm max}$ is now only a few keV, and we see from Fig.~\ref{fig:form-factor} that the $q_{\rm typ}$ for $\Delta E_e \sim \mathcal{O}$(keV) is higher. Note that for such small $m_{\chi}$, there is barely any kinetic energy in the DM to give recoil energies larger than $|\delta|$, which is already around keV. So the spectrum sharply cuts off above $\Delta E_e \sim |\delta|$. 
For $F_{\rm DM}\propto 1/q^2$, the spectrum is peaked roughly at $\Delta E\sim |\delta|$ (due to the enhancement of the integrand of Eq.~\eqref{eq:scatrate} at $q\to 0$).  
However, for $F_{\rm DM}\propto 1$ and $F_{\rm DM}\propto q$, the peak (for a fixed $|\delta|$) occurs at energies less than $|\delta|$.  
The reason is that for $\Delta E_e < |\delta|$, both $q_{\rm min}$ and $q_{\rm max}$ increase, and at some $\Delta E_e$ below $|\delta|$, the allowed values of $q$ cross the peak of the form factor. Hence, we see a peak in the spectrum for $\Delta E_e$ below $|\delta|$. Moreover, the spectrum is not as sharply peaked as it is for heavier DM, since for heavier DM the allowed values of the momentum transfer are always $q\sim \mathcal{O}$(keV)$\sim q_{\rm typ}$.   
In order to have the spectrum peak near 2.5~keV (which gives a good fit to the \xe data), one then needs larger values of $|\delta|$.  However, for $|\delta|\gtrsim 4.9$~keV, the 2s- and 2p-shells can also be excited, leading to additional peaks in the spectrum.  The precise spectrum thus depends sensitively on the DM mass and splitting.  In our parameter scans below, we do not attempt to cover the entire (sub-)MeV-scale DM parameter space.  


\begin{figure}[t]
\centering
\includegraphics[width=0.32\textwidth]{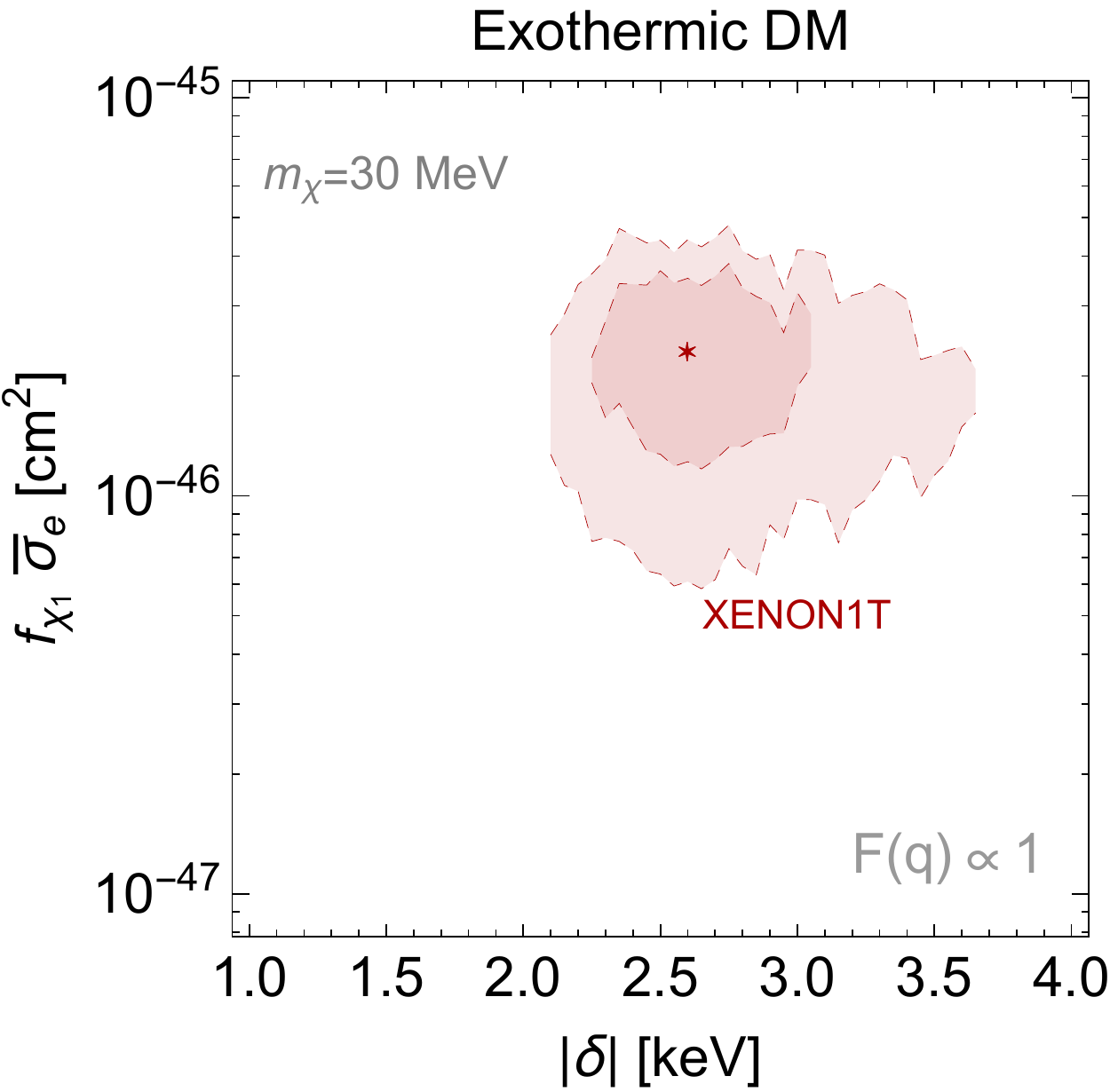}
\includegraphics[width=0.32\textwidth]{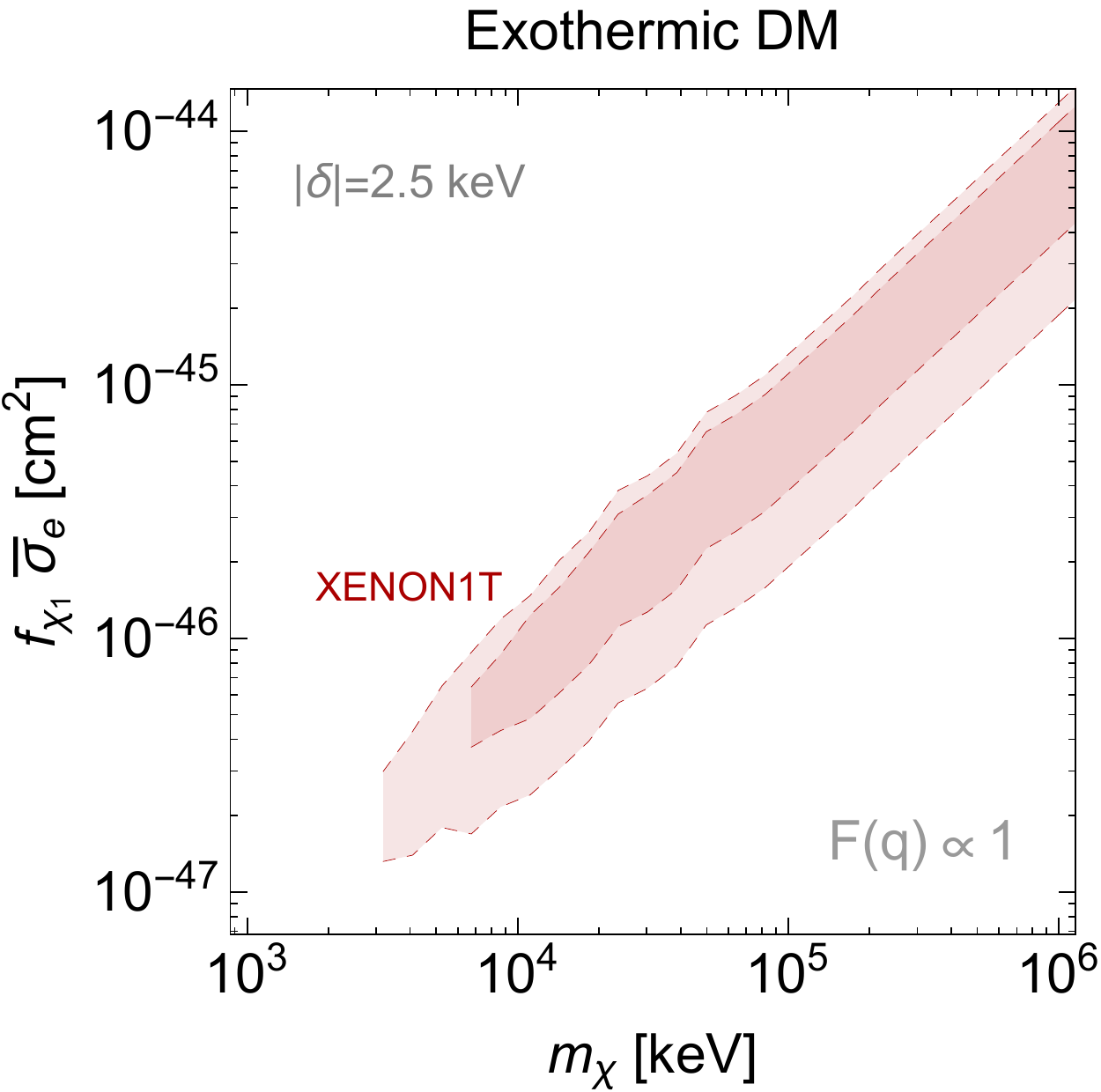}
\includegraphics[width=0.32\textwidth]{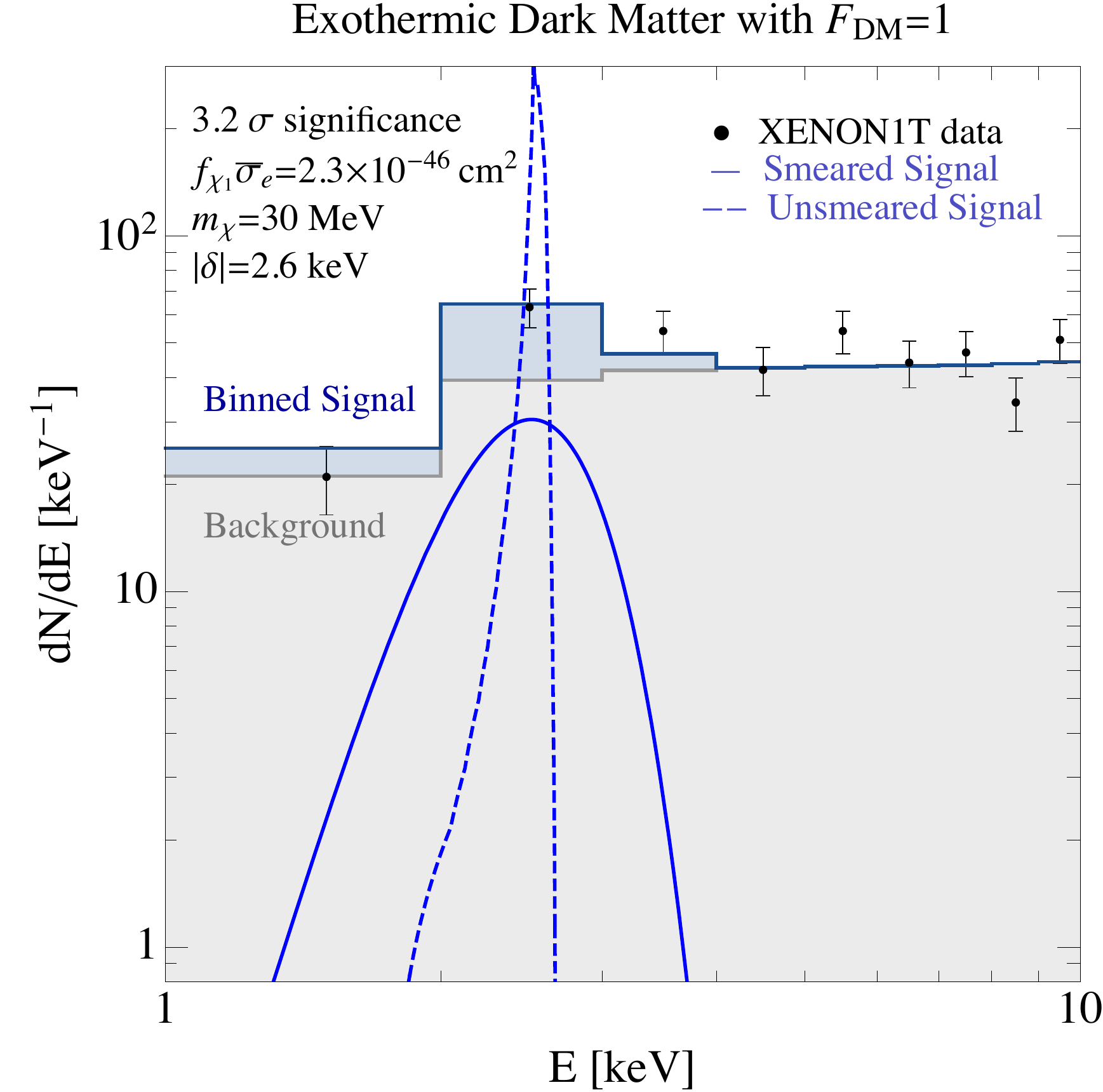}
\caption{The $1\sigma$ and $2\sigma$ best-fit regions that explain the \xe excess for exothermic DM and a heavy mediator ($F_{\rm DM}=1$) 
in the $f_{\chi_1} \overline\sigma_e$ versus $\delta$ plane for $m_\chi=30$~MeV ($f_{\chi_1}=\frac{n_{\chi_1}}{n_{\chi_1}+n_{\chi_2}}$) (\textbf{left}) , and in the $f_{\chi_1} \overline\sigma_e$ versus $m_{\chi}$ plane for $|\delta|=2.5$~keV (\textbf{middle}).  In the \textbf{right} plot,  we show an example of the predicted spectrum for the best-fit value with $|\delta| \leq 4.9~$keV in Eq.~\eqref{eq:bestfitexot}. The {\bf dashed} and {\bf solid} lines show the signal spectrum before and after detector smearing effects, respectively.   The measured \xe data is shown as {\bf black dots} while the {\bf gray-shaded} and {\bf blue-shaded} regions are the expected binned background and signal respectively. 
\label{fig:spectra-twodelta-fits}}
\end{figure}

\begin{figure}[t]
\centering
\includegraphics[width=0.32\textwidth]{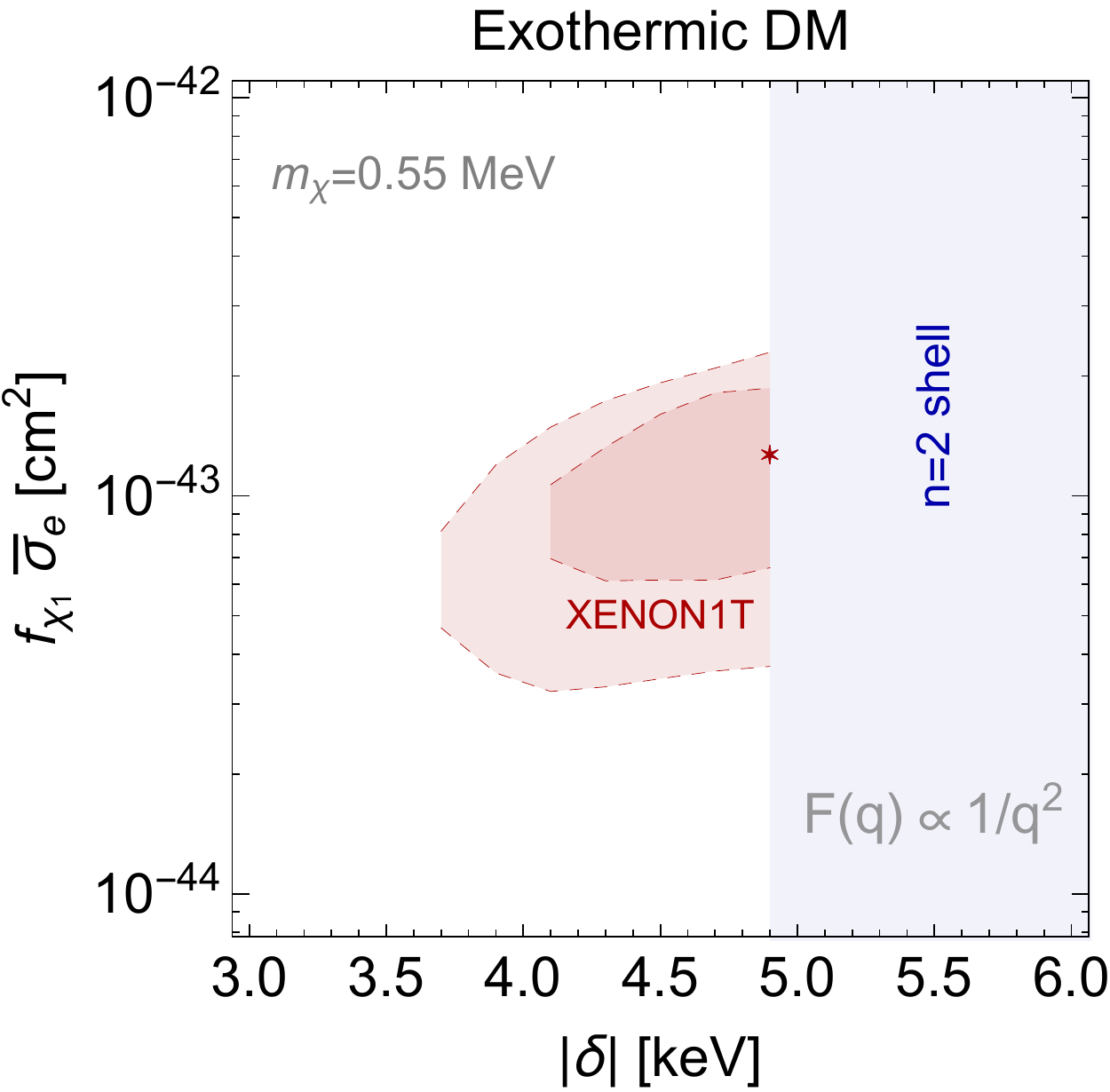}
\includegraphics[width=0.32\textwidth]{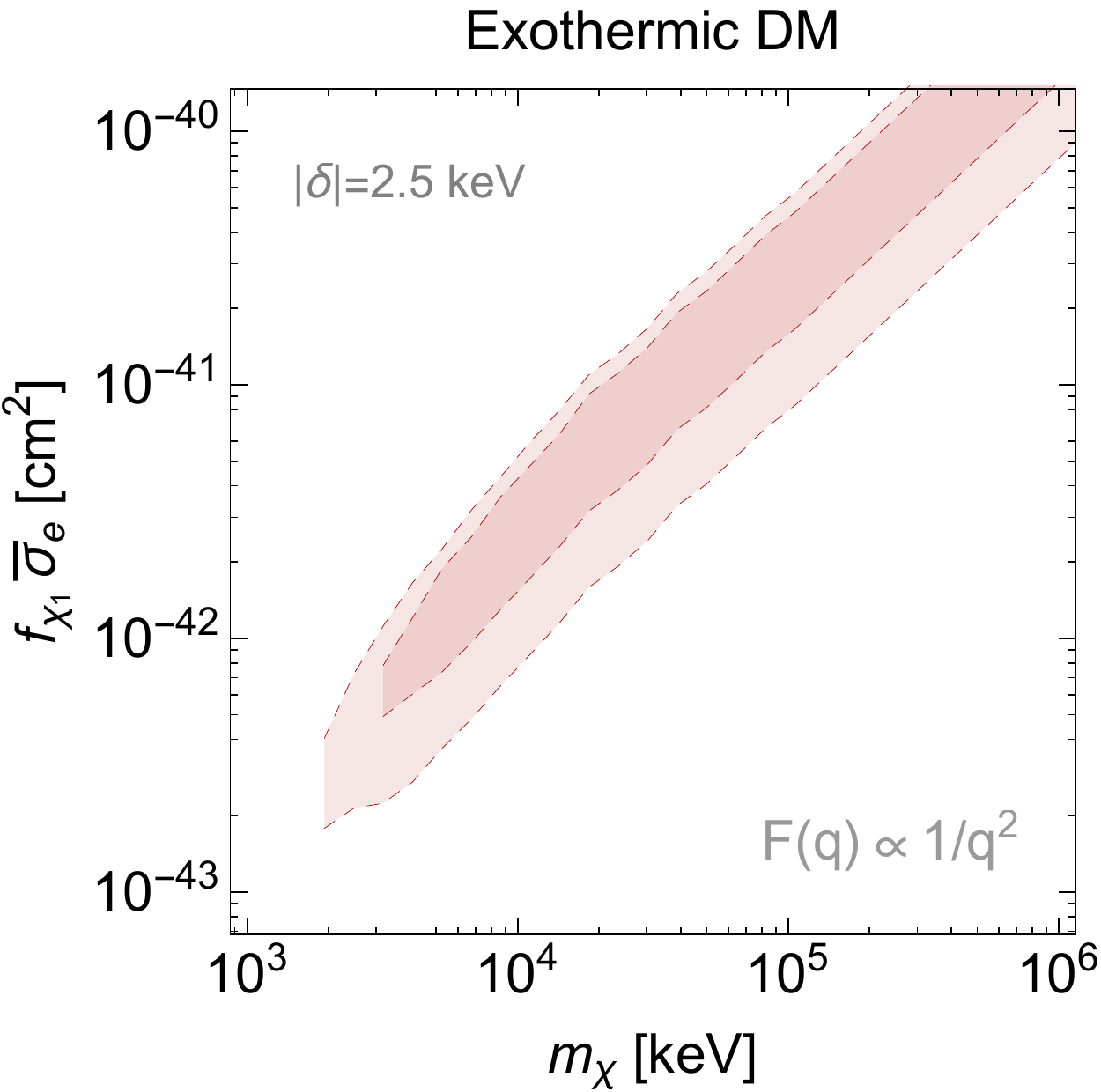}
\includegraphics[width=0.32\textwidth]{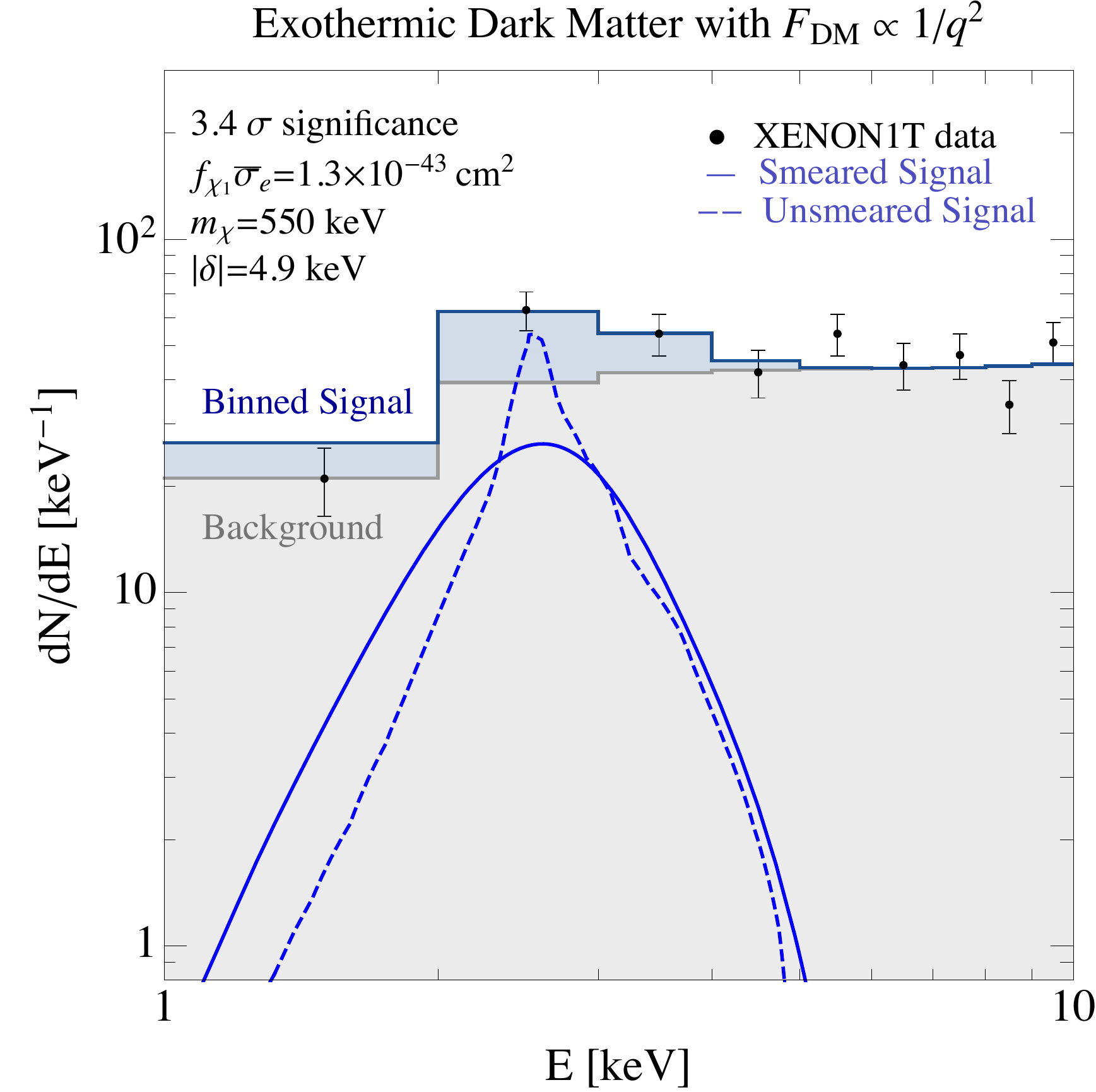}
\caption{The $1\sigma$ and $2\sigma$ best-fit regions that explain the \xe excess for exothermic DM and a light mediator ($F_{\rm DM}=(\alpha m_e/q)^2)$) 
in the $f_{\chi_1} \overline\sigma_e$ versus $\delta$ plane for $m_\chi=0.55$~MeV ($f_{\chi_1}=\frac{n_{\chi_1}}{n_{\chi_1}+n_{\chi_2}}$) (\textbf{left}), and in the $f_{\chi_1} \overline\sigma_e$ versus $m_{\chi}$ plane for $|\delta|=2.5$~keV (\textbf{middle}). The {\bf blue-shaded}  region has not been included in our scan because the second xenon shells would be excited (see text for details). In the \textbf{right} plot, we show an example of the predicted spectrum for the best-fit value with $|\delta| \leq 4.9~$keV
from Eq~\eqref{eq:bestfitexotFDMq2}. The {\bf dashed} and {\bf solid} lines show the signal spectrum before and after detector smearing effects, respectively.   The measured \xe data is shown as {\bf black dots} while the {\bf gray-shaded} and {\bf blue-shaded} regions are the expected binned background and signal respectively. 
\label{fig:spectra-twodelta-fits-FDMq2} }
\end{figure} 

\begin{figure}[t]
\centering
\includegraphics[width=0.32\textwidth]{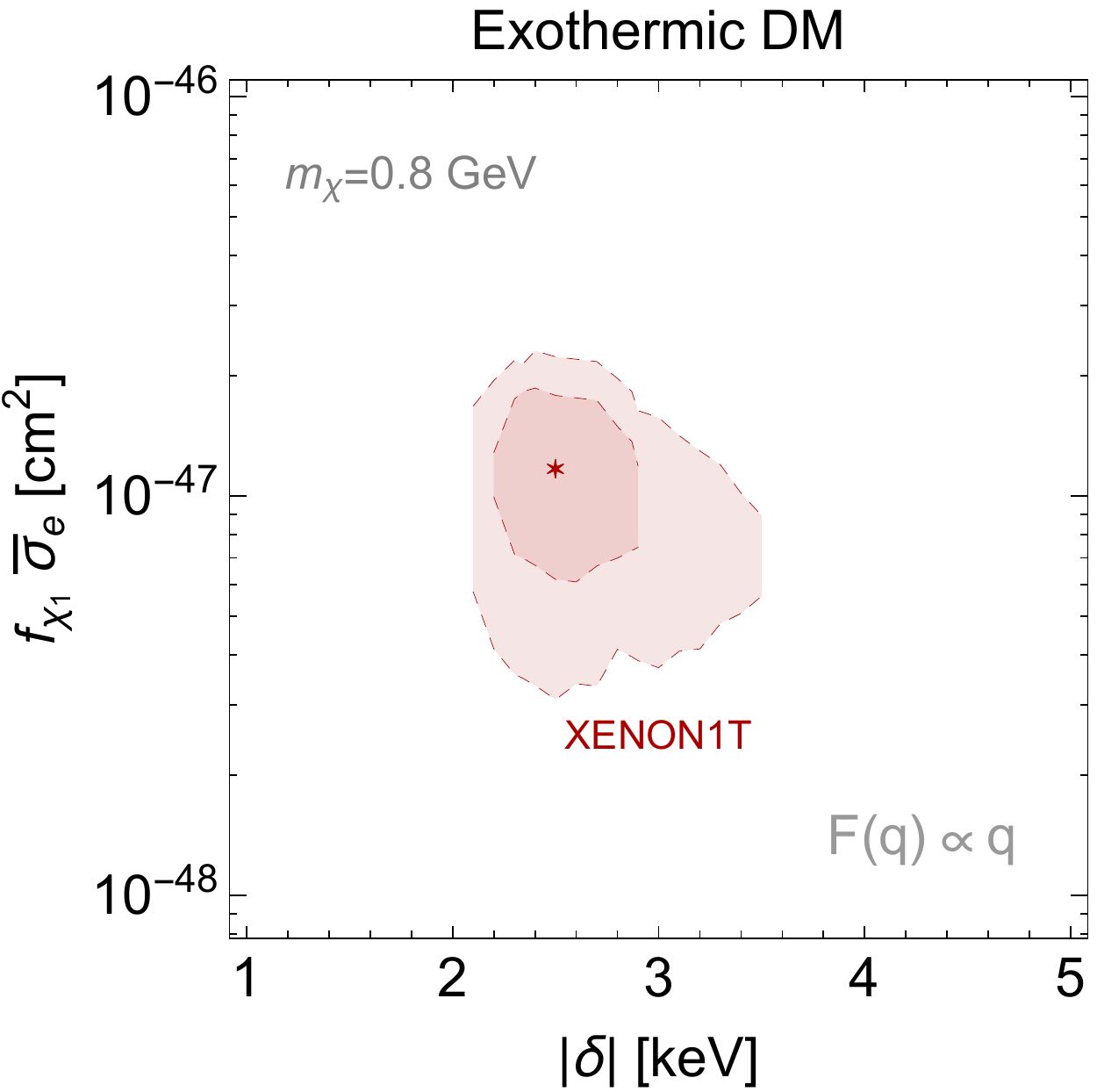}
\includegraphics[width=0.32\textwidth]{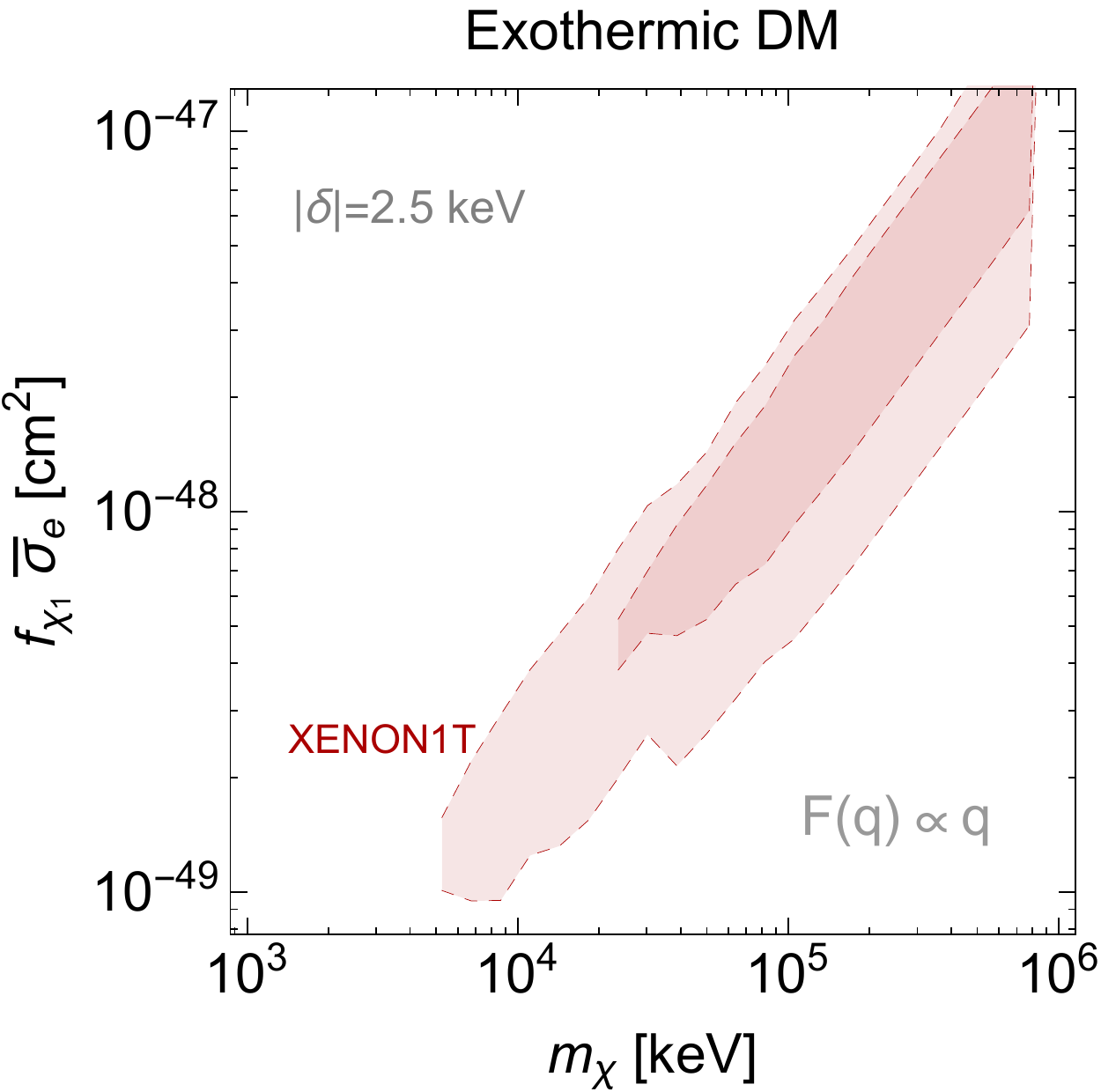}
\includegraphics[width=0.32\textwidth]{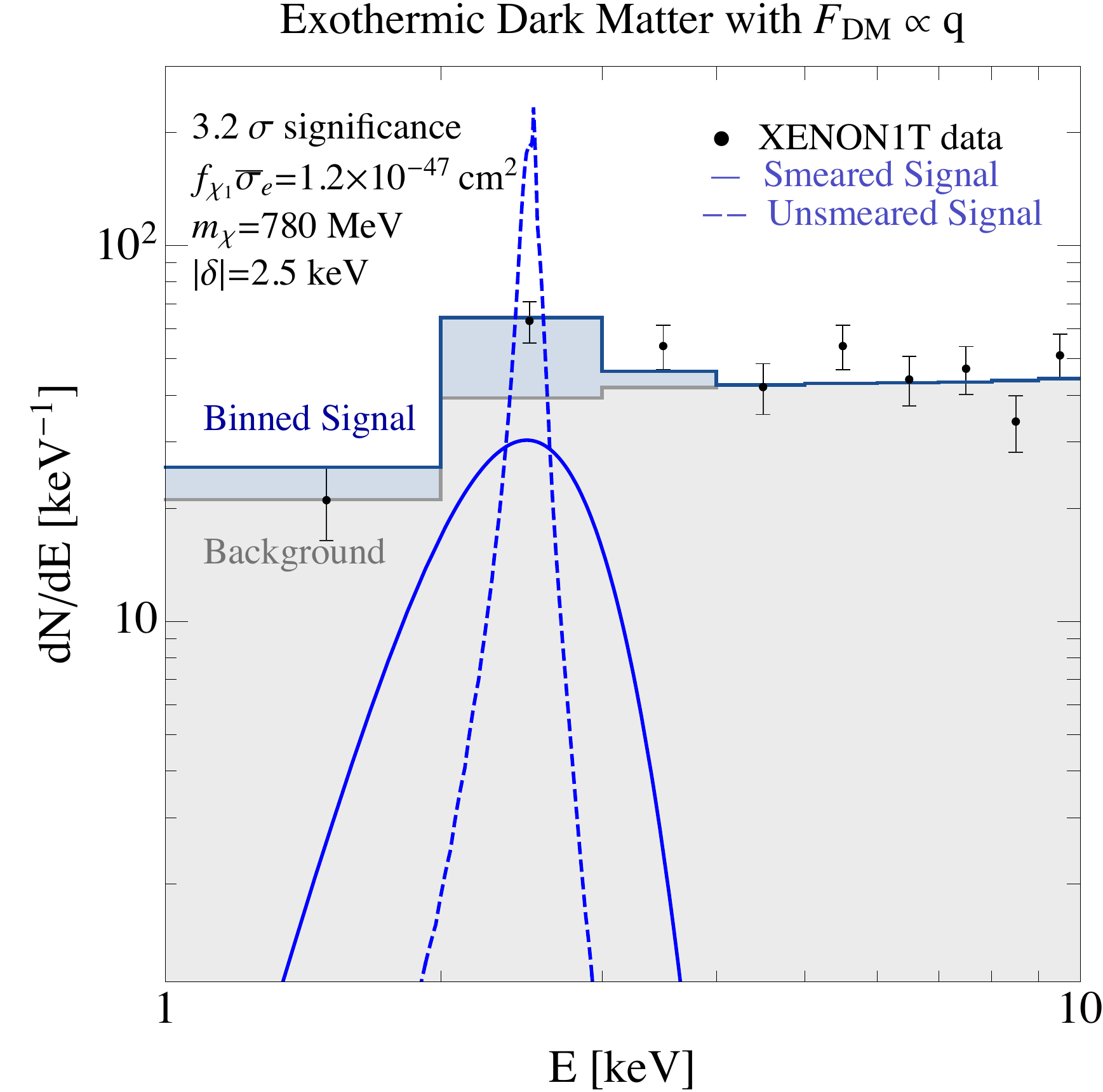}
\caption{The $1\sigma$ and $2\sigma$ best-fit regions that explain the \xe excess for exothermic DM and a momentum-dependent heavy mediator ($F_{\rm DM}=(q/\alpha m_e)$) in the $f_{\chi_1} \overline\sigma_e$ versus $\delta$ plane for $m_\chi=780$~MeV ($f_{\chi_1}=\frac{n_{\chi_1}}{n_{\chi_1}+n_{\chi_2}}$) (\textbf{left}), and in the $f_{\chi_1} \overline\sigma_e$ versus $m_{\chi}$ plane for $|\delta|=2.5$~keV (\textbf{middle}).  In the \textbf{right} plot, we show an example of the predicted spectrum for the best-fit value with $|\delta| \leq 4.9~$keV 
from Eq~\eqref{eq:bestfitexotFDMq}. The {\bf dashed} and {\bf solid} lines show the signal spectrum before and after detector smearing effects, respectively.   The measured \xe data is shown as {\bf black dots} while the {\bf gray-shaded} and {\bf blue-shaded} regions are the expected binned background and signal respectively. 
\label{fig:spectra-twodelta-fits-FDMq}}
\end{figure}

In Fig.~\ref{fig:spectra-twodelta-fits}, we show the $1\sigma$ and $2\sigma$ best-fit regions that explain the \xe excess for a heavy mediator ($F_{\rm DM}=1$) in the $\delta$-$m_{\chi_1}$ plane (left) and the  $\overline\sigma_e$-$m_{\chi_1}$ plane (middle). 
We see that the best-fit point (with $|\delta| \leq 4.9~$keV)  is given by 
\begin{equation} 
m_{\chi}\simeq 0.55 \text{ MeV}\,,\quad |\delta|\simeq 4.9~\text{ keV} ,\quad f_{\chi_1}\overline
\sigma_e\simeq 1.3 \times 10^{-43}~\text{ cm}^2, \quad 2\log(\mathcal{L}_{S+B}/\mathcal{L}_{B}) \simeq 16.7\, .
\label{eq:bestfitexot}
\end{equation}
In the right plot, we show how the signal at the best-fit point compares with the \xe data and background model.  
In Fig.~\ref{fig:spectra-twodelta-fits-FDMq2}, we show the corresponding plots for a light mediator ($F_{\rm DM}\propto1/q^2$). Here the best-fit point is given by 
\begin{equation} 
	m_{\chi}\simeq 30 \text{ MeV}\,,\quad |\delta|\simeq 2.6~\text{ keV} ,\quad f_{\chi_1}\overline
	\sigma_e\simeq 2.3 \times 10^{-46}~\text{ cm}^2, \quad 2\log(\mathcal{L}_{S+B}/\mathcal{L}_{B}) \simeq 15.7 \, .
\label{eq:bestfitexotFDMq2}
\end{equation}
Finally, in Fig.~\ref{fig:spectra-twodelta-fits-FDMq}, we show the corresponding plots for a $q$-dependent heavy mediator ($F_{\rm DM}\propto q$); here the best-fit point is given by 
\begin{equation} 
m_{\chi}\simeq 780 \text{ MeV}\,,\quad |\delta|\simeq 2.5~\text{ keV} ,\quad f_{\chi_1}\overline
\sigma_e\simeq 1.2\times 10^{-47}~\text{ cm}^2, \quad 2\log(\mathcal{L}_{S+B}/\mathcal{L}_{B}) \simeq 15.8 \, .
\label{eq:bestfitexotFDMq}
\end{equation}
We see that exothermic DM can explain well the observed \xe ER spectrum. 

\noindent We now make a few comments:  
\vspace{-2mm}
\begin{itemize}[leftmargin=0cm,itemindent=.5cm,labelwidth=\itemindent,labelsep=0cm,align=left]
\item The inclusion of relativistic corrections when calculating the atomic form factors is not essential for exothermic scattering, since $q$ is not forced to be large to obtain a large $\Delta E_e$ and the form factors typically peak at values of $q$ below which relativistic corrections become important.  We therefore neglect relativistic corrections in our calculations.  
\item While $m_\chi\sim1$~GeV provides an adequate fit to the \xe excess, one can obtain an even better fit for heavy DM 
by imagining that DM consists of three or more states. 
For example, for three states $\chi_1$, $\chi_2$, and $\chi_3$, with mass splitting $\delta_{21}\equiv m_{\chi_2}-m_{\chi_1}$, 
$\delta_{31}\equiv m_{\chi_3}-m_{\chi_1}$, and $\delta_{32}\equiv m_{\chi_3}-m_{\chi_2}$, with $\delta_{21}$, $\delta_{31}$, $\delta_{32}$ 
all negative, the electron recoil spectrum would show up to three peaks.  Of course, the actual size of the various peaks will depend sensitively on the relic abundances of the three DM states, and hence depend sensitively on the model parameters. 
\item If the DM couples also to nuclei (for example, if the mediator is a dark photon), DM could scatter exothermically off nuclei.  
We can contrast the kinematics for exothermic DM scattering off electrons with the kinematics for exothermic DM scattering off nuclei. 
For exothermic scattering off nuclei, the mean recoil energy is 
$\langle E_R\rangle \sim \frac{|\delta|\mu_{{\chi_1},N}}{m_N}$, where $m_N$ is the mass of the nucleus and $\mu_{{\chi_1},N}$ is the reduced mass of $\chi_1$ and the nucleus; the spread in energy around the mean recoil energy is given by
$\Delta E_R \sim \frac{\mu_{{\chi_1},N}}{m_N} \sqrt{8|\delta|\mu_{{\chi_1},N} v^2}$~\cite{Essig:2010ye,Graham:2010ca}.  For $\chi_1$ scattering off a xenon atom, with $m_{\chi_1}\sim 1$~GeV and $\delta \sim 1$~keV, $\langle E_R\rangle \sim 8$~eV, while the typical spread in energy around the mean recoil energy for the same parameters and a DM velocity of $v\sim 10^{-3}$ is $\Delta E_R \sim 21$~eV.  This is below the XENON1T and many other experimental thresholds, although not below the threshold achieved by CRESST-III; we will discuss this further in \S\ref{subsubsec:relicEXO}. 
\item It is possible to obtain electron recoils from the Migdal effect when DM scatters exothermically off nuclei; this could lead to additional constraints, which requires a careful study that we leave to future work.  
\item As mentioned above, the fractional abundance of the heavier state after freeze-out in the early universe will depend sensitively on the precise DM-mediator interaction strength and the DM and mediator masses.  Moreover, in a concrete model there will typically also be other constraints from searches at beam dumps, fixed-target experiments, and colliders. 
We investigate two concrete models in \S\ref{subsubsec:relicEXO}. 
\end{itemize}

\subsubsection{Exothermic Dark Matter-Electron Scattering: Relic Abundance for Concrete Models}\label{subsubsec:relicEXO}
In the case of exothermic DM, it is of crucial importance to calculate the relic abundance of the heavier state after freeze-out in the early Universe. We consider the case where the standard freeze-out is dominated by the annihilations into SM fermions $\chi_1 \chi_2 \leftrightarrow f \bar{f}$. After the dark states freeze out from the SM bath, they continue to be in chemical equilibrium and scatter with each other, $\chi_2 \chi_2 \leftrightarrow \chi_1 \chi_1$, driving the relative abundance to the equilibrium value
\begin{equation}
f^*\equiv n_{\chi_2}/n_{\chi_1} \simeq e^{-|\delta|/T_{\chi}^*}\ ,
\end{equation}
where $T_{\chi^*}$ is the temperature of the dark sector at which the DM-DM scattering decouples~\cite{Batell:2009vb,Finkbeiner:2009mi,TuckerSmith:2001hy}. If $T_{\chi^*}$ is much lower than the mass splitting $|\delta|$, the relative fraction of the excited states will be exponentially suppressed. In what follows, we compute the temperature of chemical decoupling $T_{\chi^*}$ in two explicit simple models of exothermic DM.\footnote{For the cases of interest here, the DM-DM scattering will always decouple after the scattering of DM with electrons, and hence the DM-DM scattering will set the relative abundance in the dark sector.}

We study models where the coupling between the SM and the DM sector arises via the kinetic mixing of the dark photon with the SM photon. We discuss both the cases of a complex scalar DM and a pseudo-Dirac fermion DM, which have very similar parametrical dependence on the physical quantities up to numerical factors. The cosmology of the pseudo-Dirac DM and its implications for the \xe excess have been recently considered in~\cite{Baryakhtar:2020rwy, Bramante:2020zos}. Our treatment of the cosmology here agrees with the one first presented in~\cite{Baryakhtar:2020rwy}.\footnote{We thank Hongwan Liu for correspondence and a thorough comparison of our results.} For the cosmology of the scalar case, we follow a similar treatment, obtaining results that agree with~\cite{Harigaya:2020ckz}. The final allowed parameter space compatible with the \xe excess will be somewhat different than the previous analyses because of our improved statistical analysis of the \xe data and a different treatment of the atomic form factors.

 The interaction Lagrangian between the SM and the dark sector reads
\begin{equation}
\mathcal{L} \supset g_D A'^{\mu} J_\mu^{\text{DM}} + \frac{\epsilon}{2}F_{\mu\nu}F'^{\mu\nu},
\end{equation} 
where $F'_{\mu\nu}$ is the field strength of the dark U(1), $g_D\equiv\sqrt{4\pi \alpha_D}$ is the dark photon coupling, and $J_\mu^{\text{DM}}$ the dark matter current given by 
\begin{align}
&\text{scalar DM:}\qquad J_\mu^{\text{DM}}=-i\left(\phi\partial_\mu\phi^{\ast}\right)=\chi_1\partial_\mu \chi_2-\chi_2\partial_\mu \chi_1\ ,\\
&\text{fermionic DM:}\qquad J_\mu^{\text{DM}}=-i\left(\bar{\psi}\gamma_\mu\psi\right)=-i \left(\chi_1^{\dagger}\bar{\sigma}_\mu\chi_2-\chi_2^{\dagger}\bar{\sigma}_\mu\chi_1\right)\ ,
\end{align}
where we write the complex scalar current in terms of the real scalar components $\phi=(\chi_1+i\chi_2)/\sqrt{2}$ and the Dirac fermionic current in terms of its Weyl components $\bar{\psi}=(\chi_2,\chi_1^{\dagger})$. The mass terms in the two models can be written as 
\begin{align}
&\text{scalar DM:}\qquad \mathcal{L}_{m}=m^2\phi^{\ast}\phi+y_D^2 H_D^2\phi^2+\text{h.c}\ ,\\
&\text{fermionic DM:}\qquad \mathcal{L}_{m}= m\bar{\psi}\psi+y_D H_D\psi^2+\text{h.c}\ ,\
\end{align}
where the VEV of the dark Higgs $H_D$  breaks the $U(1)$ in the dark sector, generating a mass splitting between $\chi_1$ and $\chi_2$ 
\begin{equation}
\text{scalar DM:}\quad\delta\simeq \frac{y_D^2 \langle H_D\rangle^2}{m}\ ,\qquad \text{fermion DM:}\quad \delta\simeq y_D\langle H_D\rangle\ .
\end{equation}
The splitting can be easily suppressed compared to the mass $m$. As long as $y_D$ is small enough, the mass of the dark Higgs can be made arbitrarily heavy and, hence, will be neglected in the rest of our discussion.
   
The freeze out of the $\chi_1 \chi_2 \leftrightarrow f \bar{f}$ interactions fixes the total number density of dark sector states $n_{\chi_1}+n_{\chi_2}\simeq n_{\chi_1}$. This can be chosen to match the DM abundance today if $y=\alpha_D \epsilon^2\left( m_\chi/m_{A'}\right)^4 $ is fixed as a function of the DM mass.  In the limit $m_A'\gtrsim m_{\chi_1}\gtrsim m_e$, we find roughly  
\begin{align}
&\text{scalar DM:} \quad y\sim 10^{-8}\left(\frac{m_{\chi_1}}{100 \text{ MeV}}\right)^2\ ,\\
&\text{fermionic DM:}  \quad  y\sim 10^{-10}\left(\frac{m_{\chi_1}}{100 \text{ MeV}}\right)^2\ .
\end{align} 
The precise relation dependence of $y$ on the DM mass needs to be extracted numerically~\cite{Izaguirre:2015yja} (see also~\cite{Boehm:2003hm,Essig:2015cda}).  

The compute the relative abundance of $n_{\chi_2}/n_{\chi_1}$, we need to consider the rate of the process $\chi_2 \chi_2 \leftrightarrow \chi_1 \chi_1$
\begin{equation}
\Gamma_{\chi_2 \chi_2 \leftrightarrow \chi_1 \chi_1} = e^{-|\delta|/T_{\chi}}n_{\chi_1} \langle\sigma_{\chi_2 \chi_2 \leftrightarrow \chi_1 \chi_1} v\rangle,
\end{equation}
where the thermally averaged scattering cross section reads  
\begin{equation}
\langle\sigma_{\chi_2 \chi_2 \leftrightarrow \chi_1 \chi_1} v\rangle \simeq \frac{16\kappa \sqrt{2}\pi \alpha_D^2 m_{\chi}^{3/2}}{m_{A'}^4}\text{max}\Big(2T_{\chi}/\pi, \delta\Big)^{1/2}\ ,
\label{eq:crossexo}
\end{equation}
and we introduce the numerical coefficient $\kappa=(1,1/4)$ to distinguish the scalar and fermionic cases. This formula agrees with the results of~\cite{Baryakhtar:2020rwy} and~\cite{Harigaya:2020ckz} in the fermionic and scalar case, respectively. The dependence on $\delta$ arises from the threshold velocity for the scattering process, and in most of the parameter space of interest $\vert\delta\vert \gtrsim  2T/\pi$. The chemical decoupling temperature $T_{\chi}^*$ is then defined as $ \Gamma_{\chi_1\chi_2}(T_{\chi}^*) \simeq H(T_{\chi}^*)$.

\begin{figure}[t]
\centering
\includegraphics[width=0.46\textwidth]{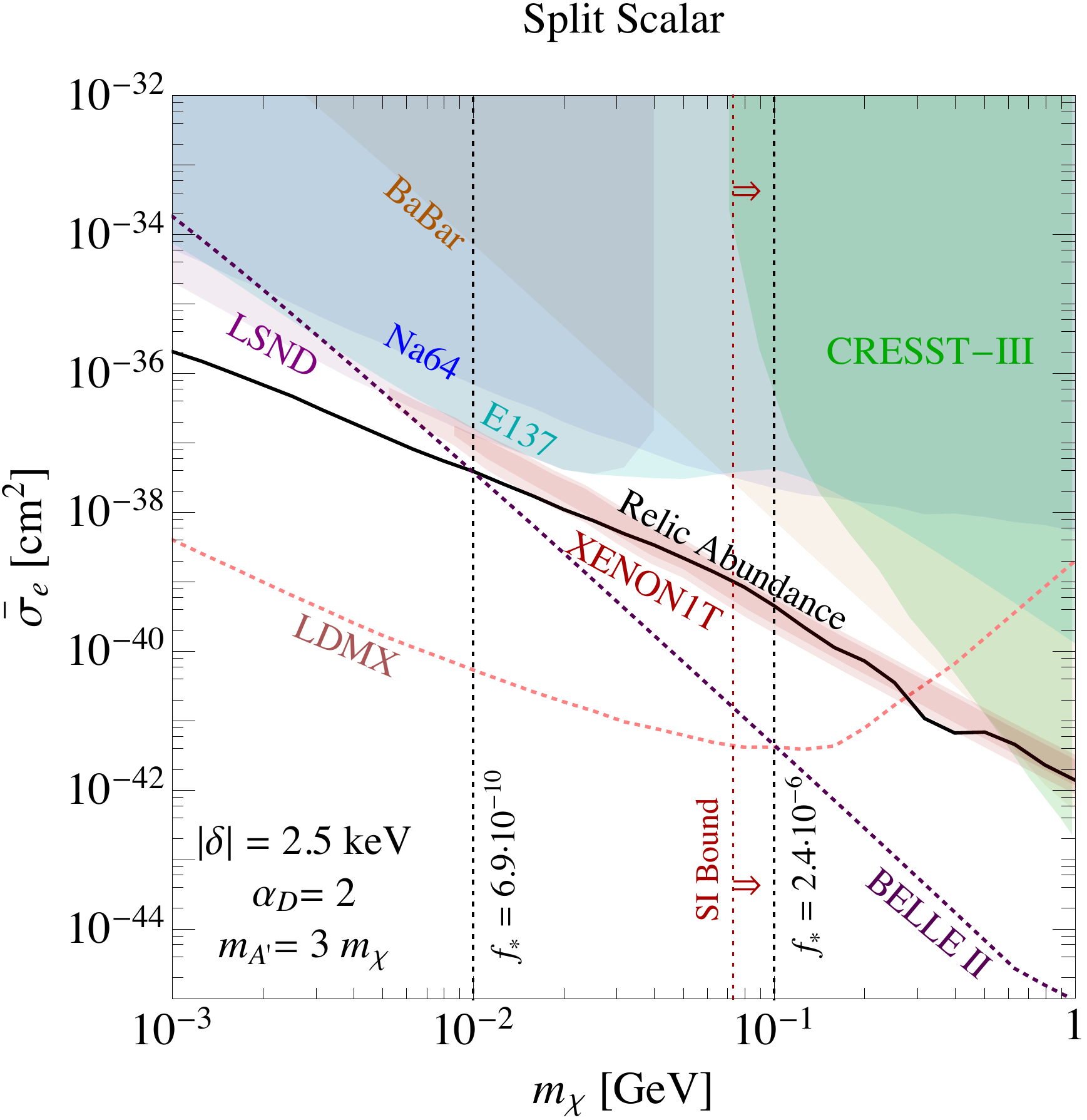}\hfill
\includegraphics[width=0.46\textwidth]{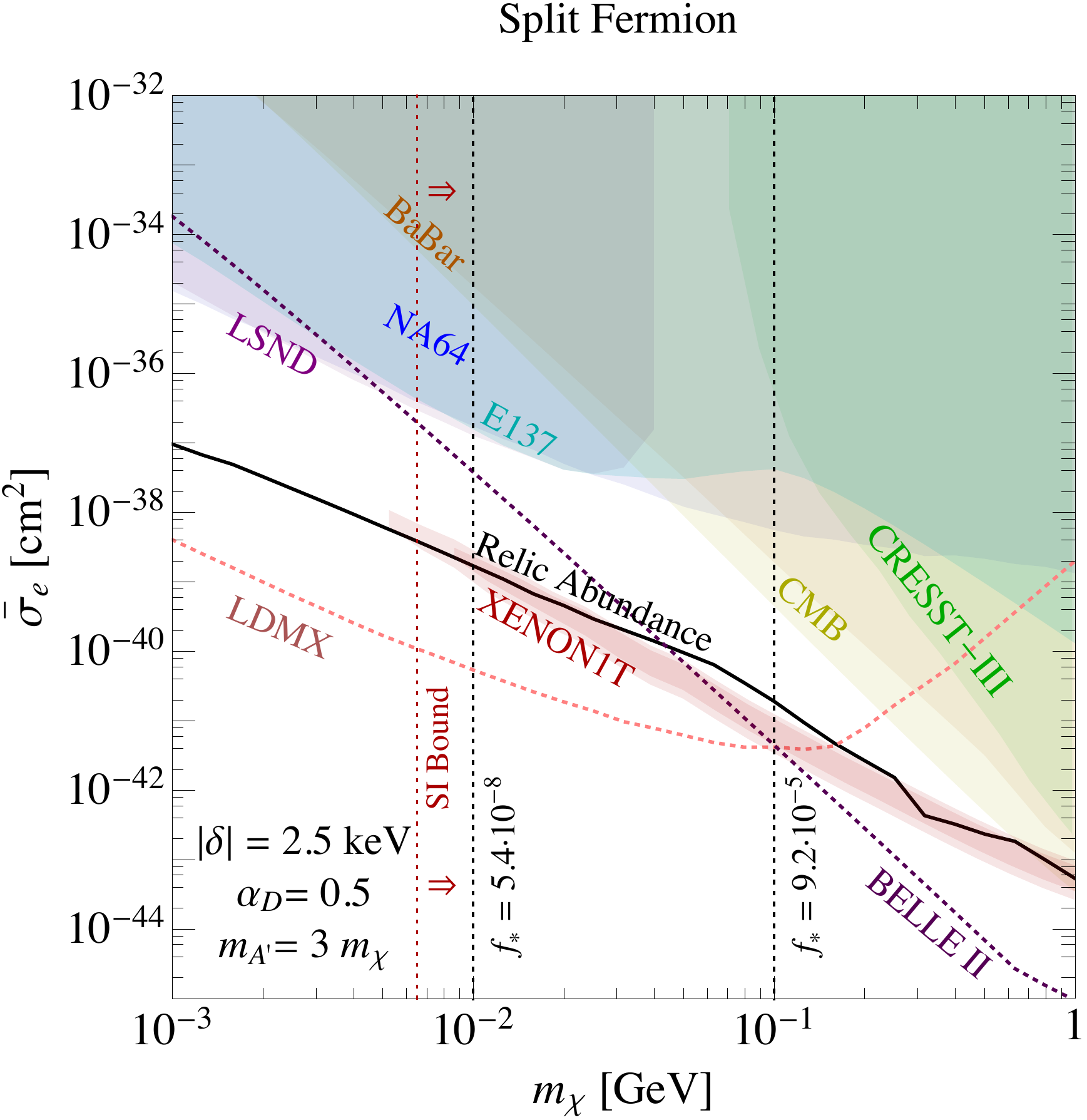}
\caption{The \textbf{red} regions indicate the $1\sigma$ and $2\sigma$ best-fit regions for exothermic DM in which the DM states consist of two scalars (\textbf{left}) or two fermions (\textbf{right}) that are split in mass by an amount $\delta=2.5$~keV, and interact through a dark photon mediator with mass $m_{A'}= 3 m_{\chi}$ and coupling $\alpha_D=2$ (\textbf{left}) and $m_{A'}= 3 m_{\chi}$ and $\alpha_D=0.5$ (\textbf{right}).  The \xe signal is produced from the heavy state (down)scattering to the lighter state. 
We show also the bounds on accelerator-based searches: in \textbf{blue} the bound from the NA64 experiments~\cite{NA64:2019imj}, in \textbf{dark orange} the BABAR constraint~\cite{Lees:2017lec,Essig:2013vha}, in \textbf{purple} the bound from LSND~\cite{deNiverville:2011it, Batell:2009di}, in \textbf{cyan} the bound from E137~\cite{Batell:2014mga}. The {\bf dotted light-red} line shows the reach of LDMX~\cite{Berlin:2018bsc}, while the dotted dark-red line shows the reach of Belle-II~\cite{Essig:2013vha,BELLEII-talk-Hearty}. 
The \textbf{light green} shaded region is the bound from the heavy state scattering exothermically off nuclei in CRESST-III~\cite{Abdelhameed:2019hmk,Abdelhameed:2019mac}.  We also show the limit from self interaction (\textbf{dotted red vertical} line)~\cite{Tulin:2017ara} and from CMB distortion due to energy injection (\textbf{dark yellow}) for the fermionic case~\cite{Madhavacheril:2013cna,Aghanim:2018eyx}.
\textbf{Dotted black vertical} lines indicate the fractional relic abundance, $f^*$, of the excited state. }  
\label{fig:split_mAp3}
\end{figure} 
In Fig.~\ref{fig:split_mAp3}, we show in red the $1\sigma$ and $2\sigma$ best-fit regions for the case of a heavy dark photon mediator with mass $m_{A'}= 3 m_{\chi}$ and $\alpha_D = 0.5$ ($\alpha_D = 2$) for the fermion (scalar) case. The required electronic interaction cross section $\bar{\sigma}_e$, defined in Eq.~\eqref{eq:sigma-scatter}, needs to be larger for lower DM masses because the fraction of primordial excited states rapidly decreases for lower DM mass. Indeed, the rate of de-excitation scales as $\Gamma_{\chi_1\chi_2} \sim m_{\chi}^{-7/2}$, which then implies $f^* \sim m_{\chi}^{7/2}$. 

In the fermionic model, for values of the cross section that explain the \xe excess it is easy to get the correct DM relic abundance for perturbative values of $\alpha_D$. This result is in agreement with previous analysis~\cite{Baryakhtar:2020rwy, Bramante:2020zos}. Conversely, in the scalar case, fitting the \xe excess pushes the parameter space with the relic abundance to values for $\alpha_D$ that are at the boundary of perturbativity~\cite{Davoudiasl:2015hxa}. Alternatively, one could explore a region of parameter space where the DM mass is almost degenerate with the dark photon mass, and resonant effects enhance the annihilation cross section~\cite{Feng:2017drg}.  Of course, this is a specific feature of the simple models presented here and adding further annihilation channels within the dark sector could open the parameter space substantially at the price of a less minimal model. 

In Fig.~\ref{fig:split_mAp3}, we show the accelerator constraints on the dark photon decaying ``invisibly'' to DM: the BABAR constraint from a monophoton search $e^+e^-\to \gamma A'$ ~\cite{Lees:2017lec,Essig:2013vha} (dark orange solid line),  from an electron-beam-dump missing-energy search induced by dark photon bremsstrahlung (to DM) at NA64~\cite{NA64:2019imj} (blue solid line), and the projections from LDMX~\cite{Berlin:2018bsc} and Belle-II~\cite{Essig:2013vha,BELLEII-talk-Hearty}. We show the bound on a DM beam produced in high-intensity beam dump experiments such as E137~\cite{Batell:2014mga} or neutrino experiments such as LSND~\cite{deNiverville:2011it, Batell:2009di}. We display the self interaction constraints (dotted red vertical line)~\cite{Tulin:2017ara} $\sigma_{\text{SI}}/m_{\chi} \lesssim 10 \text{ cm}^2/\text{g}$; in this case, the relevant scattering cross section is the elastic scattering of the light state with itself, which is loop-suppressed. This bound is relevant only at low masses and reads $\alpha_D \lesssim 0.06 \, \Big(\text{MeV}/m_{\chi}\Big)^{1/2}\Big(m_{A'}/10 \, \text{MeV}\Big)^2$~\cite{Schutz:2014nka, Izaguirre:2015yja}. 

In the fermionic case, important constraints can be derived from CMB distortion due to energy injection (dark yellow)~\cite{Madhavacheril:2013cna,Aghanim:2018eyx}. In fact, residual co-annihilations into SM states can reionize hydrogen and distort the high-$\ell$ CMB power spectrum. Planck observations limit the cross section for annihilation to electromagnetic final state to be $f^* \sigma v \lesssim\text{ pb }\times \Big(m_{\chi}/ 60 \, \text{GeV} \Big)$, which gives a bound on $\bar{\sigma}_e$ as 
\begin{equation}
\bar{\sigma}_e \lesssim\frac{\text{ pb}}{f^*}\Big(\frac{\mu_{\chi e}}{m_{\chi}}\Big)^2\Big(\frac{m_{\chi}}{60 \, \text{GeV}}\Big)\,.
\end{equation}
For the scalar case instead, the CMB bounds are weakened, because the co-annihilation to leptonic final states is p-wave suppressed. 

Last,  we include direct detection bounds on DM inelastic scattering off nuclei. The most relevant searches here are from CRESST-III, which obtained a very low energy threshold of 19.7~eV in~\cite{Angloher:2017sxg} and 30.1~eV in~\cite{Abdelhameed:2019hmk,Abdelhameed:2019mac}. Since the exposure is on the order of gram-hours in~\cite{Angloher:2017sxg} and of order kg-days in~\cite{Abdelhameed:2019hmk,Abdelhameed:2019mac}, we calculate only the bound from~\cite{Abdelhameed:2019hmk,Abdelhameed:2019mac}. To derive this bound, we reproduce Yellin's optimum interval method~\cite{Yellin:2002xd}, which is the method used by the CRESST collaboration for their own bounds. More details about the recasting of low-threshold nuclear recoil bounds for different DM models will be given elsewhere~\cite{toappear}. 

\section{Accelerated Dark Matter}
\label{sec:acceleratedDM}
A fraction of DM could be accelerated to high velocities, producing an energetic DM flux that impinges on the Earth~\cite{An:2017ojc,Yin:2018yjn,Emken:2017hnp, Bringmann:2018cvk,Ema:2018bih,Cappiello:2019qsw}. Such an accelerated component may then be detected with experiments such as XENON1T, allowing for sensitivity to very light DM, which otherwise cannot be probed without sub-keV threshold experiments.  
Specifically, two distinct mechanisms have been suggested.  In the first, DM interacts with the solar interior to produce a significantly harder spectrum~\cite{An:2017ojc,Emken:2017hnp}.  However, for standard DM with $F_{\rm DM} = 1$, the resulting flux ends at around 2~keV, thereby naively disfavoring a simple fit to the \xe data.
A second energetic DM flux is generated through interactions with cosmic rays (CRs)~\cite{Bringmann:2018cvk,Ema:2018bih,Cappiello:2019qsw}.   This was used to derive world-leading limits on DM-electron couplings for DM in the eV to few keV mass range using the Super-K experiment~\cite{Bays:2011si}.   

Naively, DM acceleration from CRs cannot address the \xe anomaly either, for the following reason.  For the previously studied DM-electron interactions with trivial form factor ($F_{\rm DM} = 1$), the predicted accelerated DM spectrum does not vary by more than an order of magnitude between keV and 100~MeV.  However, the Super-K analysis uses 176 kt-years of data (with a $\sim 100$ MeV threshold),  about five orders of magnitude larger than the 0.65 tonne-year exposure available in XENON1T.  Consequently, any signal at \xe would be naively excluded by Super-K.    

The above argument does not hold for a DM interacting with electrons via a light mediator.  Indeed, in such a case, both the produced flux and scattering rate predict a steeply falling spectrum towards higher energies, thereby easily compensating for the relative low exposure of \xe with its significantly lower threshold.  However, experiments with lower thresholds may then be more constraining.  
We now study this possibility in detail.

\begin{figure}[t]
\centering
\includegraphics[width=0.46\textwidth]{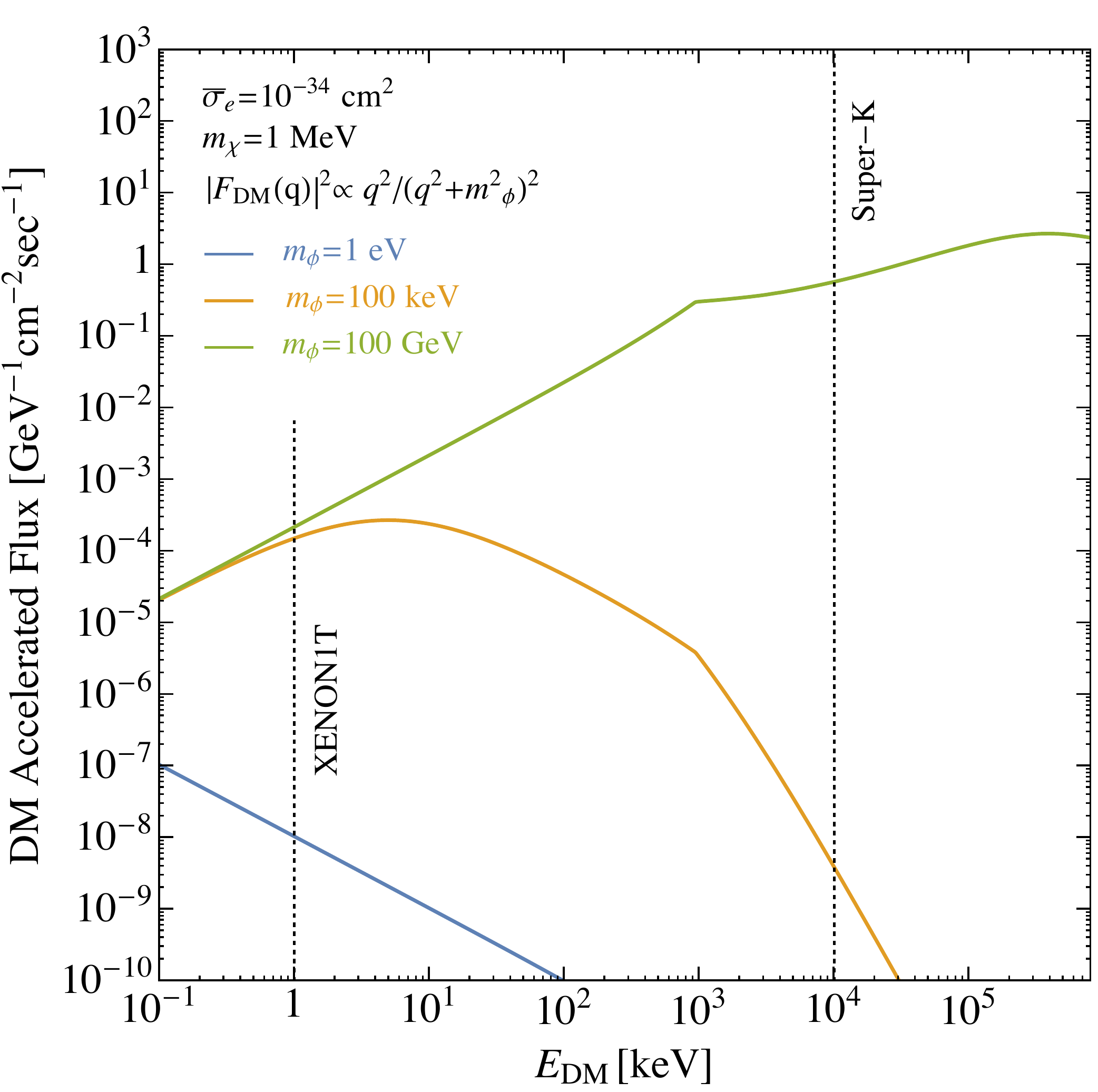}\hfill
\includegraphics[width=0.46\textwidth]{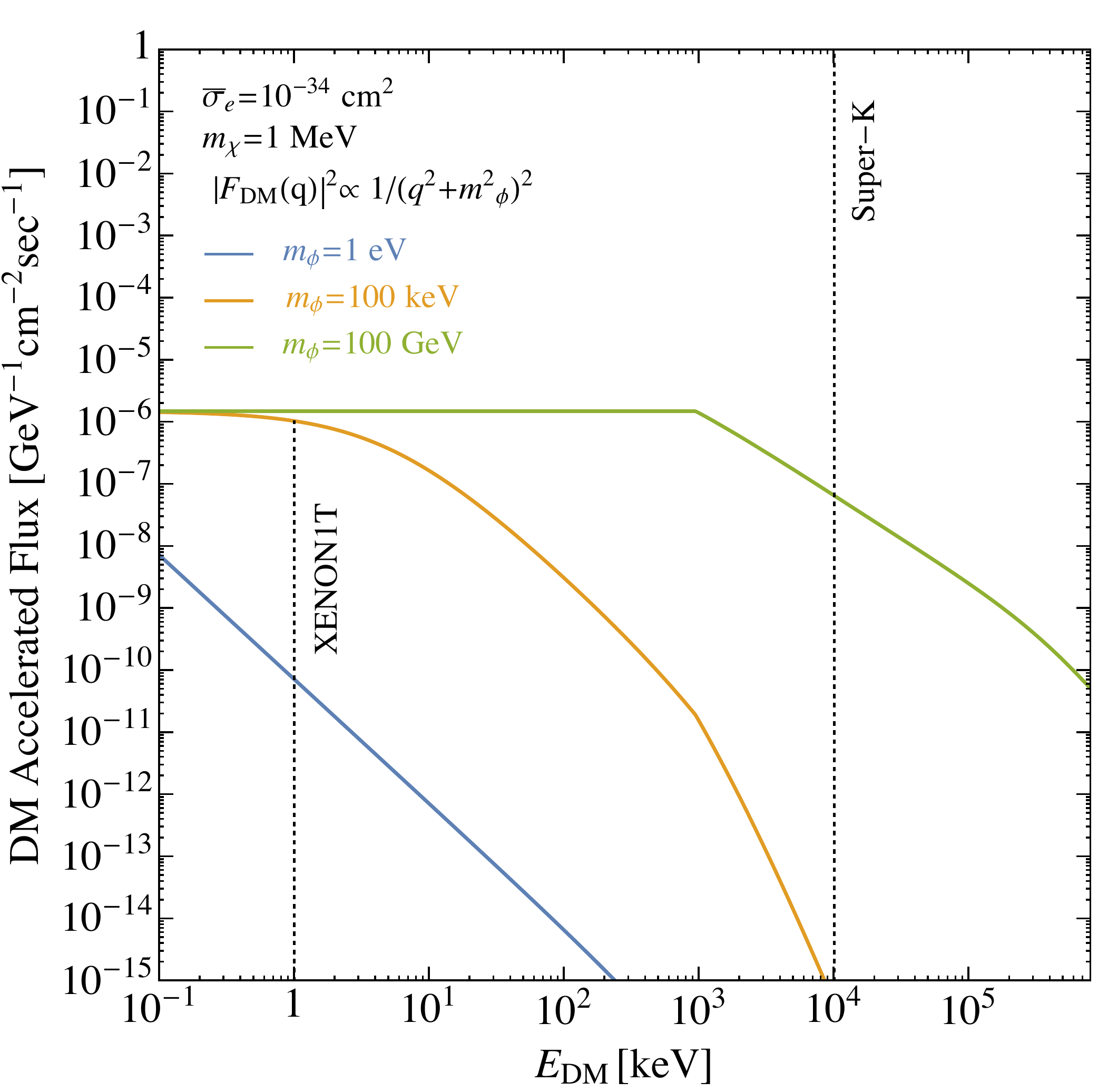}\\
\includegraphics[width=0.46\textwidth]{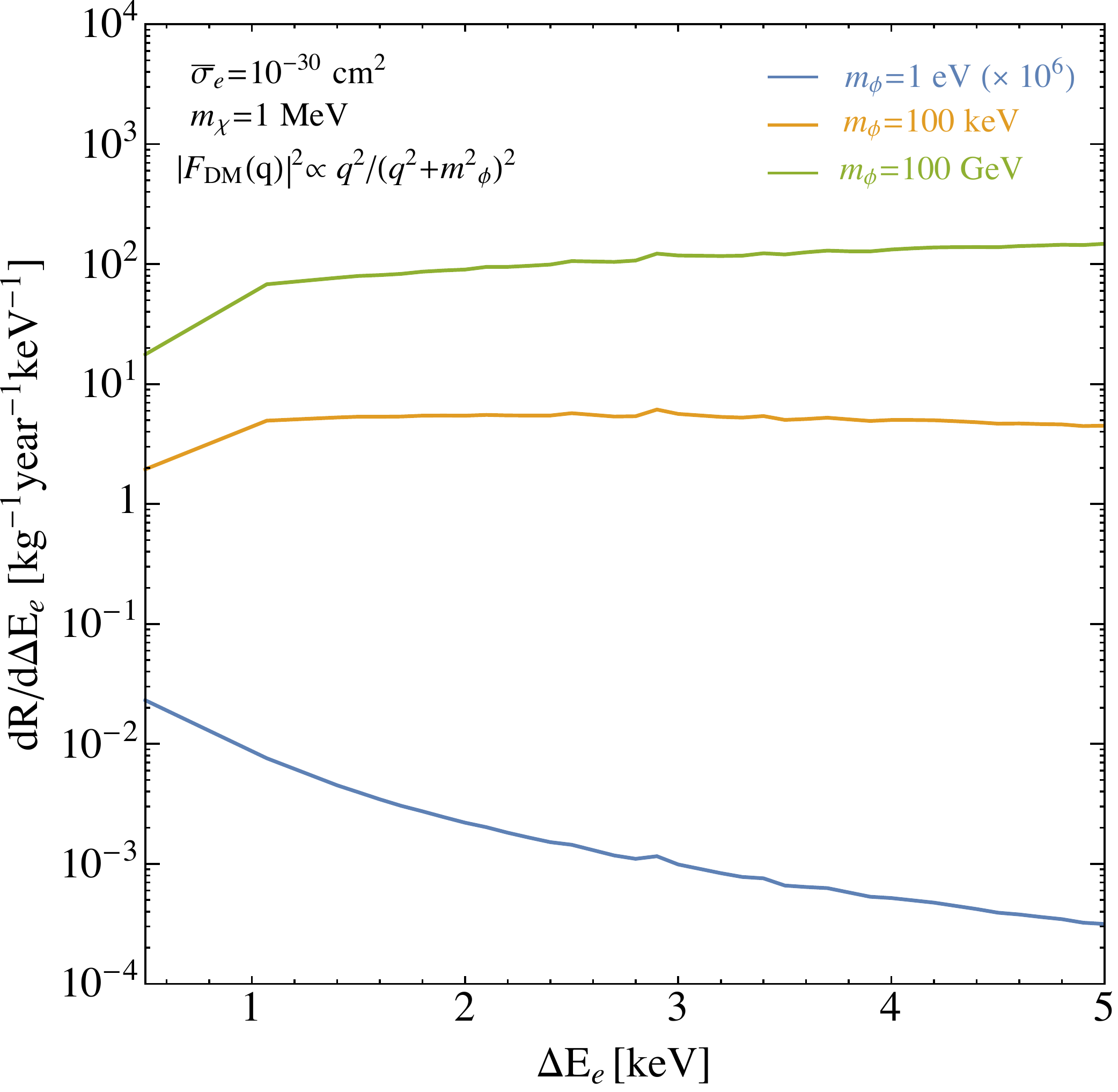}\hfill
\includegraphics[width=0.46\textwidth]{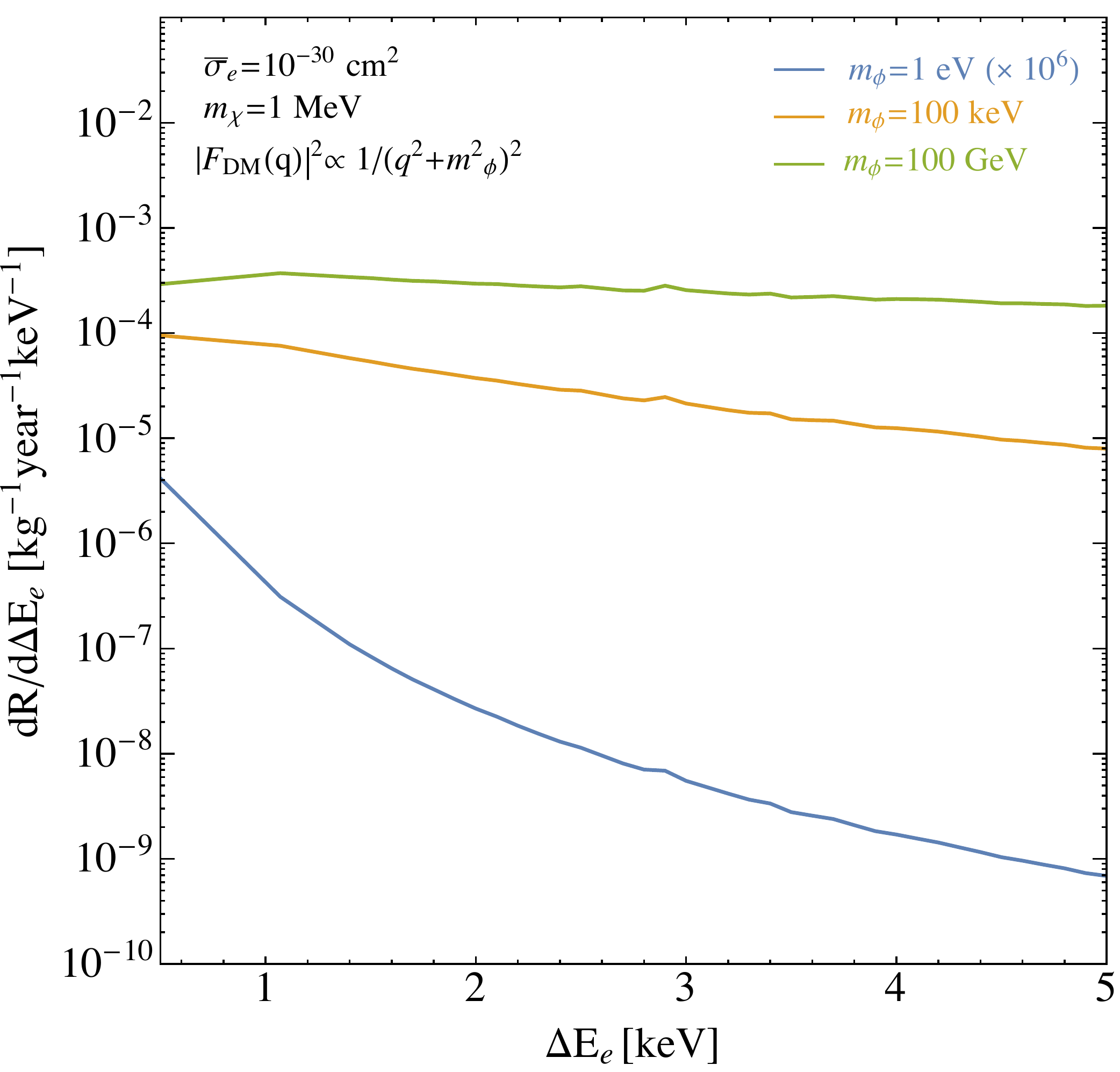}
\caption{{\bf Top:} Accelerated dark matter flux due to interactions with cosmic ray electrons.   The flux is shown for two different DM form factors: $|F_{\rm DM}(q)|^2 \propto q^2/(q^2 + m_{\phi}^2)^2$ (\textbf{left}) and $|F_{\rm DM}(q)|^2 \propto 1/(q^2 + m_{\phi}^2)^2$ (\textbf{right}).  The three different  solid colored lines show the flux for varying values of the mediator mass, $m_{\phi}$=1 eV (\textbf{blue}), $m_{\phi}$=100 keV (\textbf{orange}), and $m_{\phi}$=100~GeV (\textbf{green}).  In these plots the DM mass is set to 1~MeV and the DM-electron cross-section is taken to be $\bar\sigma_e = 10^{-30} \units{cm^2}$.  The {\bf black dashed} lines indicate the energy thresholds for the Super-K and \xe experiments. {\bf Bottom:} Electron recoil spectra from cosmic ray accelerated DM flux for $|F_{\rm DM}(q)|^2 \propto q^2/(q^2 + m_{\phi}^2)^2$~(\textbf{left}) and $|F_{\rm DM}(q)|^2 \propto 1/(q^2 + m_{\phi}^2)^2$~(\textbf{right}). Three  different values of the mediator mass, $m_{\phi}$=1 eV (\textbf{blue}), $m_{\phi}$=100 keV (\textbf{orange}), and $m_{\phi}=100$~GeV (\textbf{green}) are shown.  The DM mass is fixed as $m_\chi=1$~MeV, and the DM-electron cross section is taken to be $10^{-30}\units{cm^2}$.   As discussed in the text, only the $|F_{\rm DM}(q)|^2 \propto q^2/(q^2 + m_{\phi}^2)^2$  with an intermediate mediator mass can viably  address the \xe data.} 
\label{fig:CRDMflux}
\end{figure} 

We consider DM that interacts solely with electrons via a light mediator.  In order to fit the \xe anomaly, the mediator mass must be i) lighter than a few MeV or else the benefit of having a low-threshold experiment in comparison to the Super-K experiment is lost ii) heavier than roughly 1~keV in order to evade the S2-only analysis of \xe\cite{Aprile:2019xxb}. In this mass range, and for the range of electron couplings we consider, the mediator coupling to the SM model ends up being excluded by complementary searches for the light mediator (see e.g.~\cite{Knapen:2017xzo} for a summary). We expect this feature to be quite generic in all the models of DM accelerated by cosmic rays.  A complementary study to this scenario with significantly lighter mediator masses is upcoming~\cite{Bringmann:2020}. 

The DM flux obtained from interactions with CRs is given by,
\begin{equation}
\label{eq:DMflux}
\frac{d\Phi_\chi}{d{\cal E}_\chi} = \int d{\cal E}_e \frac{d\Phi_\chi}{d{\cal E}_e} \frac{1}{{\cal E}_\chi^{\rm max}({\cal E}_e)}\Theta\left[{\cal E}_\chi^{\rm max}({\cal E}_e) - {\cal E}_\chi\right]\,,
\end{equation}
where ${\cal E}_{\rm \chi}$ and ${\cal E}_{\rm e}$ are the DM's and CR-electrons' kinetic energy, 
\begin{equation}
{\cal E}_\chi^{\rm max}=\frac{2m_\chi({\cal E}_e^2+2m_e{\cal E}_e)}{(m_e+m_\chi)^2+2m_\chi{\cal E}_e}\,,
\end{equation}
and 
\begin{equation}
\label{eq:CRflux}
\frac{d\Phi_\chi}{d{\cal E}_e} = \int \frac{d\Omega}{4\pi} \int_{l.o.s.} dl \,\bar\sigma_{\chi e} |F_{\rm DM}(q)|^2\frac{\rho_\chi}{m_\chi} \frac{d\Phi_e}{d{\cal E}_e}\,.
\end{equation}
Here $\sigma_{\chi e}$ is the DM-electron (momentum-dependent) cross-section, $\rho_\chi$ is the DM density profile, taken to be an NFW profile~\cite{Navarro:1996gj} with a scale radius of $r_s = 20$~kpc, a local density of $\rho_{\odot} = 0.4\units{GeV/cm^3}$, and $d\Phi_e/d{\cal E}_e$ is the CR-electron flux.  

In order to derive limits from the low threshold ($\sim$150~eV) S2-only \xe analysis, one crucially needs to know the CR flux down to ${\cal O}({\rm keV})$ energies.  However, measurements only provide the spectrum down to MeV energies~\cite{Boschini:2018zdv} and therefore an extrapolation must be used.   Strictly speaking, this implies that systematic uncertainties hinder the possibility of using the S2-analysis to exclude the CR-accelerated DM solution.   In what follows, we thus simply assume that the CR flux drops to zero below MeV energies.   

 \begin{figure}[t]
\centering
\includegraphics[width=0.33\textwidth]{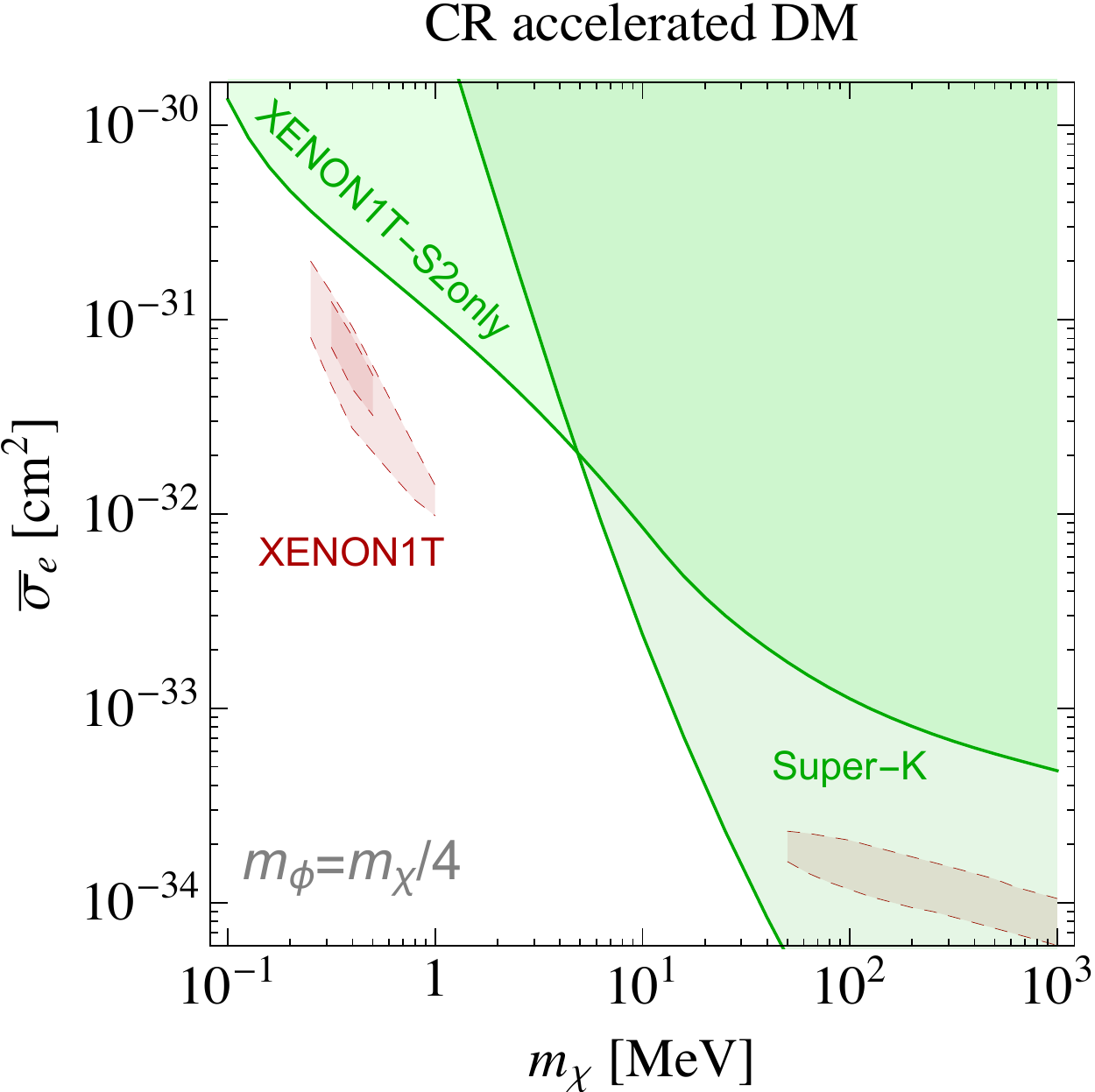}\hfill
\includegraphics[width=0.33\textwidth]{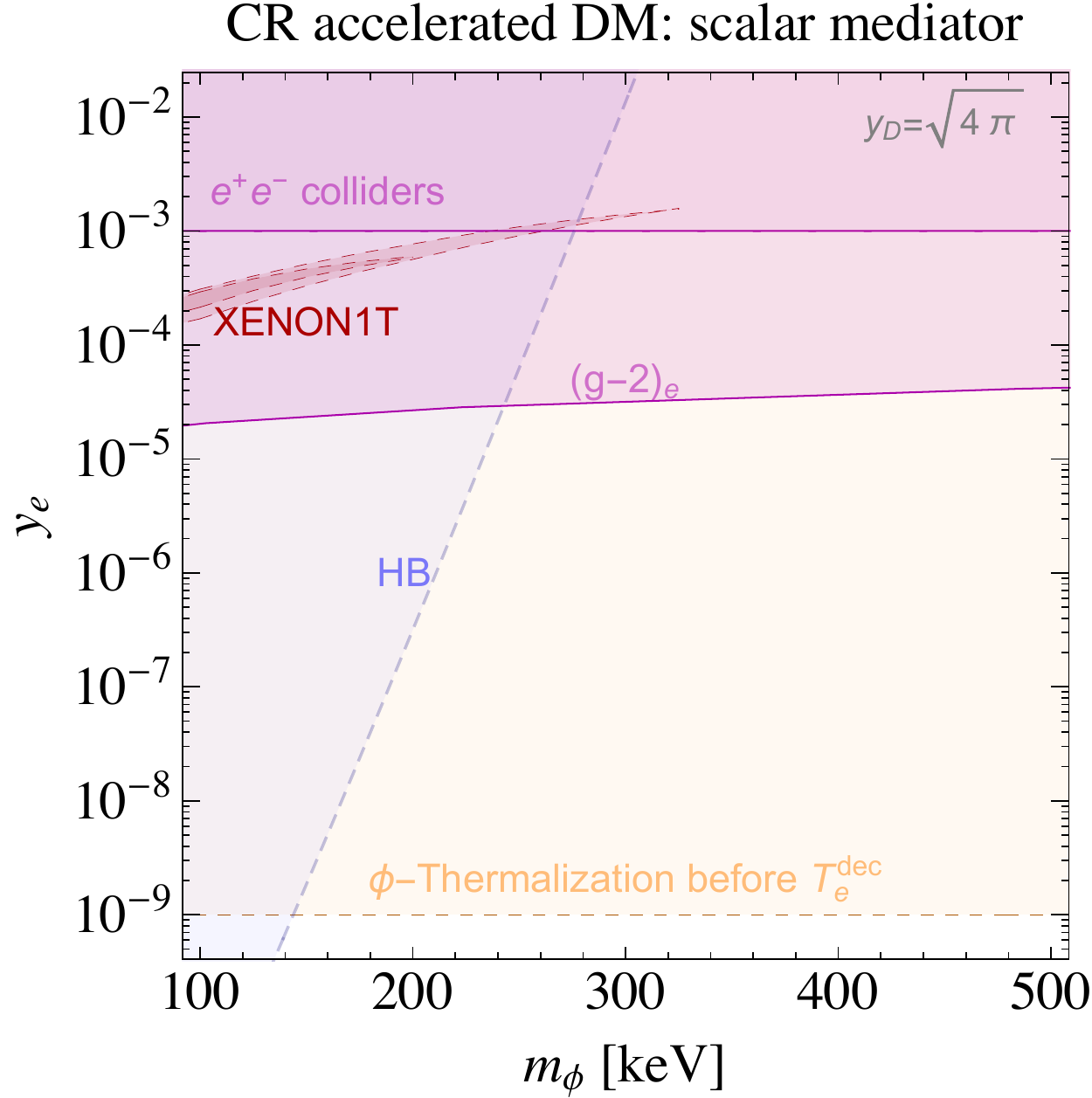}\hfill
\includegraphics[width=0.33\textwidth]{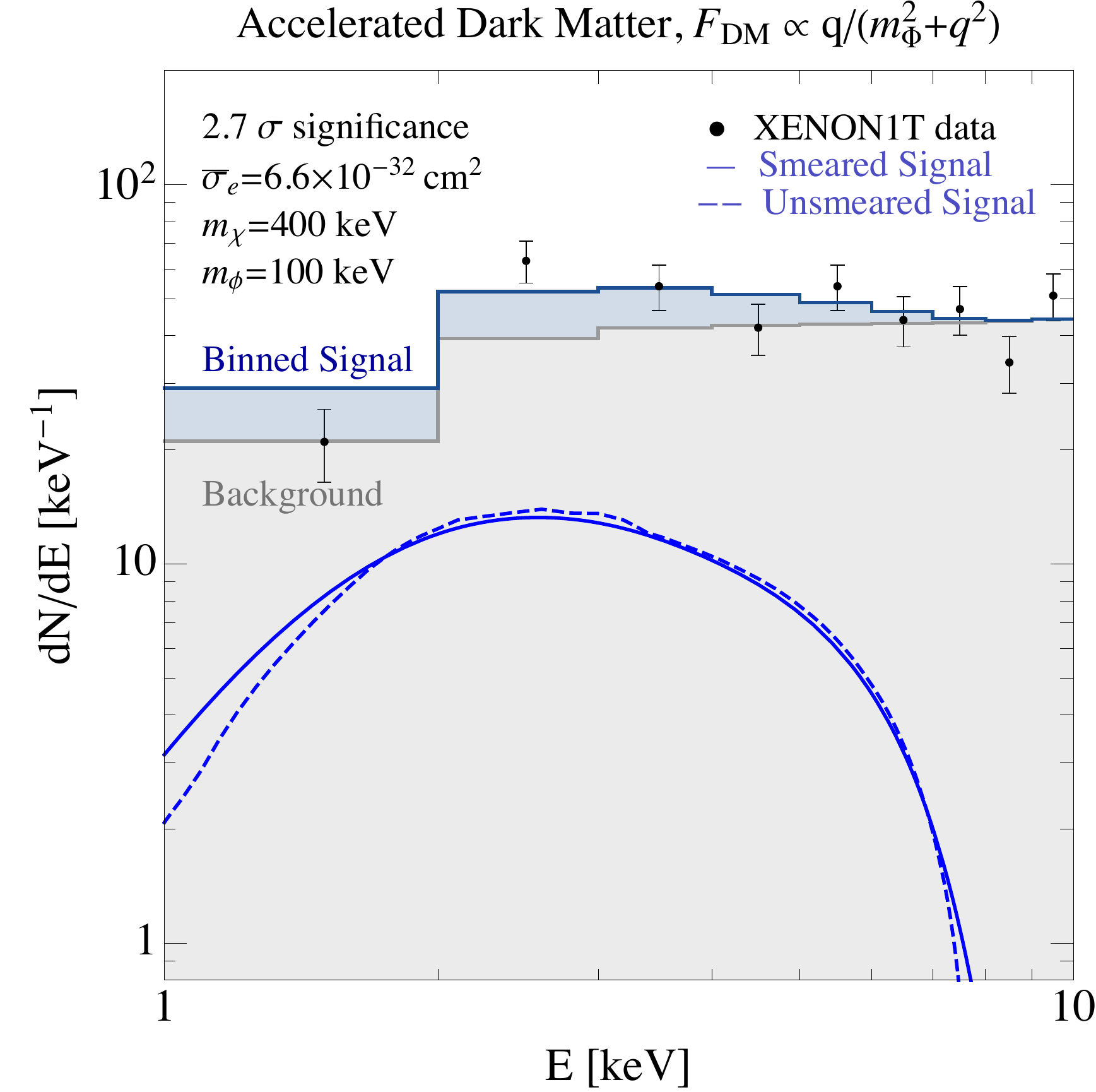}
\caption{
{\bf Left}: The $1\sigma$ and $2\sigma$ best-fit regions ({\bf red})  for cosmic-ray accelerated DM that interacts with electrons via a form factor $|F_{\rm DM}(q)|^2 \propto q^2/(q^2 + m_{\phi}^2)^2$.  Exclusion regions due to the \xe S2-only analysis are shown by the {\bf light green-shaded} region, while the limits from the Super-K data~\cite{Cappiello:2019qsw} are shown in {\bf darker shaded green}. Various complementary constraints strongly bound the light-mediator coupling to electrons. To illustrate this, in the {\bf middle} plot we show the best fit region for the mediator-electron coupling $y_e$ as a function of the mediator mass, $m_\phi$. The {\bf blue} region is excluded by HB cooling~\cite{Hardy:2016kme}, while in the {\bf orange} region the mediator thermalizes before electron decoupling, and may therefore suffer from (model-dependent)  limits on $N_{\rm eff}$.  The {\bf magenta} shaded regions is excluded by the ($g-2$) measurement of the electron~\cite{Liu:2016qwd,Liu:2016mqv}, and the region at higher $y_e$ is excluded by collider constraints from monophoton searches~\cite{Essig:2013vha,Lees:2014xha}. {\bf Right}: The signal spectral shape for the best fit point for this model. The {\bf black dots} are the \xe data, the {\bf gray shaded} region is the expected background, the {\bf blue solid} line is the signal shape after detector smearing, and the {\bf blue dotted} line is the signal before smearing.  The {\bf blue shaded} region is the resulting signal plus background distribution.
 } 
\label{fig:CRBestFit}
\end{figure}

The model discussed here has three independent parameters:  the DM mass, $m_\chi$, the mediator mass, $m_\phi$, and the DM-electron cross-section, $\bar\sigma_e$.   In Fig.~\ref{fig:CRDMflux}, we show the predicted accelerated DM flux (top) and expected electron spectra induced by the accelerated DM flux in xenon (bottom) for three different values of $m_\phi$ and for two different form factors, fixing $m_\chi = 1$~MeV and $\bar \sigma_e = 10^{-40}\units{cm^2}$: 
\begin{itemize}
\item The left panels of Fig.~\ref{fig:CRDMflux} with $|F_{\rm DM}(q)|^2 \propto q^2/(q^2 + m_{\phi}^2)^2$ corresponds to a scalar-pseudoscalar or vector-pseudovector interaction, where the spin-dependent interaction is on the DM side so that the mediator has scalar or vector coupling to the SM. As expected, for lighter mediator masses, the DM flux peaks  at lower energies. From the bottom left panel of Fig.~\ref{fig:CRDMflux}, we see that when choosing appropriate mediator masses, we can get a decreasing spectrum at lower energies below the mediator mass for $|F_{\rm DM}(q)|^2 \propto q^2/ m_{\phi}^4$, and at higher energies compared to the mediator mass, where the suppression $|F_{\rm DM}(q)|^2 \propto 1/q^2$ enhances the suppression from the atomic form factor.
\item The right panel of Fig.~\ref{fig:CRDMflux} with $|F_{\rm DM}(q)|^2 \propto 1/(q^2 + m_{\phi}^2)^2$ corresponds to a scalar-scalar or vector-vector interaction, which can be obtained by the exchange of a leptophilic scalar or a dark photon mixing with the SM photon. Here the predicted spectrum does not flatten out at low energies, and thus we find it to be excluded by the \xe S2-only analysis~\cite{Aprile:2019xxb}.  
\end{itemize} 

We perform a wide scan of the accelerated DM parameter space for different DM masses and mediator masses. The best fit point we find is 
\begin{equation}
m_{\chi}= 0.4 \text{ MeV}\,, \quad m_\phi/m_\chi=4\,, \quad \overline\sigma_e=6.6\times 10^{-32}~\text{ cm}^2\,,
\quad 2\log(\mathcal{L}_{S+B}/\mathcal{L}_{B})=12.1 \,, \label{eq:bestfit1over4}
\end{equation}
and in Fig.~\ref{fig:CRBestFit} (right) we show its expected signal. The dashed and solid blue lines show the unsmeared and detector-smeared spectra.  The gray  region is the expected binned background, while the blue shaded regions show the contribution of the binned signal.  
For simplicity, we fix $m_\phi=m_\chi/4$, and we explore the parameter space as a function of the DM mass. Our conclusions do not depend much on this choice.  In Fig.~\ref{fig:CRBestFit} (left), we show the corresponding $1\sigma$ and $2\sigma$ best-fit regions in red and the best fit point in Eq.~\eqref{eq:bestfit1over4} as a red star. In agreement with the previous discussion, we see how in this setup the best-fit region lies close to the boundary of the region excluded by the \xe S2-only analysis~\cite{Aprile:2019xxb}, and the Super-K experiment~\cite{Cappiello:2019qsw} gives an upper bound on the DM mass (which is related to the mediator mass). 

Since the mediator mass must be small enough for the spectrum to be suppressed at high energies, the relevant parameter space is subject to severe constraints that  we illustrate in the middle of Fig.~\ref{fig:CRBestFit}.  We plot the mediator coupling to electrons, $y_e$, as a function of the mediator mass, fixing $y_D$ to the maximal value allowed by unitarity.  We include stellar cooling constraints from HB stars~\cite{Hardy:2016kme},  the electron $g-2$~\cite{Liu:2016qwd,Liu:2016mqv} constraints, and a line that shows the coupling required for the mediator $\phi$ to thermalize before electron decoupling. This region is likely to be subject to (model-dependent) BBN constraints~\cite{Boehm:2013jpa}. At higher mediator masses, the required coupling to electrons is so high to be robustly excluded by direct production of the light mediator at colliders through $e^+e^-\to \gamma \phi$~\cite{Essig:2013vha,Lees:2014xha}. In summary, it seems that this explanation of the \xe excess is robustly excluded.  
 
Before closing, a remark is in order.  In deriving the limits above, we used non-relativistic form factors for the DM-electron interactions.  Corrections that arise in the relativistic limit can be found in~\cite{Roberts:2016xfw}.  To understand why it is justified to neglect relativistic corrections, we note that  a DM with mass around 1~MeV must be accelerated to velocities above $\sim0.03c$.  Using Eq.~\eqref{eq:vmin} and since $E_e \simeq \units{keV}$, one finds $q \gtrsim 30$ keV.   Since the atomic and DM form factors both dominate at low $q$, this justifies neglecting the relativistic corrections, which become important only at significantly higher values of $q$.

\section*{Acknowledgements}
We  thank T. Bringmann, R. Budnik, H. Kim, H. Liu, P. Meade, G. Perez, J. Pradler, G. Rossi, F. Sala,  Y. Soreq, L. Ubaldi, and N. Weiner for useful discussions. We also thank G.~Alonso-Alvarez, L.~Calibbi, F.~Ertas, J.~Huang, J.~Jaeckel, F.~Kahlhoefer, G.~Marquez-Tavares, P.~Panci, E.~Salvioni, M.~Szydagis, L.J.~Thormaehlen, K.~Van Tilburg for many useful feedbacks on the draft. IB is grateful for the support of the Alexander Zaks Scholarship, The Buchmann Scholarship, and the Azrieli Foundation. AC acknowledges support from
the ``Generalitat Valenciana'' (Spain) through the ``plan GenT'' program (CIDEGENT/2018/019), as well as national grants FPA2014-57816-P, FPA2017- 85985-P.
RE and MS are supported in part by DoE Grant DE-SC0017938 and Simons Investigator in Physics Award~623940.  
TV is supported by the Israel Science Foundation-NSFC (grant No.~2522/17), by the Binational Science Foundation (grant No.~2016153) and by the European Research Council (ERC) under the EU Horizon 2020 Programme (ERC-CoG-2015 - Proposal n. 682676 LDMThExp).

\vskip 5mm 
\noindent 
{\bf Note added:}  While this work was in progress,~\cite{Takahashi:2020bpq,Kannike:2020agf,Alonso-Alvarez:2020cdv,Boehm:2020ltd,Fornal:2020npv,Su:2020zny,Bally:2020yid,Harigaya:2020ckz,Du:2020ybt,Choi:2020udy,Chen:2020gcl,AristizabalSierra:2020edu,Bell:2020bes,Paz:2020pbc,DiLuzio:2020jjp,Buch:2020mrg,Dey:2020sai,Cao:2020bwd,Khan:2020vaf,Nakayama:2020ikz,Primulando:2020rdk,Lee:2020wmh} appeared.  A number of these papers have partial overlap with the study presented here.  Another model that could explain the excess is~\cite{Smirnov:2020zwf}.

\bibliographystyle{apsrev4-1.bst}
\bibliography{XenonDM.bib}
		
\end{document}